\documentclass[12pt]{article}
\usepackage{graphicx,wrapfig}
\usepackage{xcolor}
\usepackage{enumerate}
\usepackage{natbib}
\usepackage{url} 
\usepackage{float}
\usepackage{amsmath,amssymb,amsfonts,amsthm}
\usepackage{multirow}
\usepackage{array}
\usepackage{makecell}
\usepackage{enumitem}
\usepackage{soul}
\usepackage{rotating}
\usepackage{fancyvrb}
\usepackage[font=scriptsize,labelfont=bf,skip=5pt]{caption}
\usepackage[normalem]{ulem}
\usepackage{subcaption}
\usepackage{titlesec}

\newcolumntype{P}[1]{>{\centering\arraybackslash}p{#1}}

 \bibpunct[, ]{(}{)}{,}{a}{}{,}%

\renewcommand{\arraystretch}{0.65}

\titleformat{\section}
  {\normalfont\fontsize{14}{15}\bfseries}{\thesection}{1em}{}

\titleformat{\subsection}
  {\normalfont\fontsize{13}{14}\bfseries}{\thesubsection}{1em}{}  

\theoremstyle{definition}
\newtheorem{definition}{Definition}[section]

\DeclareMathOperator*{\argmin}{argmin}
\DeclareMathOperator{\Tr}{Tr}

\RequirePackage{fix-cm}

\begin{document}

\def\spacingset#1{\renewcommand{\baselinestretch}%
{#1}\small\normalsize} \spacingset{1}

\newcommand{\comment}[1]{{\textcolor{red}{(\emph{#1})}}}
\newcommand{\add}[1]{{\textcolor{red}{\emph{#1}}}}

\title{\bf Statistical Emulations of Human Operational Motions in Industrial Environments}
\author{Yanliang Chen\thanks{
The authors gratefully acknowledge \textit{the support from the National Science Foundation (grants \# 2132311, 2428742, 2413748) and the Air Force Office of Scientific Research (grant \# FA9550-23-1-0673).}}\hspace{.2cm}\\
{Dept of Statistics, Florida State University}\\
Anuj Srivastava \\
{Dept of Applied Math and Statistics, Johns Hopkins University}\\
Chiwoo Park \\
{Dept of Industrial and Systems Engineering, University of Washington}}
\maketitle

\bigskip
\begin{abstract}
This paper tackles the challenging problem of developing emulators for human operational motions in industrial workplaces. We represent human motion as time-indexed sequences of body shapes and formulate a statistical generative model for these shape sequences. The sequences are modeled as continuous-time stochastic processes on a Riemannian shape manifold. Key challenges include the manifold’s nonlinearity, variability in motion execution rates, the infinite-dimensional nature of the processes, and population-level variability across action classes. Deep learning methods are ineffective due to the small training samples typically available in this domain. To address these issues, we integrate a number of tools: temporal alignment via time warping, Riemannian geometry for handling nonlinearities, and shape- and functional-PCA for dimensionality reduction. A Gaussian model is then imposed on the reduced Euclidean spaces to emulate random motion sequences, which are then evaluated in representative industrial scenarios. We utilize a number of metrics to validate randomly generated shape sequences. 
\end{abstract}

\noindent%

\spacingset{1.8} 

\section{Introduction}\label{sec1}
\begin{figure}[htbp]
    \centering
    \subfloat[]{
    \includegraphics[width=0.18\textwidth]{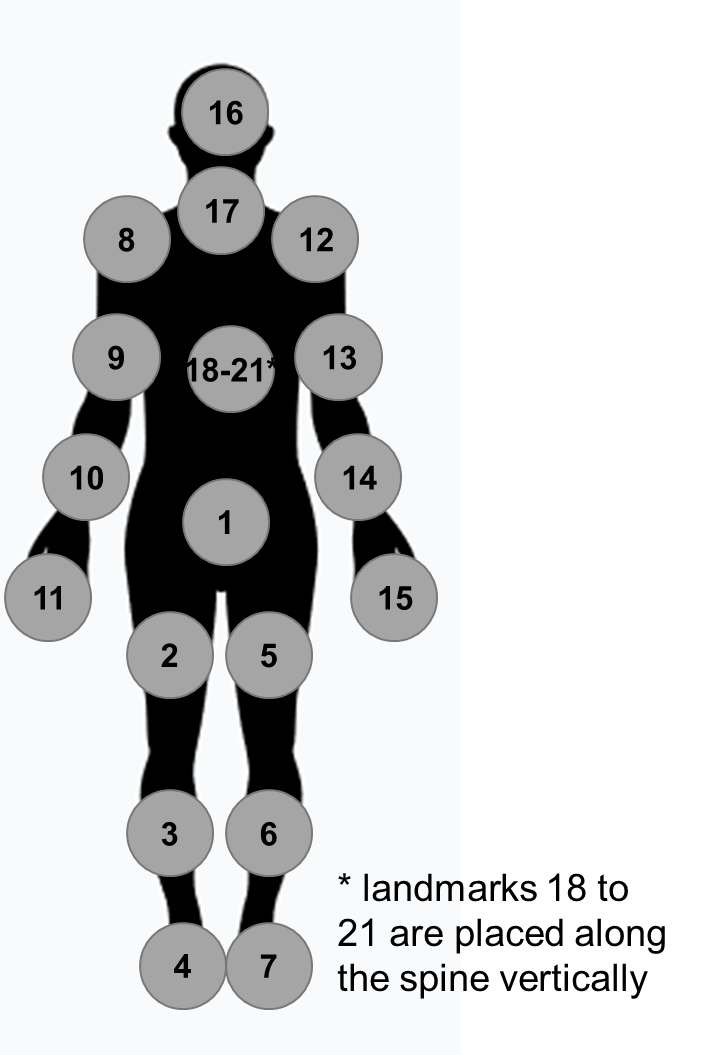}}
    \hspace{1em}
    \subfloat[]{\includegraphics[width=0.123\textwidth]{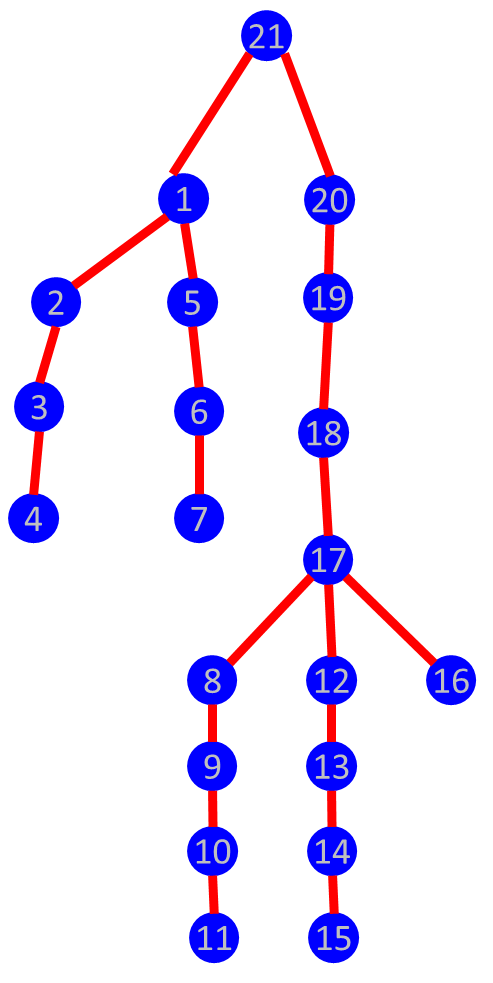}}          
    \subfloat[]{\includegraphics[width=0.65\textwidth]{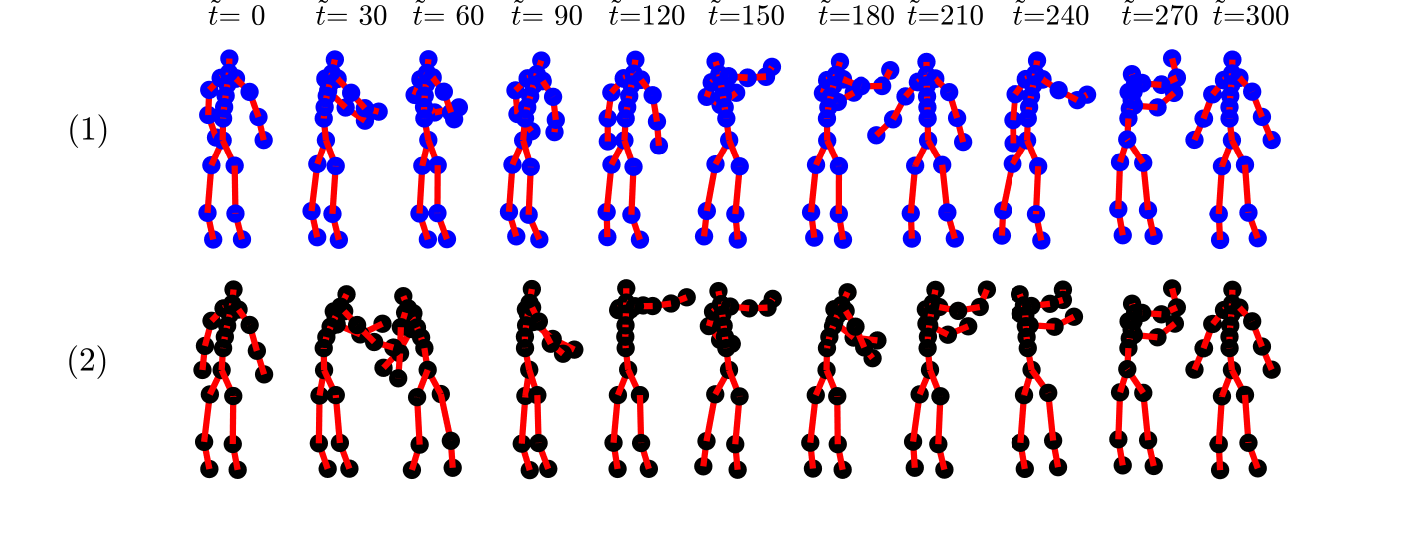}}
    \caption{Illustration of landmarks on the human body. (a) is the landmarks on a human body; (b) is the tree hierarchy of landmarks; (c) is two identical posture sequences,  $\alpha\circ\gamma_1$ and $\alpha\circ\gamma_2$, with different execution rates.}
    \label{fig:3-1}
\end{figure}

This paper presents statistical models and algorithms to simulate human operational motions using models trained from past data. The goal is to capture the statistical variations inherent in human motion data, providing an understanding of human motion~\citep {barnes1949motion}, and enabling the simulation of new data. 
Such simulators are crucial for creating digital twins of industrial operations in factories, hospitals, and other workplaces that involve human workers. While robots and machines have automated numerous tasks, most workplaces still rely significantly on human operators. Human operations often introduce substantial process variations due to different levels of skills, fatigue, errors, and  training~\citep{urgo2019human}. The operational uncertainties in manual operations are seldom quantified and integrated into digital twin models for operations planning and optimization~\citep{park2022data}. The proposed digital emulator would provide a tool to incorporate human-related process variations into a digital twin of systems involving human workers. In this paper, we utilize motion data from five classes of manufacturing operations from an industrial dataset~\citep{park2022data} to study the motion emulator. Fig.~\ref{fig:3-1}(c) shows two example sub-sequences from this dataset. 

There are two broad approaches to simulating human motions. One uses physical or biomechanical models that, in turn, rely on detailed dynamical systems with explicit physical constraints and input forces to predict or generate human motion~\citep{da2008simulation,xiang2012hybrid,park2019learning}. These physical approaches rely on our prior understanding of physical interactions, often in the form of constrained differential equations. While these solutions are realistic, the approaches are not easily generalizable. The second approach is the data-driven approach, which utilizes and analyzes past observations of human operators' motions. Several studies, including \citet{zhang2014arima,wang2018arima,xu2019human}, take this approach to address problems such as segmenting actions, modeling locomotion, or classifying actions. However, simulating long, realistic sequences of complex human postures, including several segments performing diverse tasks, remains a distant goal. A long sequence observation is a concatenation of several smaller actions performed by an agent. Machine learning techniques, especially those involving deep neural networks, are proficient in learning complex patterns from input data and reproducing them via generative models. However, these techniques typically require abundant data for successfully learning and generating complex observations. While there are several datasets for studying short-length human actions and activities, there is a paucity of longer datasets in the current problem context. 
We will study shape sequences as single processes, rather than splitting them into smaller sequences representing distinct actions.
\color{black}

In this paper, we develop a data-driven approach to learning, modeling, and emulating long sequences, {\it i.e.} lasting several seconds,  of human operational movements. Toward that goal, we utilize a shape representation of human posture introduced in \cite{park2022data}. As shown in Fig.~\ref{fig:3-1}, the device records several landmark locations on the human body. We then extract a posture or shape representation of these landmarks by discarding irrelevant information, such as translation, size, and rotation. Specifically, we define a shape space ${\cal Y}$, whose elements represent the shapes or postures of the human body; details are presented in Sec.~\ref {subsec3-1}. This shape space is a nonlinear, non-Euclidean space that is endowed with a Riemannian structure. Thus, we view the human motion as a time-indexed trajectory in this shape manifold. 
While the literature has frequently studied static (individual) shapes or short sequences of shapes, the analysis and modeling of longer shape sequences present significant challenges. One exception is ~\cite{LESA-Zhang:2022} that uses sparsely sampled data of human brain anatomical structures to estimate individual and population level shape evolutions over long time periods. However, this study only addressed the problem of densification and modeling of the sparse shape data, which is distinct from our goals of simulating and evaluating new sequences.

We are faced with multiple modeling choices to overcome technical challenges in representing long shape sequences in the shape space ${\cal Y}$. Viewing a human motion sequence as a stochastic process $X(t)$ taking values in ${\cal Y}$, one can either take a discrete-time approach by building a time-series model or a continuous-time approach by using a stochastic differential equation or a functional representation. Similarly, one can choose a parametric approach, {\it e.g.}, a Gaussian model,  or a non-parametric approach, where each sample path is treated as an unstructured function. For modeling a process in the nonlinear manifold ${\cal Y}$, there are two broad approaches: (1) modeling a shape sequence directly on ${\cal Y}$ using its intrinsic Riemannian geometry, and (2) approximately flattening ${\cal Y}$ into a Euclidean space (often a tangent space of ${\cal Y}$) that allows the use of traditional (Euclidean) statistical models. In addition to the shape variability in human postures, due to several human factors, the sequences contain temporal variability caused by the different work speeds of different operators. Accounting for temporal misalignments is known to  improve classification tasks ~\citep{Veeraraghavan2009timewarping,amor2015action} and should improve emulation performance.

Lastly, different combinations of these choices lead to different solutions and performances. One needs to evaluate these modeling choices and select the best model for our application. We will assess these models in multiple ways: using visualizations, statistical likelihoods, non-parametric hypothesis tests, physical plausibility, and customized ideas. We apply these ideas to a manufacturing and a workout dataset and present our findings. We also provide the strengths and limitations of different models. 

The main contributions of this paper are: It develops a comprehensive framework for modeling and simulating long-duration human motion sequences for use in industrial settings.
 It combines several flattening-based mathematical representations of shape sequences, including: (i) {\it Integrated, Single-Hop Transported Vector Field}, or IS-TVF, and (ii) {\it Single Inverse-Exponential Map}, or SIEM, with parametric modeling choices for efficient modeling.
 It utilizes multiple metrics to evaluate the performance of different modeling choices and the quality of the simulated motion sequences. In summary, it provides a realistic modeling approach, a first one in the literature, to capture and simulate long, structured motion sequences corresponding to the shapes of human operators. In the context of data paucity for long human-motion sequences, our emulator can also serve as a piece in a larger AI pipeline, taking small datasets and generating plenty of simulated sequences for training AI engines.

\section{Scientific Questions, Problem Challenges, and Data Sets}\label{sec2}

In this problem context, some scientific questions of interest are: 
  (1)  Can the essential statistical structure of long human skeletal motion sequences be accurately learned from limited observations ($\sim$ 100s of sequences)?
(2) When only limited motion data are available, do analytically tractable, geometry-based statistical models provide more reliable and interpretable emulators than deep-learning frameworks?
(3) How should the variability in human motion be represented mathematically and modeled statistically?
(4) Can these emulators generate sequences with summaries of dynamical features, such as jerk and acceleration, that match real sequences? 
 (5) Can low-dimensional representations capture the dominant modes of motion variability?

\color{black}


There are four main challenges in answering these questions.
{\bf Nonlinearity of Shape Manifolds}: Each human posture resides in a non-Euclidean shape manifold. A specialized representation is required to preserve its underlying structure and ensure that the generated postures remain within the natural posture space.
  {\bf High-dimensionality in Space and Time}: The complexity of the shape manifold is further compounded by the high temporal dimensions of the motion sequences. With a limited sample size, capturing the joint dynamics of these high-dimensional motions becomes more difficult.
 {\bf Handling Execution Rate Variability}: The observed sequences represent activities that are executed at an arbitrary rate and one needs to remove this phase variability during modeling.  
  {\bf Complexity of Temporal Patterns in Long Sequences}: Unlike monotonic or repetitive actions, our datasets present highly non-linear and unstable dynamics dictated by the underlying instructions. Modeling these complex temporal patterns exceeds the capabilities of the standard temporal models.

\noindent {\bf Data Collection \& Description}: 
We utilize two motion-capture datasets. The first is the \textit{Worker Motion} dataset collected in an industrial manufacturing environment~\citep{park2022data}. It contains five operations performed at fixed workstations along a conveyor line, primarily involving upper-body motions such as tool handling and workpiece manipulation. For each operation, an experienced worker first demonstrated a standard procedure as a reference motion. Two additional workers then repeatedly performed the same operation, producing 60 sequences per operation and 300 sequences in total. Motion was recorded using the Perception Neuron Motion Capture\texttrademark{} system at $n=21$ body landmarks; Each shape is $3\times(21-1)=60$ dimensional. Each sequence lasts approximately 30 seconds and is originally sampled at $T=1000$ time points. To reduce redundancy and computational cost, we down-sample each sequence to $T=301$ time points.

The second dataset, \textit{Exercise Motion}, contains 99 full-body exercise sequences based on a 100-second segment of an equipment-free YouTube workout video.\footnote{\url{https://www.youtube.com/watch?v=7GkMHPe_OXw}} The routine consists of quiet jacks, squats, and alternative reverse lunges. Two participants followed the same routine, with each participant completing 50 recordings over five days within one week. After excluding one unusable recording, the dataset contains 99 sequences. Motion was captured using OptiTrack Motive: Body with participant-specific skeletons containing $n=41$ landmarks. Each sequence was recorded at 120 frames per second, yielding approximately $T=12{,}000$ time points, with minor variation in duration due to manual recording. For analysis, the sequences were down-sampled to $T=800$ time points and interpolated to a common skeleton with $n=21$ landmarks.

\color{black}

\begin{figure}[htbp]
\centering
    \includegraphics[width=0.8\textwidth]{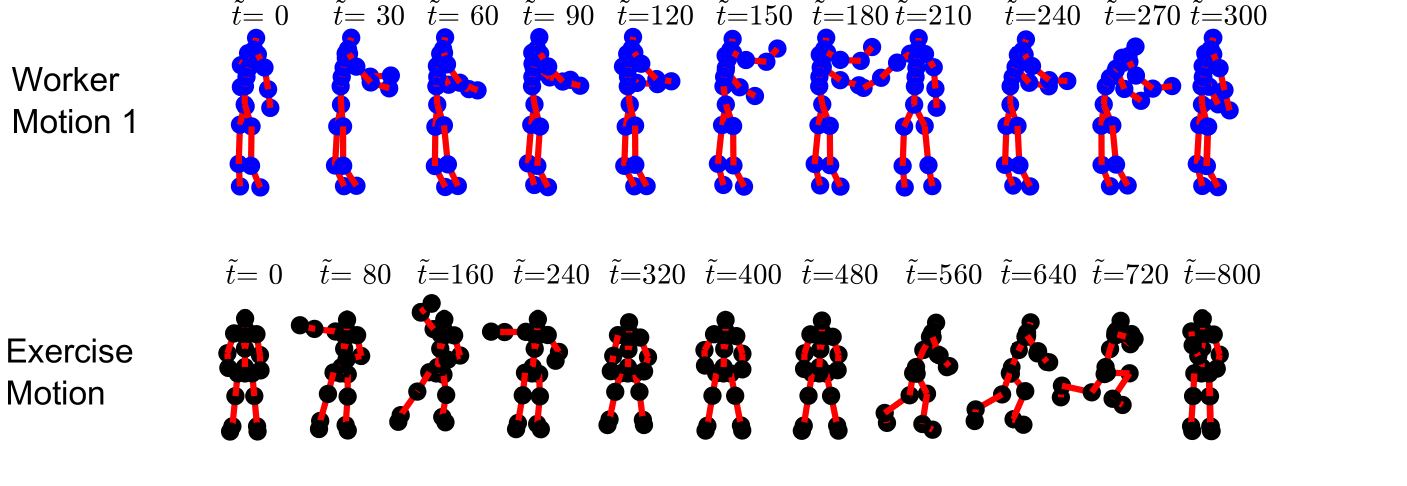}
    \caption{Examples of the skeleton postures of the Worker Motion dataset (top) and the Exercise Motion dataset (bottom).}
    \label{fig:2-1}
\end{figure}

\section{Related Works}\label{subsec1-1}

While most existing research focuses on short-term activities or single actions~\citep{zhang2014arima,wang2018arima}, we focus on methods that handle long sequences. Since the shapes of human postures naturally reside on a nonlinear manifold ${\cal Y}$, another focus is on statistical modeling and analysis on manifolds.

\noindent {\bf Shape-Sequence Representation Space}:  The shape space ${\cal Y}$ of human postures is a nonlinear, Riemannian manifold. A common intrinsic approach to modeling stochastic processes on ${\cal Y}$ is cross-sectional~\citep{sidenbladh2002prob} neglecting temporal correlations. The other common approach is to flatten the nonlinear shape space ${\cal Y}$ to a Euclidean space, either through a {\it Single Inverse Exponential Map} ~\citep[SIEM]{yi2012brownian} or using velocity vector fields \citep{zhang2015video, DENG-PR:2025}, and then apply traditional vector-space modeling. To reconcile velocity vectors at different locations,  \cite{su2014tsrvf} parallel transported them to a reference point in a single-hop and called them {\it Transported Square-Root Velocity Field} (TSRVF). These single-hop transports can significantly distort the data when the data are spread far from the reference point. Later, \cite{zhang2015video} addressed this distortion issue by utilizing a multi-hop parallel transport. We refer to them as {\it single-hop TSRVF} (S-TSRVF) and {\it multi-hop TSRVF (M-TSRVF)}, respectively. A multi-hop approach is prone to accumulating numerical errors that can become substantial for long sequences. \cite{DENG-PR:2025} used a multi-hop {\it transported velocity field} (M-TVF) (without the square-root transformation) for pre-aligned sequences. This representation is closely related to the {\it unrolling} and {\it unwrapping} ideas of \cite{jupp1987unrolling} and \cite{le2003unrolling}.

\noindent {\bf Temporal Alignments of Sequences}: Shape sequences can depend on the execution rates of actions being observed. Human operators working at different speeds introduce temporal misalignments in shape sequences. Several methods have been proposed for temporal alignments of sequences. \cite{raptis2013keyframe} used landmarks for alignment and \cite{Veeraraghavan2009timewarping} used dynamic time warping (DTW). \cite{su2014tsrvf,amor2015action,park2022data} used the TSRVF framework, which exploits the invariance properties of the $\mathbb{L}^2$ metric on the TSRVFs to time warping. The TSRVF approach has been extended to other manifolds and metrics \citep{celledoni2016lie, le2019discrete}.

\noindent {\bf Spatio-Temporal Dimension Reduction}:  Shape spaces are high-dimensional, and one needs to capture motion variability efficiently in a smaller space. A flattening approach leads to Euclidean spaces where dimension-reduction techniques such as linear principal component analysis (PCA) and nonlinear methods like autoencoder (AE)~\citep{Hinton2006AE} and variational autoencoder (VAE)~\citep{kingma2013vae} are available.  
The issue of high temporal dimensions (due to long sequences) remains. One can use techniques from fPCA (functional PCA) to control temporal dimensions. Alternatively, one can minimize spatio-temporal dimension jointly using tensor PCA (see~\cite{Lu2008MPCA}). \cite{LESA-Zhang:2022} utilizes a combination of spatial and temporal PCA to reduce representation space and to perform statistical inferences from given shape data. 

\noindent {\bf Stochastic Models for Motion}: After flattening ${\cal Y}$ into a vector space, several ideas for stochastic process modeling become available. First, we can consider a discrete-time model. A classical model for (discrete) time-series data is the autoregressive model. ~\cite{zhang2014arima} and ~\cite{wang2018arima} used the {\it autoregressive integrated moving average} (ARIMA) model to predict the movement of certain body parts. \cite{zhang2015video,DENG-PR:2025} used the {\it vector autoregressive} (VAR) model on the flattened 2D shape sequences. One can use a more general Markov chain designed to capture human motions, such as an implicit probabilistic model~\citep{sidenbladh2002prob} and the {\it hidden Markov model} (HMM)~\citep{Lehrmann2014nonlinearhmm}. In addition to traditional statistical models, deep neural networks, such as {\it recurrent neural networks} (RNNs), provide an alternative approach for predicting and simulating human motion~\citep {guo2020rnn,pavllo2020modeling}. While effective for predicting short-term repetitive movements, these time-series approaches struggle to simulate extended sequences of several distinct actions. Furthermore, the use of time-series models on nonlinear manifolds, especially for long sequences, remains undeveloped. 

Human motion can also be viewed as a continuous-time process on ${\cal Y}$. From this perspective, \cite{amor2015action} applied the TSRVF framework to classify and analyze the motion sequences.
Some studies have used diffusion models~\citep{rasul2021autoregressive,bilovs2023modeling}, adding random noise to the observed functions and employing autoregressive recurrent neural networks for denoising. Stochastic models such as Brownian motion are also used for motion analysis~\citep{yi2012brownian, lyu2021learning}. These methods have primarily been applied to classification rather than simulations of new sequences. 

\section{Background: Shape Representation and Sequence Registration}\label{sec3}

We view human motion as a time-indexed sequence of human postures. A posture, in turn, is a set of 3D coordinates of fixed, pre-determined landmarks on the human body. In such motion analysis, one considers variables such as locations, body sizes, and execution rates as nuisance variables. Therefore, we need data analysis to be invariant to these variables. 

\subsection{Skeletal Shape or Posture Representation}\label{subsec3-1}
Let $\mathbf{X}\in\mathbb{R}^{n\times 3}$ be the matrix of 3D coordinates of $n$ pre-determined landmarks on the human body, as shown in Fig.~\ref{fig:3-1}(a). For any landmark $i$, the relative coordinates based on its parent landmark $j$ (Fig.~\ref{fig:3-1}(b)) can be defined as $\tilde{\mathbf{x}}_i=\mathbf{x}_i-\mathbf{x}_j$ where $\mathbf{x}_i\in\mathbb{R}^3$ is the coordinate of the $i$th landmark. The posture data $\mathbf{X}$ is the root location $\mathbf{x}_n$ and the relative coordinates $\tilde{\mathbf{X}}=(\tilde{\mathbf{x}}_1,\tilde{\mathbf{x}}_2,\ldots,\tilde{\mathbf{x}}_{n-1})$. We discard the root location $\mathbf{x}_n$ and normalize each relative coordinate $\tilde{\mathbf{x}}_i$ as $\mathbf{y}_i=\frac{\tilde{\mathbf{x}}_i}{\|\tilde{\mathbf{x}}_i\|}$, resulting in a final posture representation $\mathbf{Y}=(\mathbf{y}_1,\mathbf{y}_2,\ldots,\mathbf{y}_{n-1})$ that is invariant to locations, body size, and body ratio.

The space of all postures is
$\mathcal{Y}=\underbrace{\mathbb{S}^2\times\mathbb{S}^2\times\ldots\times\mathbb{S}^2}_{\text{$(n-1)$ times}}$, 
where $\mathbb{S}^2\in\mathbb{R}^3$ is a unit sphere. The distance between any two postures $\mathbf{Y},\mathbf{Z}\in\mathcal{Y}$ is defined as
$d_{\mathcal{Y}}(\mathbf{Y},\mathbf{Z})=\sum_{i=1}^{n-1}d_{\mathbb{S}^2}(\mathbf{y}_i,\mathbf{z}_i)$,
where $\mathbf{y}_i, \mathbf{z}_i\in\mathbb{S}^2$ are the labeled $i$th elements of $\mathbf{Y}$ and $\mathbf{Z}$, and $d_{\mathbb{S}^2}(\mathbf{y}_i,\mathbf{z}_i)=\cos^{-1}(\mathbf{y}_i^{\top}\mathbf{z}_i)$ is the Riemannian distance on $\mathbb{S}^2$. We can also define other geometry operations, such as the inverse exponential map $\exp^{-1}_{\mathbf{Y}}(\mathbf{Z})$ and the exponential map $\exp_{\mathbf{Y}}(\mathbf{V})$, element-wise on the shape space $\mathcal{Y}$. A detailed description is provided in the Supplementary Material Sec.~S1.

\subsection{Temporal Alignment Based on S-TSRVF}\label{subsec3-2}
Although each posture is invariant to location, body size, and body ratio, a posture sequence indexed by the execution time may have phase variability, {\it i.e.}, different execution rates of activities across different agents (\textit{e.g.}, Fig.~\ref{fig:3-1}(c) shows two identical posture sequences performed at different rates). When considering motion models, this variability becomes a nuisance and needs to be separated from the sequence data. In other words, we need to temporally register different sequences. 
In the following, we will assume that all sequences have similar starting and ending
postures. In general, one has to register even the end points of the sequence, but this issue does not arise here in the datasets used.

\color{black}
 
We can express the posture sequence as a smooth map on a normalized time interval $[0,1]$, $\alpha:[0,1]\rightarrow \mathcal{Y}$. Let $\mathcal{A}$ denote the space of all posture sequences, $\mathcal{A}=\{\alpha:[0,1]\rightarrow \mathcal{Y},\ \alpha~~\mbox{is smooth}\}$. The distance between two sequences $\alpha_1$ and $\alpha_2$ can be defined as 
\begin{equation}\label{eq:4}
    d_{\cal A}(\alpha_1,\alpha_2)=\int_{[0,1]}d_{\mathcal{Y}}(\alpha_1(t),\alpha_2(t))dt \approx \frac{1}{T}\sum_{t=1}^T d_{\mathcal{Y}}(\alpha_1(t),\alpha_2(t)).
\end{equation}
We can time warp a posture sequence $\alpha$ by composing a time warping $\gamma:[0,1]\rightarrow[0,1]$ according to $\alpha \mapsto \alpha\circ\gamma\in\mathcal{A}$. Let $\boldsymbol{\Gamma}$ denote the set of all time-warping functions, $\boldsymbol{\Gamma}=\{\gamma:[0,1]\rightarrow[0,1]|\gamma\text{ is a diffeomorphism, }\gamma(0)=0,\gamma(1)=1\}$. 
To register shape sequences,  we represent them by their transported square-root vector fields (TSRVF) with respect to a reference point $\mathbf{Y}_{R}\in\mathcal{Y}$. The TSRVF of a sequence $\alpha$ is given by $h_{\alpha}=\frac{\dot{\alpha}(t)_{\alpha(t)\rightarrow\mathbf{Y}_R}}{\sqrt{\|\dot{\alpha}(t)\|}} \in T_{\mathbf{Y}_R}(\mathcal{Y})$,
where $\dot{\alpha}(t)=\exp^{-1}_{\alpha(t)}(\alpha(t+1))\in T_{\alpha(t)}(\mathcal{Y})$ is the derivative of $\alpha$ at time $t$, $\|\cdot \|$ is the Euclidean norm on the tangent space $T_{\alpha(t)}(\mathcal{Y})$, and $\dot{\alpha}(t)_{\alpha(t)\rightarrow\mathbf{Y}_R}$ denotes the parallel transport of the shooting vector $\dot{\alpha}(t)$ from its original tangent space  $T_{\alpha(t)}(\mathcal{Y})$ to $T_{\mathbf{Y}_R}(\mathcal{Y})$ (along a geodesic in $\mathcal{Y}$). Due to a single transport used here (from $\alpha(t)$ to $\mathbf{Y}_{R}$), we call this the Single-Hop TSRVF or S-TSRVF. The S-TSRVF of a time warped shape sequence $\alpha\circ\gamma$ is $h_{\alpha\circ\gamma}=h_\alpha(\gamma(t))\sqrt{\dot{\gamma}(t)} $. 
The distance between any two S-TSRVFs $h_{\alpha_1}$ and $h_{\alpha_2}$ is defined as
$d_\mathcal{H}(h_{\alpha_1},h_{\alpha_2})=\sqrt{\int_0^1 \|h_{\alpha_1}(t)-h_{\alpha_2}(t)\|^2 dt}$, where $\|\cdot\|$ is the Euclidean norm. 
Then, we solve for the optimal time-warping using
$\gamma^* =\argmin_{\gamma\in\boldsymbol{\Gamma}}d_{\mathcal{H}}(h_{\alpha_1\circ\gamma},h_{\alpha_2})$~\citep{srivastava2016functional}. Given a reference sequence $\alpha_R$, all motion sequences $\{\alpha_m\}$ are aligned pairwise  to $\alpha_R$ by 
$\gamma^*_{\alpha_m}=\argmin_{\gamma\in\boldsymbol{\Gamma}}d_{\mathcal{H}}(h_{\alpha_m\circ\gamma},h_{\alpha_R})$, removing the phase variability from the data. From here onwards, we will assume that all sequences have been pre-registered. 

\section{Mathematical Representations and Stochastic Modeling}\label{sec4}
We develop a pipeline for processing, modeling, and synthesizing human motion sequences by flattening the representation space ${\cal Y}$ to a Euclidean space, applying dimension reduction (\textit{e.g.}, PCA), and fitting generative models on the resulting low-dimensional Euclidean representations. We evaluate the effectiveness of these models in several ways, including using shape metrics, visualizations, and non-parametric tests. 

\begin{figure}[htbp]
    \centering
    \includegraphics[width=1\textwidth]{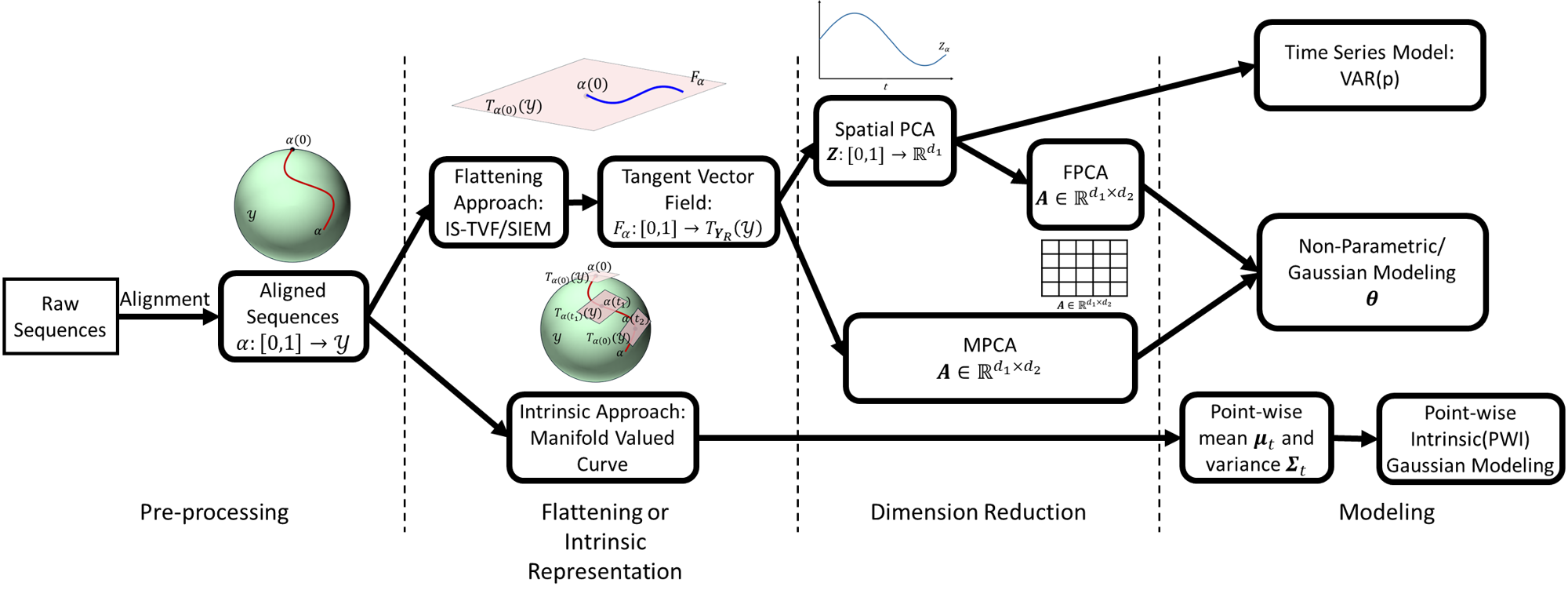}
    \caption{Pipeline for modeling motion sequences. The raw sequences are temporally aligned during the pre-processing, flattened into Euclidean time series, projected into small spaces using PCA, and then modeled using statistical models. One path follows a point-wise, intrinsic Gaussian (PWI) on the shape manifold directly. The pipeline is invertible: random quantities from statistical models on the right can be mapped back to full motion sequences.}
    \label{fig:4-1}
\end{figure}

As shown in Fig.~\ref{fig:4-1}, we pre-align the motion sequences using S-TSRVFs and discretize them to length $T$. The resulting sequences are now invariant to location, body scale, and execution rate. However, these sequences lie on a non-linear manifold $\mathcal{Y}^T$. The next step is to either map the sequences to a Euclidean space, followed by dimension reduction (sequential or joint PCA) and VAR/Gaussian modeling, or model them as $\mathcal{Y}^T$-valued curves using a point-wise Gaussian model. Finally, we generate random samples from these stochastic models and reconstruct new shape sequences. We describe these components next.

\subsection{Mathematical Representation}\label{subsec4-1}

This section discusses three representations of shape sequences: two that globally flatten ${\mathcal{Y}}$ into a Euclidean space and one that locally flattens ${\mathcal{Y}}$.

\noindent 1. {\bf Global Flattening Approach 1 --  Integrated Single-Hop Transported Vector Field (IS-TVF)}: 
We introduce a novel representation called {\it integrated single-hop transported vector field} (IS-TVF). This is inspired by the transported velocity field (TVF), which parallel transports the shooting vector along the path to a reference point. There are two transport schemes: (1) a single hop from the current point to a reference point and (2) multiple hops along the path from the starting point to a reference point. The multi-hop method accumulates significant numerical errors as the number of parallel transports to a reference point increases for longer sequences. In contrast, a single-hop transport is numerically more accurate, but may introduce distortions due to inverse exponential mapping.

\begin{figure}[htbp]
    \centering
    \subfloat[]{
    \includegraphics[width=0.4\textwidth]{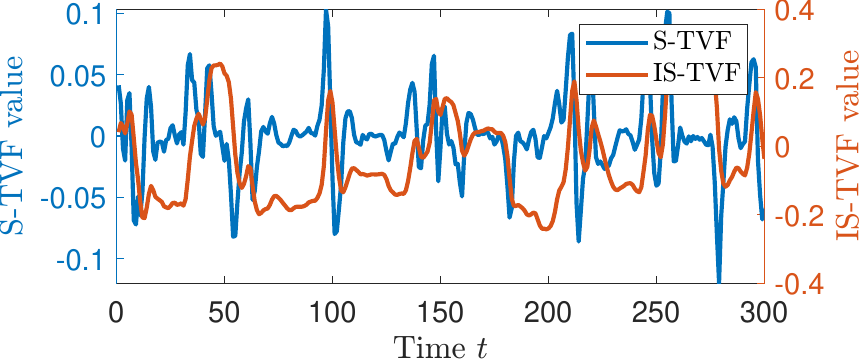}}   
    \subfloat[]{        \includegraphics[width=0.4\textwidth]{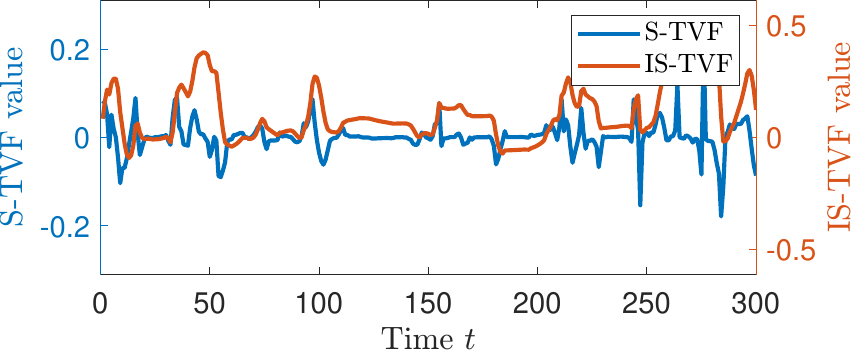}}
    \caption{Examples of S-TVF (blue lines) and IS-TVF (red lines) of two motion sequences.}
    \label{fig:4-3}   
\end{figure}

For a sequence $\alpha$, the single-hop transported velocity field (S-TVF) is defined as $\mathbf{F}_\alpha: [0,1] \to T_{\mathbf{Y}_R}(\mathcal{Y})$ with
$    \mathbf{F}_\alpha(t)=\dot{\alpha}(t)_{\alpha(t)\rightarrow\mathbf{Y}_R}\in T_{\mathbf{Y}_R}(\mathcal{Y})$.
A multiple-hop transported velocity field (M-TVF) is defined similarly, but with the key difference that it successively applies parallel transports of the shooting vector $\dot{\alpha}(t)$, from $T_{\alpha(t)}(\mathcal{Y})$ to $T_{\alpha(t-1)}(\mathcal{Y})$, then to $T_{\alpha(t-2)}(\mathcal{Y})$, and so on until we reach $T_{\alpha(0)}(\mathcal{Y})$. 
For a fixed reference $\mathbf{Y}_{R}$, the pair $(\alpha(0), \mathbf{F}_\alpha)$ is bijective to $\alpha$, which means that the posture sequence could be reconstructed by the exponential map:
$\alpha(t)=\exp_{\alpha(t-1)}(\mathbf{F}_\alpha(t)_{\mathbf{Y}_R\rightarrow\alpha(t-1)})$, $t=1,\dots,T-1$. By mapping all the sequences to the shared tangent space $T_{\mathbf{Y}_R}(\mathcal{Y})$, the S-TVF $\mathbf{F}_\alpha$ initially lies in  $\mathbb{R}^{3(n-1)\times (T-1)}$ (reflecting $T-1$ shooting vectors for $T$ time points and $n$ landmarks on the body). By projecting onto a fixed orthonormal basis, we simplify the representation to $\mathbf{F}_{\alpha_m} \in\mathbb{R}^{2(n-1)\times (T-1)}$. 
A practical challenge in using $\{\mathbf{F}_{\alpha_m}\}$ for analysis is that they represent velocities of the motion, which could be noisy and sparse (the sparsity implies that the velocity data tend to concentrate near zero, reflecting the intermittent and balanced nature of human motion), as illustrated in Fig. \ref{fig:4-3}(a). To reach a denser, smoother representation, we utilize the integral of the S-TVF in practice; call it the {\it integrated} S-TVF (IS-TVF).

\begin{figure}[htb]
    \centering
    \begin{minipage}{0.28\textwidth}           
    \subfloat[]{\includegraphics[width=\textwidth]{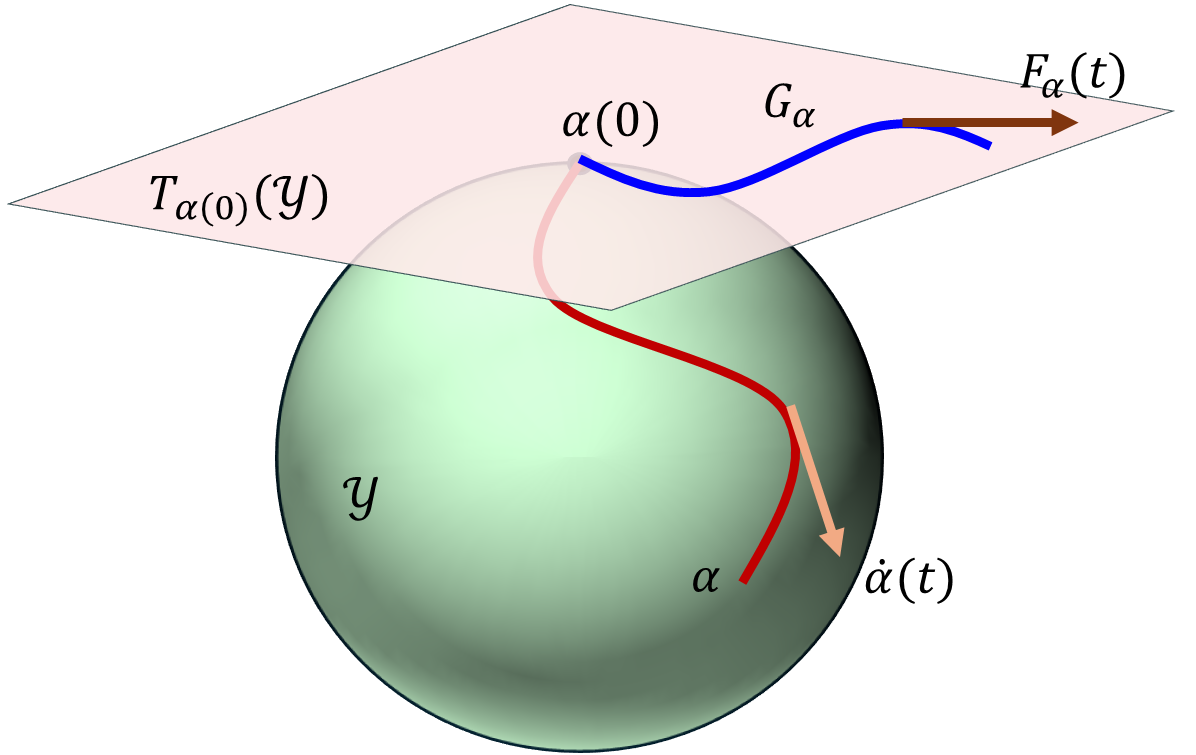}}   
    \end{minipage}
    \hfill
    \begin{minipage}{0.28\textwidth}
    \subfloat[]{\includegraphics[width=\textwidth]{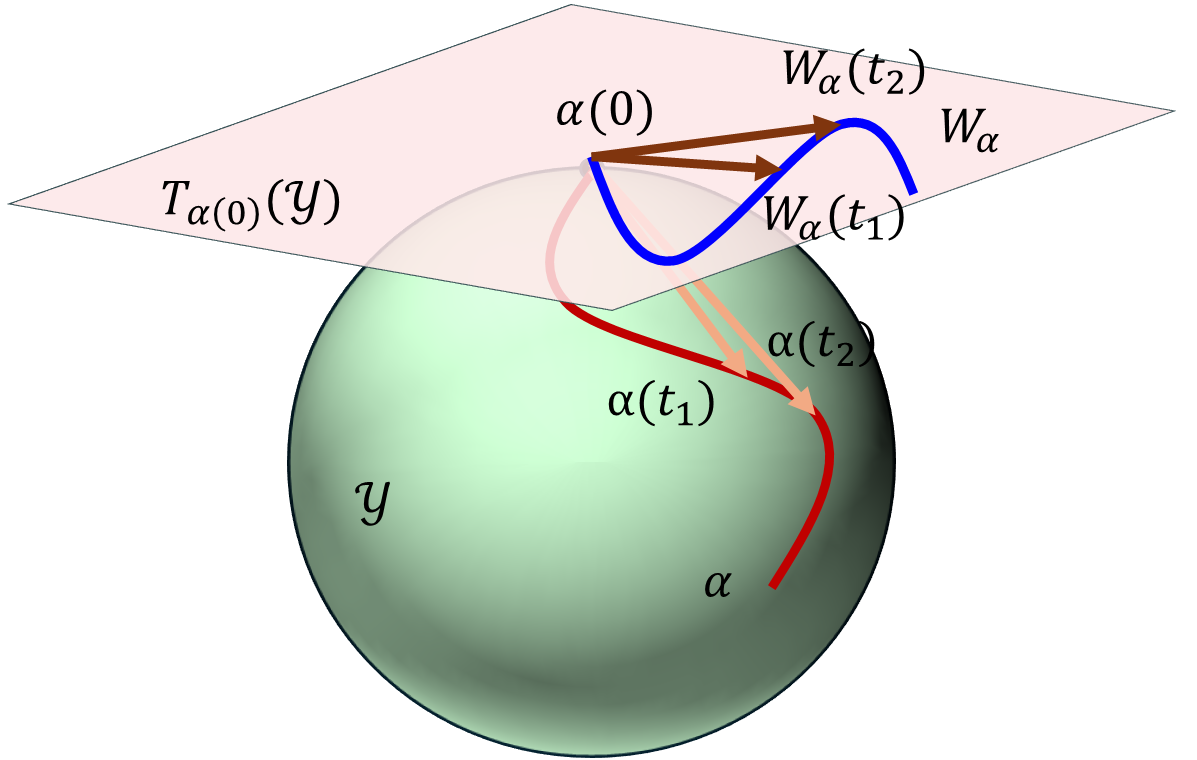}}
    \end{minipage}
    \hfill
    \begin{minipage}{0.28\textwidth}
    \hspace{2em}
    \subfloat[]{\includegraphics[width=0.6\textwidth]{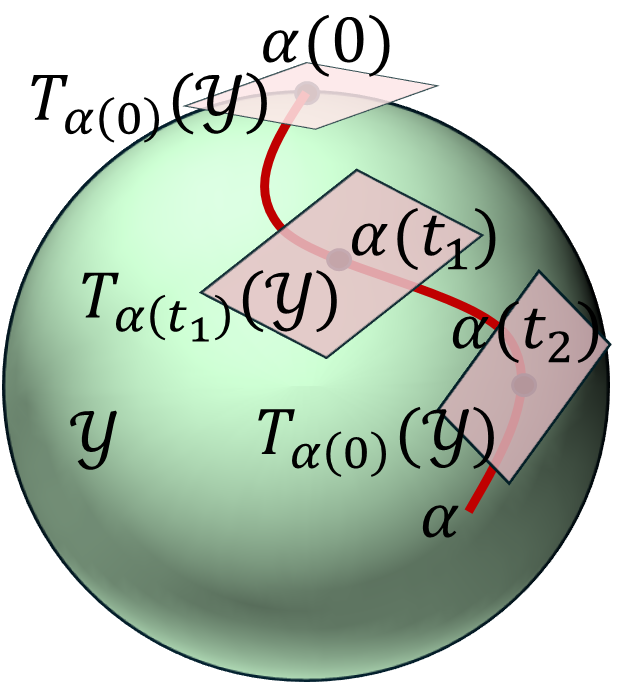}} 
    \end{minipage}
    
    \begin{minipage}{0.5\textwidth}
        \subfloat[]{
        \includegraphics[width=\textwidth]{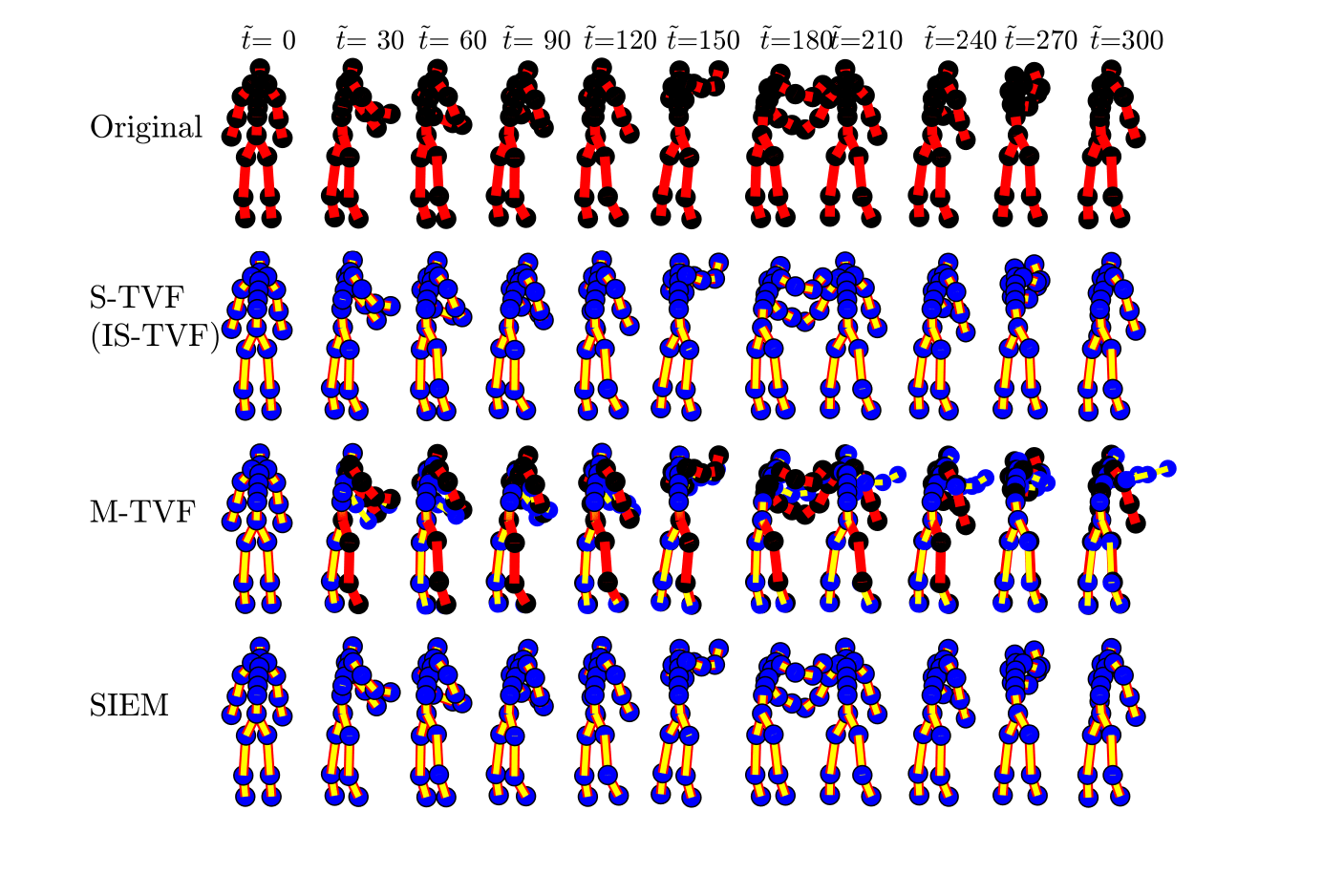}}  
    \end{minipage}
    \hspace{1em}
    \begin{minipage}{0.26\textwidth}
        \subfloat[]{
        \includegraphics[width=\textwidth]{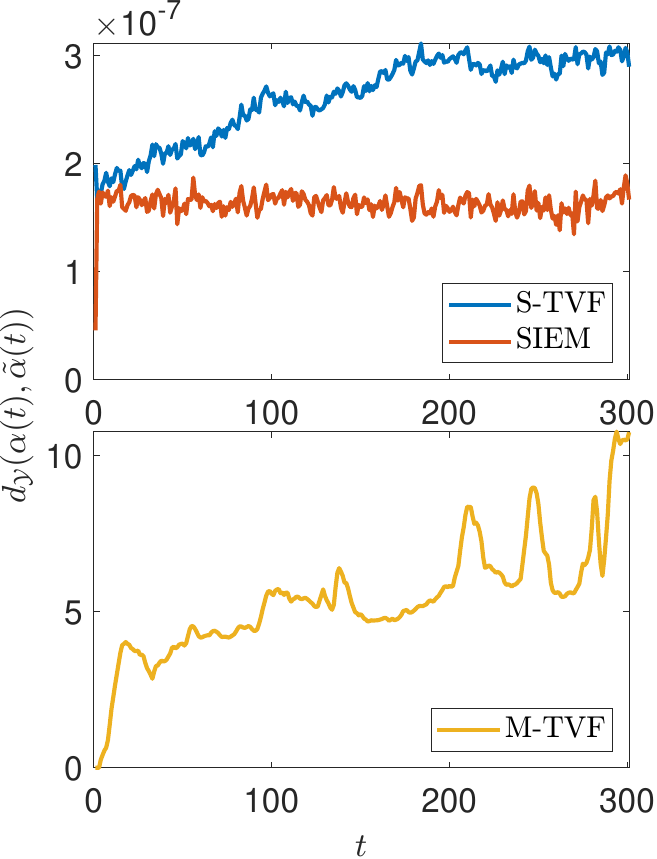}}
    \end{minipage}  
    
    \caption{(a): Cartoon illustration of IS-TVF, (b) SIEM, and (c) PWI. For IS-TVF, the velocity vectors at each time are transported to the tangent space of the reference point and then integrated to become trajectories in the tangent space. For SIEM, every point is directly mapped into the tangent space of a reference point. For PWI, the postures are analyzed in the individual tangent space of $T_{\mathcal{Y}}(\boldsymbol{\mu}(t))$ at each time $t$.
    (d): Reconstruction examples under different flattenings. The black and red skeletons represent actual observations, and the blue and yellow skeletons represent reconstructed postures. (e): The reconstruction error, $e_{\mathcal{Y}}(t) = d_{\mathcal{Y}}(\alpha(t),\tilde{\alpha}(t))$, of one sequence as an example, over time using S-TVF, SIEM, and M-TVF.}
    \label{fig:4-2}   
\end{figure}

\begin{definition} (Integrated Single-Hop TVF)
The integrated single-hop TVF (IS-TVF) is defined as $\mathbf{G}_\alpha: [0,1] \to T_{\mathbf{Y}_R}(\mathcal{Y})$, where $\mathbf{G}_\alpha(t)=\int_0^t\mathbf{F}_\alpha(s) ds \approx \frac{1}{T-1}\sum_{s=1}^t \mathbf{F}_\alpha(s)$. 
\end{definition}
\noindent The IS-TVF represents positions rather than velocities of the motion sequence. Fig.~\ref{fig:4-2}(a) shows a cartoon illustration of the IS-TVF. The shooting vector $\dot{\alpha}(t)$ at time $t$ is parallel transported to the tangent space of the reference point $\mathbf{Y}_R$ along a geodesic and then integrated to become a smooth curve on the tangent space. We can map IS-TVF back to S-TVF using finite differences. Both S-TVF and IS-TVF are $2(n-1)$-dimensional sequences with length $(T-1)$, {\it i.e.}, $\mathbf{F}_\alpha,\mathbf{G}_\alpha \in \mathbb{R}^{2(n-1)\times(T-1)}$. 
Fig.~\ref{fig:4-3} presents the S-TVFs and IS-TVFs of two example motion sequences. The S-TVF functions (in blue lines) appear sparse and noisy, while IS-TVF functions (in red lines) are much smoother due to integration. 

Given the set $(\mathbf{Y}_{R},\alpha(0), \mathbf{G}_\alpha)$, we reconstruct a posture sequence using partial sums and parallel transports. We will denote the reconstructed sequence by $\tilde\alpha$. 
Fig.~\ref{fig:4-2}(d) shows examples of reconstructions of a shape sequence from different flattened representations. The top row shows the original sequence, and the next three rows are reconstructions from S-TVF, M-TVF, and SIEM (Single Inverse Exponential Map). In these rows, the reconstructions (blue) are overlaid on the original sequences (black) to highlight the differences. The M-TVF approach accumulates more distortion along the process. Since IS-TVF and S-TVF curves overlap perfectly in this experiment, we have not drawn the former.

\noindent 2. {\bf Global Flattening Approach 2 --  Single Inverse Exponential Map (SIEM)}: Instead of using the shooting vectors $\dot{\alpha}(t)$, the posture at time $t$, $\alpha(t)$, can itself be mapped to the tangent space of a reference posture $\mathbf{Y}_R$ by a single inverse exponential map (SIEM). 
\begin{definition} (Single Inverse Exponential Map)
The SIEM is defined as $\mathbf{W}_\alpha(t)=\exp_{\mathbf{Y}_R}^{-1}(\alpha(t))$, 
where $\alpha(t)$ is the human posture at time $t$ and $\exp_{\mathbf{Y}_R}^{-1}(\cdot)$ is the inverse exponential map at the reference posture $\mathbf{Y}_R$.
\end{definition}
\noindent SIEM is a bijective map locally, and we can invert it using the exponential map
$\alpha(t)=\exp_{\mathbf{Y}_R}(\mathbf{W}_\alpha(t))$. Similar to IS-TVF, $\mathbf{W}_{\alpha(t)}$ lies on the tangent space $T_{\mathbf{Y}_R}(\mathcal{Y})$ at the reference point $\mathbf{Y}_R$, $\mathbf{W}_{\alpha}\in\mathbb{R}^{2(n-1)\times T}$.
$\mathbf{W}_\alpha$ is computationally simpler than the IS-TVF since it avoids computing velocity and integral. However, under SIEM, the distance between any two postures is not well preserved when the postures are far away from the reference posture. For example, Fig.~\ref{fig:4-2}(b) shows an illustration of the SIEM map, where the manifold distance between two points, say $\alpha(t_1)$, $\alpha(t_2)\in\mathcal{Y}$, $d_{\mathcal{Y}}(\alpha(t_1),\alpha(t_2))$, is different from the Euclidean distance of the corresponding points, {\it i.e.}. $\|\mathbf{W}_\alpha(t_1)- \mathbf{W}_\alpha(t_2)\|$, on the tangent space $T_{\alpha(0)}(\mathcal{Y})$. In contrast, the IS-TVF approach uses the velocity vectors $\dot\alpha(t)$ at each point (Fig.~\ref{fig:4-2}(a)), which better preserves distances between points since parallel transport preserves the norm of the velocity vector $\dot\alpha(t)$. Fig.~\ref{fig:4-2}(d) shows examples of the reconstruction. Similar to IS-TVF, the reconstructed sequences are practically identical to the original sequence. Fig.~\ref{fig:4-2}(e) shows the plots of reconstruction errors at each time to quantify the differences. The reconstruction error is defined as $e_{\mathcal{Y}}=d_{\mathcal{Y}}(\alpha(t),\tilde\alpha(t))$, where $d_{\mathcal{Y}}$ is the posture distance and $\alpha(t)$, $\tilde\alpha(t)$ are the original and reconstructed sequences at time $t$. S-TVF and SIEM use single-hop flattening and have negligible reconstruction errors ($\sim 10^{-7}$), whereas the errors in M-TVF reconstruction accumulate significantly ($\sim 10$).

\noindent 3. {\bf Pointwise Intrinsic (PWI) Process or Local Flattening}: The posture sequence can also be studied as an intrinsic $\mathcal{Y}$-valued process but with considerable challenges. Given that the sequences are already temporally aligned, we will explore a simple white noise model on $\alpha$,  i.e., choose an independent model at each time $t$. Fig.~\ref{fig:4-2}(c) shows an illustration of this approach termed {\it Pointwise Intrinsic} (PWI) process. Here we model data locally on tangent spaces $T_{\hat{\boldsymbol{\mu}}_{\mathcal{Y}}(t)}(\mathcal{Y})$, where the mean posture $\hat{\boldsymbol{\mu}}_{\mathcal{Y}}(t)=\argmin_{\mathbf{Y}\in\mathcal{Y}}\sum_{m=1}^M d_{\mathcal{Y}}(\alpha_m(t),\mathbf{Y})^2$ and $d_{\mathcal{Y}}$ is the posture distance (see Sec.~\ref{subsec3-1}). Since the postures at time $t$ are often close to the mean $\hat{\boldsymbol{\mu}}_{\mathcal{Y}}$, the distortion due to local flattening is smaller than earlier global methods. The downside is that it neglects the temporal relationship between points on the trajectory, {\it i.e.}, $\alpha(t_1)$ and $\alpha(t_2)$ are assumed independent. To incorporate spatial and temporal correlations in a $\mathcal{Y}$-valued process is considerably challenging and left for the future. 

\subsection{Space and Time Dimension Reduction}\label{subsec4-2}

Both IS-TVF ($\mathbf{G}_\alpha\in\mathbb{R}^{2(n-1)\times(T-1)}$) and SIEM ($ \mathbf{W}_\alpha\in\mathbb{R}^{2(n-1)\times T}$) are high-dimensional time series. In the dataset used here, there are $n=21$ landmarks in a posture and a length of $T=301$ for each sequence. Thus, an IS-TVF function has a total dimension of $12,000$. We explore two different PCA approaches to reduce the data dimension to a manageable size. 

\noindent {\bf Sequential Spatial and Temporal PCA}: 
In this approach, we first apply spatial PCA to reduce the spatial dimension from $2(n-1)$ to $d_1$, resulting in a vector-valued function $\mathbf{H}_{\alpha_m}\in\mathbb{R}^{d_1\times (T-1)}$. Then, we perform functional PCA on each of the $d_1$ components of  $\mathbf{H}_{\alpha_m}$ to reduce the temporal dimension from $(T-1)$ to $d_2$. This yields a matrix of the coefficient $\boldsymbol{\Phi}_m\in\mathbb{R}^{d_1\times d_2}$. A description of this process can be found in the Supplementary Material Sec.~S2. For spatial reduction, one can also use other nonlinear methods such as {\it autoencoder} (AE) and {\it variational autoencoder} (VAE). The performance comparisons between shape PCA and AE/VAE for spatial reduction are presented later.

\begin{figure}[htbp]
\centering
    \begin{minipage}{0.24\textwidth}
        \subfloat[]{
        \includegraphics[width=\textwidth]{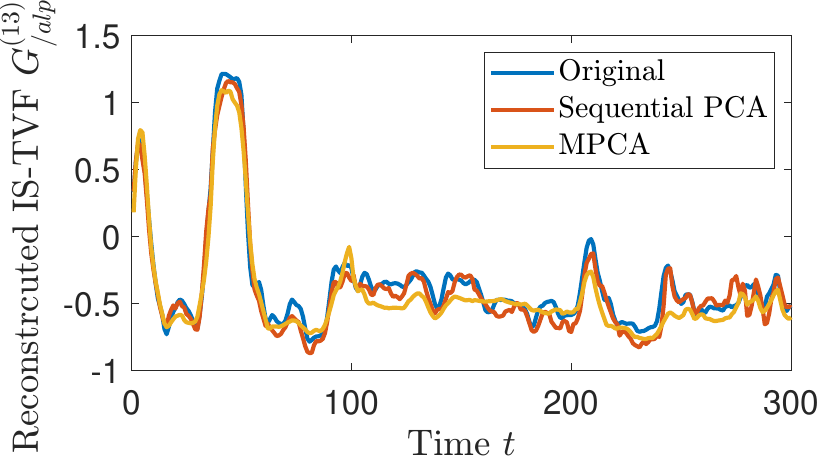}}
    
        \subfloat[]{
        \includegraphics[width=0.95\textwidth]{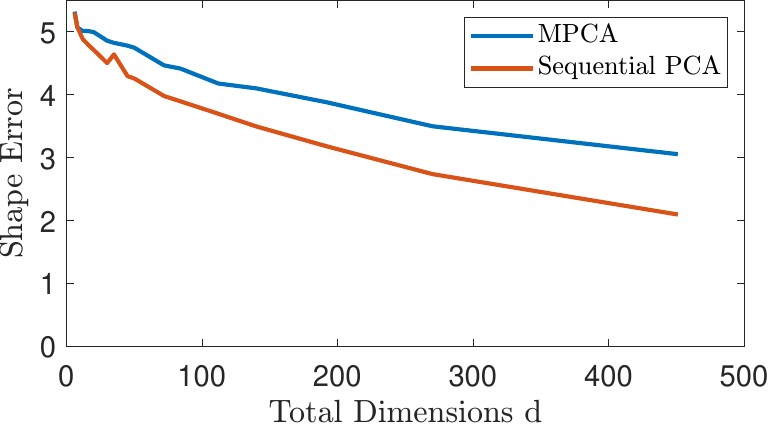}}
    \end{minipage}%
    \begin{minipage}{0.6\textwidth}
        \subfloat[]{\includegraphics[width=\textwidth]{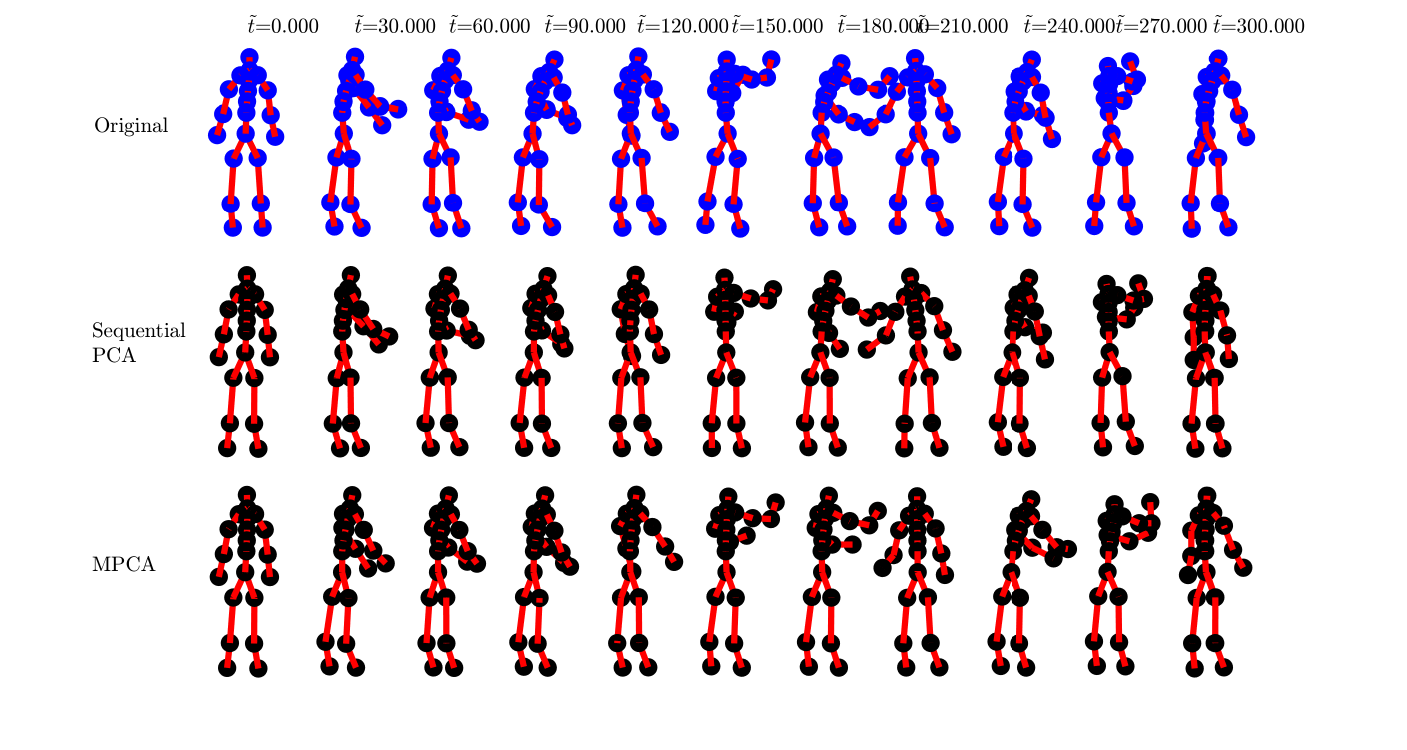}}
    \end{minipage}
    
    \caption{Comparison of the MPCA and sequential PCA. (a): an example of one of the IS-TVF elements reconstructed by MPCA and sequential PCA. (b): the reconstruction error versus the total reduced dimensions. (c): the comparison between the reconstruction of sequential PCA and MPCA with the original sequence as a reference.}
    \label{fig:4-11}   
\end{figure}

\noindent {\bf Joint Dimension Reduction}:
A more comprehensive albeit complex approach for dimension reduction is to reduce spatial and temporal dimensions simultaneously rather than sequentially. We apply the multilinear principal component analysis (MPCA) framework~\citep{Lu2008MPCA}. One can refer to \cite{Lu2008MPCA}'s paper for a detailed optimization algorithm. As a result, we can obtain feature tensors $\mathcal{Z}_{\alpha_m}\in\mathbb{R}^{d_1}\bigotimes\mathbb{R}^{d_2}$, which is similar to the coefficient matrix $\boldsymbol{\Phi}_m\in\mathbb{R}^{d_1\times d_2}$ obtained by the sequential approach. Fig.~\ref{fig:4-11}(a) shows a reconstructed IS-TVF element $\tilde{\mathbf{G}}_{\alpha_m}^{(i)}$ and Fig.~\ref{fig:4-11}(c) shows the reconstructed sequences $\tilde{\alpha}_m$.

\noindent {\bf Dimension-Reduction Performance Comparisons}:
We use the $85\%$ of the total variance as the criterion for selecting $d_1$ and $d_2$. In the sequential approach, we obtain $d_1=5$ to retain $90\%$ spatial variance and $d_2=10$ (for each spatial PCA dimension) to cover $95\%$ temporal variance. This results in $d_1 \times d_2 = 50$ total components. The joint PCA leads to $20$ components (as a tensor matrix $\mathcal{Z}_m\in\mathbb{R}^{5\times 4}$). Thus, MPCA is more effective than sequential PCA in reducing dimensions for the same retained variance. 
The reconstructed IS-TVF functions and posture sequences (shown in Fig.~\ref{fig:4-11}(c)) are visually similar for both methods. 
We also compare their reconstruction errors defined as  $e_{\mathcal{A}}=d_{\mathcal{A}}(\alpha_m,\tilde{\alpha}_m)$ (see Eqn.~\ref{eq:4} for $d_{\cal A}$) and $\alpha_m, \tilde{\alpha}_m$ are the original and the reconstructed sequences, respectively. The average reconstruction error over $60$ sequences of the motion class one decreases faster using the sequential PCA approach as the PCA dimension increases, as shown in Fig.~\ref{fig:4-11}(b). 
As for MPCA, certain essential details  ({\it e.g.}, in $\mathbf{G}_{\alpha_1}^{(1)}$, $\mathbf{G}_{\alpha_2}^{(1)}$) are lost, leading to poor reconstructions. In addition, we include the combination of AE/VAE (for spatial reduction) and FPCA (for temporal reduction) with the similar setting of $d_1=5$ and $d_2=10$ in this comparison. Tab.~\ref{tab:10} shows the average reconstruction errors using different methods. Although a nonlinear approach like VAE yields a good result, the linear spatial PCA is simpler. Thus, we use the sequential approach in the rest of this paper. 

\subsection{Statistical Modeling and Emulating Human Motion}\label{subsec4-3}
Having reached reduced Euclidean representations of motion sequences, we proceed to model them statistically to simulate new sequences. 

\begin{figure}[htbp]
    \begin{minipage}[t]{.35\linewidth}
    \vspace{0.7\baselineskip}
    \resizebox{0.9\columnwidth}{!}{%
    \setlength{\tabcolsep}{2pt}
    \footnotesize
    \centering    
    \begin{tabular}{|c|P{2cm}|P{2cm}|} 
        \hline
        \multirow{3}{*}{Method} & {Reduced Dimension} & {Average Reconstruction Error} \\
        \hline
        PCA+FPCA & $5\times 10$ & 4.96 \\
        \hline
        MPCA & $5\times 9$ & 4.78 \\
        \hline
        AE+FPCA & $5\times 10$ & 5.09 \\
        \hline
        VAE+FPCA & $5\times 10$ & 4.73 \\
        \hline
    \end{tabular}}
    \captionsetup{belowskip=10pt}
    \captionof{table}{Comparison of the dimension reduction methods.}
    \label{tab:10}
    \end{minipage} 
    \hfill
    \begin{minipage}[t]{.6\linewidth}    
    \centering
    \vspace{-0.1\baselineskip}
    \includegraphics[width=\textwidth]{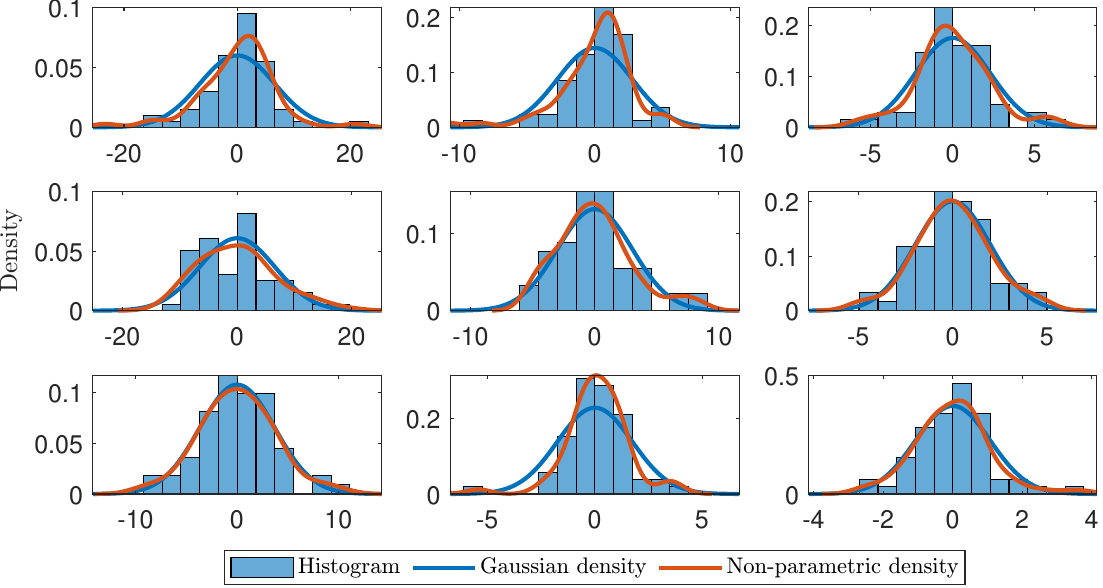} 
    \captionsetup{belowskip=10pt}
    \caption{The histogram of coefficients obtained by sequential PCA using the motion class of the Worker Motion. Each row contains the first three temporal PCA coefficients of the same spatial PCA dimension.}
    \label{fig:4-13} 
    \end{minipage}
\end{figure}

\noindent {\bf Gaussian Models of PCA Coefficients}:
We use $d=d_1\times d_2$ scalar coefficients to represent each IS-TVF or SIEM sequence and impose statistical models on them. Fig.~\ref{fig:4-13} shows histograms of individual coefficients for training sequences, overlaid with estimated Gaussian densities. We will impose a joint Gaussian distribution on the coefficient vector $\mathrm{vec}(\boldsymbol{\Phi})$ and call this a multivariate Gaussian or simply an MVG model. In case we treat the elements of $\mathrm{vec}(\boldsymbol{\Phi})$ are independent, we get an independent Gaussian or IG model. (IG model is used later in the two-level simulation section to create simulated data.) Sometimes we are limited by the amount of training data, leading to the relative instability of the estimated parameters. It is also possible to avoid making any assumptions on the distribution and impose a non-parametric model on the coefficients. For our situation, however, the marginal empirical distributions of the PCA coefficients are found to be close to the estimated Gaussian densities, as shown in Fig.~\ref{fig:4-13}. Thus, we only pursue the MVG model in the paper. Using an estimated model, one can generate a random set of coefficients $\boldsymbol{\Phi}^{*}$ and construct new posture sequences $\alpha^{*}$ from these coefficients $\boldsymbol{\Phi}^{*}$. By setting higher PCA coefficients (above $d_1$ and $d_2$ for the two PCAs), we can invert the coefficients to represent back the shape sequences.

We will use the following convention to label a scheme with choices of representation, dimension reduction, and statistical model: 
Scheme = {\it Representation/Dimension Reduction/Statistical Model}.
For example, {\it IS-TVF/Sequential-PCA/MVG} denotes using IS-TVF representation, Sequential PCA for dimensional reduction, and multivariate Gaussian for modeling. Similarly, {\it SIEM/Sequential-PCA/MVG} implies the same choices except the representation is now based on SIEM flattening; {\it SIEM/Sequential-PCA/IG} stands for an SIEM representation, Sequential PCA, and independent Gaussian models; {\it IS-TVF/Spatial-PCA/VAR} denotes IS-TVF representation, spatial PCA, and a VAR time-series model. 

\subsection{Evaluation Metrics}\label{subsec4-4}

How well do the random sequences generated from these models compare to real motion sequences? We will utilize several criteria for qualitative (visualization) and quantitative (like energy distance, cross-sectional variance, and posture validity score) evaluations.  

\noindent 1. {\bf Energy Distance}:
We can use an energy term~\citep{rizzo2016,Zhang2022test} to quantify differences between two shape distributions from their random samples. The expected value of this energy is zero {\it iff} the two underlying distributions are identical. Our experiment uses the training and the simulated sequences as these two sample sets, $A=\{\alpha_i^A\}_{i=1}^{n_A}$ and $B=\{\alpha_j^B\}_{j=1}^{n_B}$, and the posture sequence distance $d_{\mathcal{A}}(\alpha_1,\alpha_2)$ (defined in Eqn.~\ref{eq:4}) as the pairwise metric. The energy is defined as: 
$$
{\cal D}(A,B)=\frac{2}{n_A n_B}\sum_{i=1}^{n_A}\sum_{j=1}^{n_B}d_{\mathcal{A}}(\alpha_i^A,\alpha_j^B)-\frac{1}{n_A^2}\sum_{i=1}^{n_A}\sum_{j=1}^{n_A}d_{\mathcal{A}}(\alpha_i^A,\alpha_j^A)-\frac{1}{n_B^2}\sum_{i=1}^{n_B}\sum_{j=1}^{n_B}d_{\mathcal{A}}(\alpha_i^A,\alpha_j^B).
$$
The energy ${\mathcal D}(A,B)$ compares cross-group distances, $d_{\mathcal{A}}(\alpha_i^A,\alpha_j^B)$, with the within-group distances, $d_{\mathcal{A}}(\alpha_i^A,\alpha_j^A)$ and $d_{\mathcal{A}}(\alpha_i^B,\alpha_j^B)$, to compare the underlying distributions. One can use a permutation test~\citep{liu2022wilcox} to test equality of the two distributions.

\noindent 2. {\bf Cross-sectional Variance}: To capture inherent variability in a set of motion sequences, one can use the cross-sectional variance, $\sigma^2(t)$, the trace of the covariance matrix on the tangent space $T_{\hat{\boldsymbol{\mu}}_\mathcal{Y}(t)}(\mathcal{Y})$: $\sigma^2(t) = \Tr(\hat{\boldsymbol{\Sigma}}_{\mathcal{Y}}(t))=\Tr(\frac{1}{M-1}\sum_{m=1}^M V_m(t)V_m(t)^{\top})$, where $V_m(t)=\exp^{-1}_{\hat{\boldsymbol{\mu}}_{\mathcal{Y}}(t)}(\alpha_m(t))$ and $\hat{\boldsymbol{\mu}}_{\mathcal{Y}}(t)$ is the mean posture at time $t$ (defined in Sec.~\ref{subsec4-1}. This variance can be accessed by either plotting $\sigma^2$ or computing a single summary statistic such as the mean of the variance function $\mu_{\sigma^2}=\sum_t\sigma^2(t)/T$.

\noindent 3. {\bf Jerk/Acceleration}:
Jerk is the rate of change of acceleration and can be used as an additional measure of the smoothness of movements. Using the transported velocity function, $\mathbf{F}_{\alpha}$, we can compute the jerk using $J=\frac{d^2\mathbf{F}_{\alpha}}{dt^2}$. The energy of jerk is defined by the $\mathbb{L}^2$ norm of $J$:  $E_J=\|J\|^2/T$. In addition to jerk, we also compute the acceleration as $\vec{A}=\frac{d\mathbf{F}_{\alpha}}{dt}$ and acceleration energy  $E_{\vec{A}}=\|\vec{A}\|^2/T$. Smaller jerk and acceleration indicate more realistic and smoother motions.

\begin{figure}[htbp]
\centering
    \includegraphics[width=0.75\textwidth]{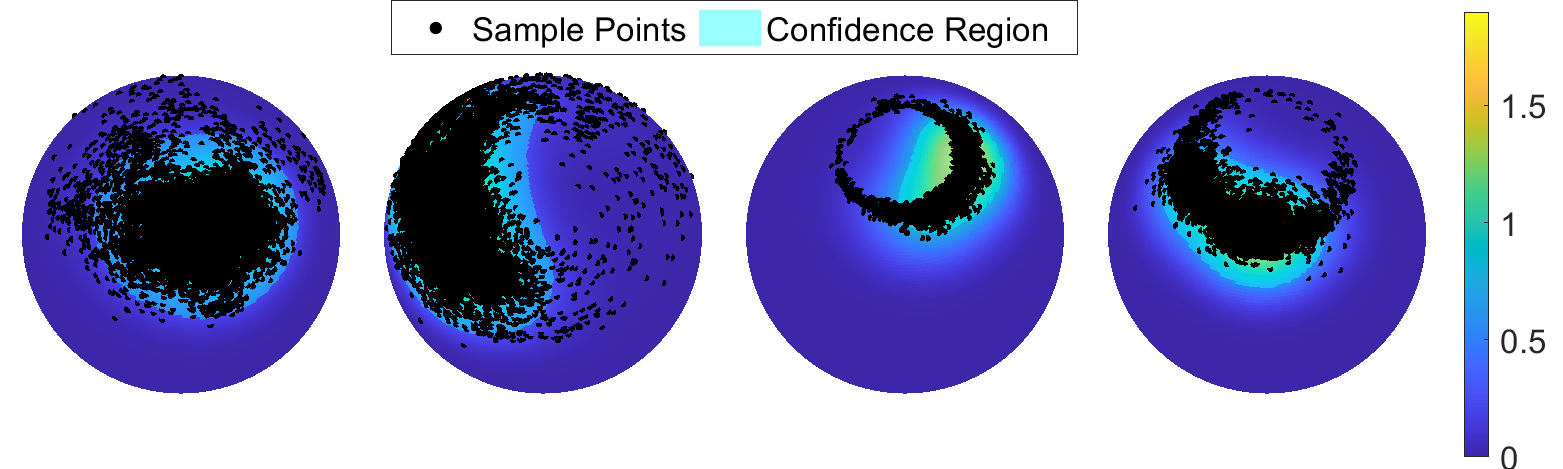}
    
    \caption{Examples of the estimated kernel distributions for some joint angles on $\mathbb{S}^2$. Black dots represent samples of the relative joint angles. The heatmaps are the fitted kernel distributions, and the light blue areas are $95\%$ confidence region. }
    \label{fig:4-16}
\end{figure}

\noindent 4. {\bf Posture Validity Score}:
To study the plausibility of the simulated postures, we also developed a posture validity score. We first compute the relative angles of some key joints, such as the shoulder, elbow, and knee, which are located on a unit sphere. 
Then we estimate a nonparametric density (on $\mathbb{S}^2$) with a von Mises-Fisher kernel for each joint using the training data and compute a confidence region that contains $95\%$ of the training data. Fig.~\ref{fig:4-16} shows some examples of the estimated distribution. Sample points within the confidence region (blue) are treated as valid positions, while points outside of the region are labeled as invalid positions. For individual posture $\mathbf{Y}$, we compute the validity score as $S_{\mathbf{Y}}=\frac{\text{\# of joints within those confidence regions}}{\text{\# of joints}}$; and for each sequence, we compute the average score over time: $S_{\alpha}=\frac{1}{T}\sum S_{\mathbf{Y}}(t)$. While a physically plausible posture may have a few joints located outside the confidence region, the average score $S_{\alpha}$ can mask these localized errors. To supplement the average score $S_{\alpha}$, we also define an integrity rate as the ratio of the postures where all joints simultaneously satisfy the confidence criteria, $\rho_{\alpha}=\frac{1}{T}\sum \mathbf{1}_{S_\mathbf{Y}=1}$.

\begin{figure}[htbp]
\centering
    \subfloat[]{\includegraphics[width=0.26\textwidth]{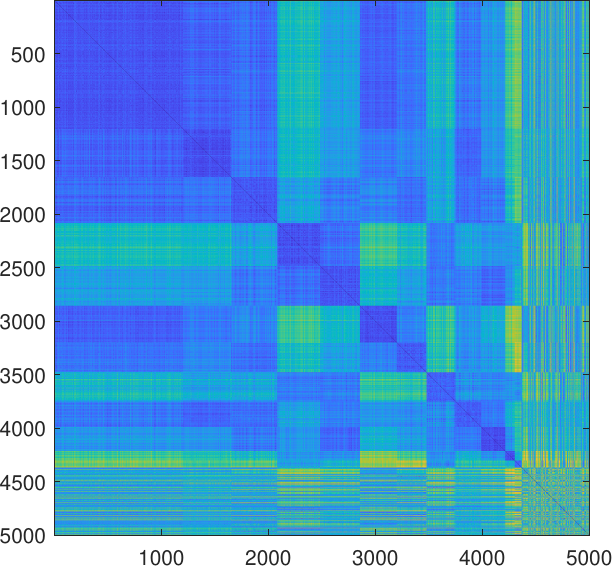}}
    \hspace{1em}
    \subfloat[]{\includegraphics[width=0.27\textwidth]{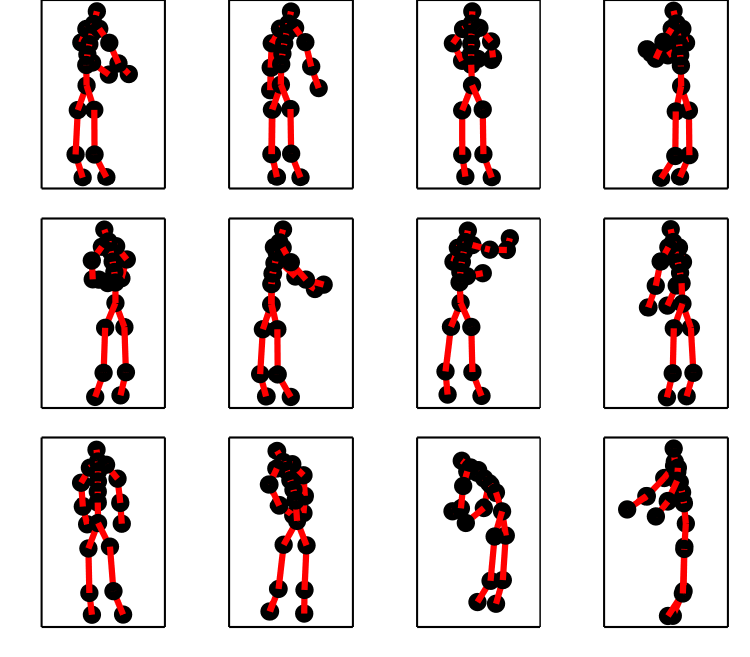}}
    \hspace{1em}
    \subfloat[]{\includegraphics[width=0.31\textwidth]{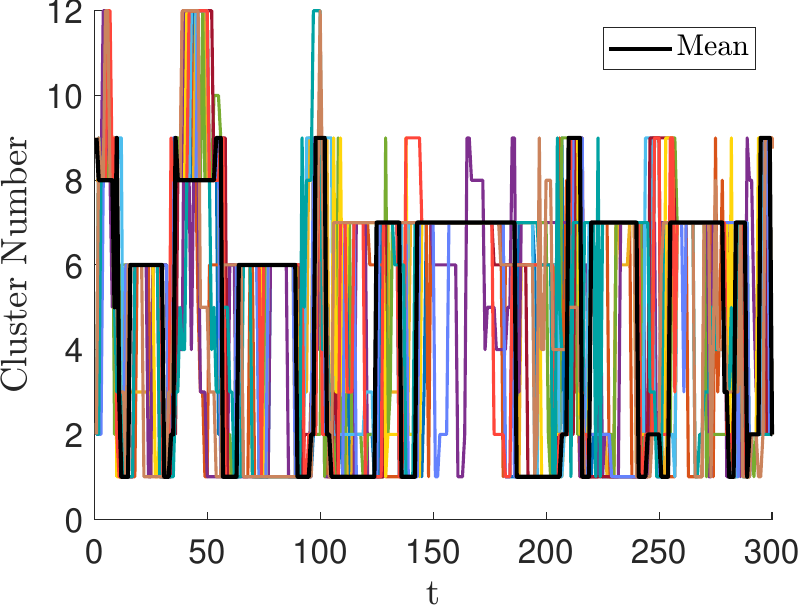}}
    \caption{Quantization of motion sequences. (a): sorted shape distance matrix; (b): twelve cluster-mean shapes for the posture data; (c): examples of quantized sequences (colored) and the quantized mean sequence (black) for a motion class.}
    \label{fig:4-15}
\end{figure}

\noindent 5. {\bf Shape Quantization}: Another possibility is to consider the underlying being tasks performed by human agents. These tasks require the workers to go through preset sequences or {\it recipes} of postures ({\it e.g.}, lifting a part, walking to the station, etc.). One can use adherence to these recipes to evaluate simulated sequences. To extract a recipe from training data, we cluster (\cite{deng2022clustering}) a large subset of training postures ($\{\alpha_m(t), m=1,\dots, M,~t=1,\dots, T\}$) into $K$ disjoint groups. This mode-based clustering technique automatically selected $K = 12$ clusters for a random sample of $5,000$ postures from the training data and provided a mode for each cluster: $\{\mathbf{Y}_{c,k}, k=1,2,\dots, 12\}$. Fig.~\ref{fig:4-15}(a), (b) show the pair-wise shape distance matrix (after clustering) and the 12 cluster modes $\{\mathbf{Y}_{c,k}\}$. 
To quantize a sequence, we assign to each posture $\alpha_m(t)$ its nearest mode and set the posture sequence $\alpha_m$ to be $\mathbf{B}_i(t) = b_t, \text{ where } b_t \in \{1,2,\ldots, K\}$ is the mode index. Let $\bar{\mathbf{B}}$  denote the quantization of the mean sequence $\hat{\boldsymbol{\mu}}_{\mathcal{Y}}(t)$ (defined in Sec.~\ref{subsec4-1}). Fig.~\ref{fig:4-15}(c) shows some examples of the individual quantized sequences and the quantized mean sequence.

Since most simulation approaches rely on mean sequences to fit their models, the means of simulated sequences are common to all the methods. Therefore, we focus on comparing the variation from the mean sequences rather than the means themselves. We use the quantized mean sequence $\bar{\mathbf{B}}$ as a reference sequence to check if an arbitrary sequence follows a similar pattern. The variation of a quantized sequence $\mathbf{B}_i$ can be measured using:
$E(m)=\frac{1}{T}\sum_{t=1}^T 1_{\mathbf{B}_m(t)\ne\bar{\mathbf{B}}(t)}$.
We can further compute the mean variation $\bar{E}$ over many simulated sequences to evaluate simulations.

\color{black}
\noindent 6. {\bf Log-likelihood}: If the underlying statistical model is known, one can also use the log-likelihood to compare the two samples. Suppose that the underlying model is Gaussian with known parameters. 
We can then compute the log-likelihood of a sequence $\alpha$ as:
\begin{equation}\label{eq:7}
    \log \left\{(2\pi)^{-\frac{d_1d_2}{2}}det(\boldsymbol{\Sigma})^{-\frac{1}{2}}\exp(\frac{1}{2}\mathrm{vec}(\mathbf{A})^{\top}\boldsymbol{\Sigma}^{-1}\mathrm{vec}(\mathbf{A}))\right\},
\end{equation}
where $\mathbf{A}$ is the PCA coefficients matrix of $\alpha$.

\section{Experiment Results and Evaluation}\label{sec5}
This section presents comprehensive results on estimating statistical models from sequence data and simulating novel human motions from these models. We also provide exhaustive evaluations of different approaches and draw conclusions about their performances. To ensure statistical robustness, we conduct 10 independent experiments. In each experiment, we simulate 100 new motion sequences. The final results represent the average of the 10 independent trials. The corresponding standard deviations for all metrics are detailed in the Supplementary Material Sec.~S4.1.

\subsection{Pre-Processing and PCA Dimensions}\label{subsec5-1}

Following Sec.~\ref{subsec3-2}, the motion data are transformed to the posture space (thus, become invariant to locations, body size, and body ratio), followed by a temporal alignment. 

\begin{figure}[htbp]
\centering
    \subfloat[]{\includegraphics[width=0.25\textwidth]{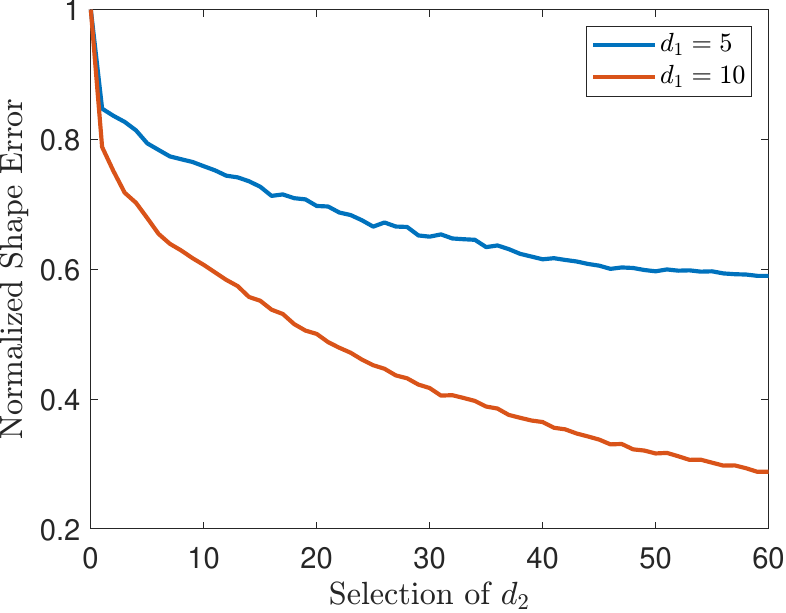}}
    \hspace{1em}
    \subfloat[]{\includegraphics[width=0.25\textwidth]{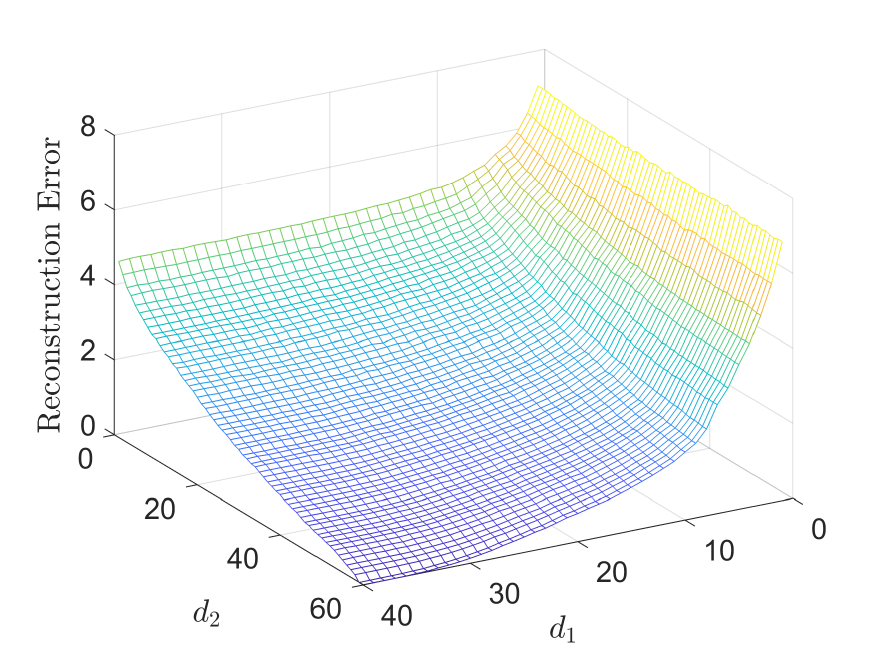}}
    \hspace{1em}
    \subfloat[]{\includegraphics[width=0.25\textwidth]{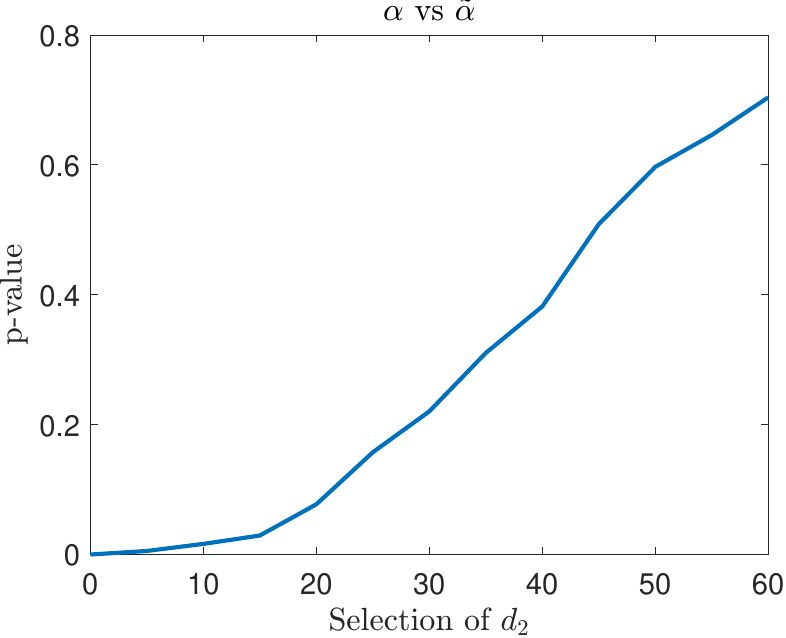}}
    \caption{PCA reconstruction experiment. (a): the average reconstruction error $e_{\mathcal{A}}$ defined in Sec.~\ref{subsec4-2} over different $d_2$ when $d_1=5$ and $10$. (b): the mesh plot of the average reconstruction error $e_{\mathcal{A}}$ over $d_1$ and $d_2$. (c): the $p$-value of the two-sample test over $d_2$ when $d_1$ is fixed at $10$.}
    \label{fig:5-2}
\end{figure}

We choose $d_1$, $d_2$ for spatial, temporal PCAs using two criteria:  (1) reconstruction error $e_{\mathcal{A}}$, 
and (2) a two-sample test between the original and the reconstructed sequences. Fig.~\ref{fig:5-2} shows the PCA result of the first motion class of the Worker Motion. Fig.~\ref{fig:5-2}(a) shows a plot of $e_{\mathcal{A}}$ versus $d_2$ (for $d_1=5,\ 10$), and Fig.~\ref{fig:5-2}(b) shows a mesh plot of reconstruction error versus $d_1$ and $d_2$. Increasing $d_1$ seems more effective in terms of reducing the reconstruction error. Hence, we chose $d_1=10$ and perform a two-sample test between the original and the reconstructed sequences. Fig.~\ref{fig:5-2}(c) shows the $p$-values of the two-sample test versus $d_2$ values, with chosen $d_2=30$. Similar results were obtained for other motion classes using $d_1 = 10$, $d_2=30$. 

\subsection{Baseline Methods for Comparison}

\noindent {\bf Intrinsic, Point-Wise Independent Process (PWI)}: There is also a possibility of modeling shapes on ${\cal Y}$ without a global flattening, following the PWI model described in Section 4.1. The point-wise (or cross-sectional) mean and covariance at each time $t \in [0,1]$ can be estimated from the training data using: 
$\hat{\boldsymbol{\mu}}_{\mathcal{Y}}(t)=\argmin_{\mathbf{Y}\in\mathcal{Y}}\sum_{m=1}^M d_{\mathcal{Y}}(\alpha_m(t),\mathbf{Y})^2$ and $\hat{\boldsymbol{\Sigma}}_{\mathcal{Y}}(t)=\frac{1}{M-1}\sum_{m=1}^M V_m(t)V_m(t)^{\top}$.
Here $d_{\mathcal{Y}}$ is the posture distance and $V_m(t)=\exp^{-1}_{\hat{\boldsymbol{\mu}}_{\mathcal{Y}}(t)}(\alpha_m(t))$ is the shooting vector defined by the inverse exponential map. With the estimated model parameters, we can generate new sequences as follows. At each time $t$, we generate a random vector from an independent Gaussian model $\mathbf{V}^{*}(t) \sim N(\boldsymbol{\mu}_{\mathcal{Y}}(t),\boldsymbol{\Sigma}_{\mathcal{Y}}(t))$ in the tangent space $T_{\mathcal{Y}}(\boldsymbol{\mu}_{\mathcal{Y}}(t))$. 
Then, we map this vector $\mathbf{V}^{*}(t)$ into a new posture $\alpha^{*}(t)$ using the exponential mapping, according to $\alpha^{*}(t)=\exp_{\boldsymbol{\mu}_{\mathcal{Y}}(t)}(\mathbf{V}^{*}(t))$. Note that this model assumes independence of $\alpha^*(t)$ from $\alpha^*(s)$ for $t \neq s$, which is a major limitation. 

\noindent {\bf Time Series VAR Model}: A classic Euclidean approach is a time-series model like the VAR (Vector Auto-Regressive) model~\citep{hamilton1994timeseries}. We will investigate how well the VAR model fits our time-series data resulting from spatial PCA. That is, the motion sequence data are flattened by IS-TVF or SIEM, followed by a spatial PCA to obtain low-dimensional vector-valued functions $\mathbf{H}_{\alpha_m}\in\mathbb{R}^{d_1\times (T-1)}$. A stationary VAR($p$) model (with $p=4$) is estimated using a randomly selected whole sequence as training data. 
Then, we simulate new sequences using the estimated model parameters.

\noindent {\bf Gaussian Process Model}: In addition to the VAR model, we employ the widely used Gaussian Process (GP) model. The motion data are first flattened by SIEM and dimension-reduced by a spatial PCA to obtain vector-valued functions $\mathbf{H}_{\alpha_m}\in\mathbb{R}^{d_1\times (T-1)}$ ($d_1=10$ in our study). We apply a sparse variational GP model with a Mat\'ern 1/2 kernel. The new motion sequences are sampled from the posterior mean and covariance of the GP model.

\noindent {\bf Long Short-Term Memory (LSTM)}: LSTM is a deep learning-based technique commonly used for time-series predictions. We fit an LSTM model on the shape space $\mathcal{Y}$ without flattening, {\it i.e.}, using the intrinsic representation. The model has two layers with a hidden size of 128, followed by a linear projection to the feature space ($D=10$). The motion is trained with a combination of position MSE loss and velocity MSE loss. 
The simulation uses the first $30$ postures of the original sequences ($\sim 10\%$) as the starting seeds to predict new postures auto-regressively. 

\noindent {\bf GCN-Transformer}: We also apply a hybrid of the Graphical Convolutional Network (GCN) and the Transformer, inspired by the STTG-NET \citep{Chen2022STTGNET}, to model shape sequences. The GCN is applied on the shape space $\mathcal{Y}$ to extract the spatial features. The extracted features are then passed to a Transformer decoder to model the long-term temporal dependencies and simulate new motions. This model is trained with an MSE loss between the predicted and ground-truth coordinates. Similar to the LSTM, we use the first $30$ postures from the real motion ($\sim 10\%$) as the starting seeds for simulation. The seeds postures are first processed by the GCN to obtain the features in the latent space and then sent to the Transformer to simulate the new motion trajectory. 

\subsection{Evaluating Simulated Sequences}\label{subsec5-3}

\begin{figure}[htbp]
    \centering 
        \subfloat{\includegraphics[width=0.45\textwidth]{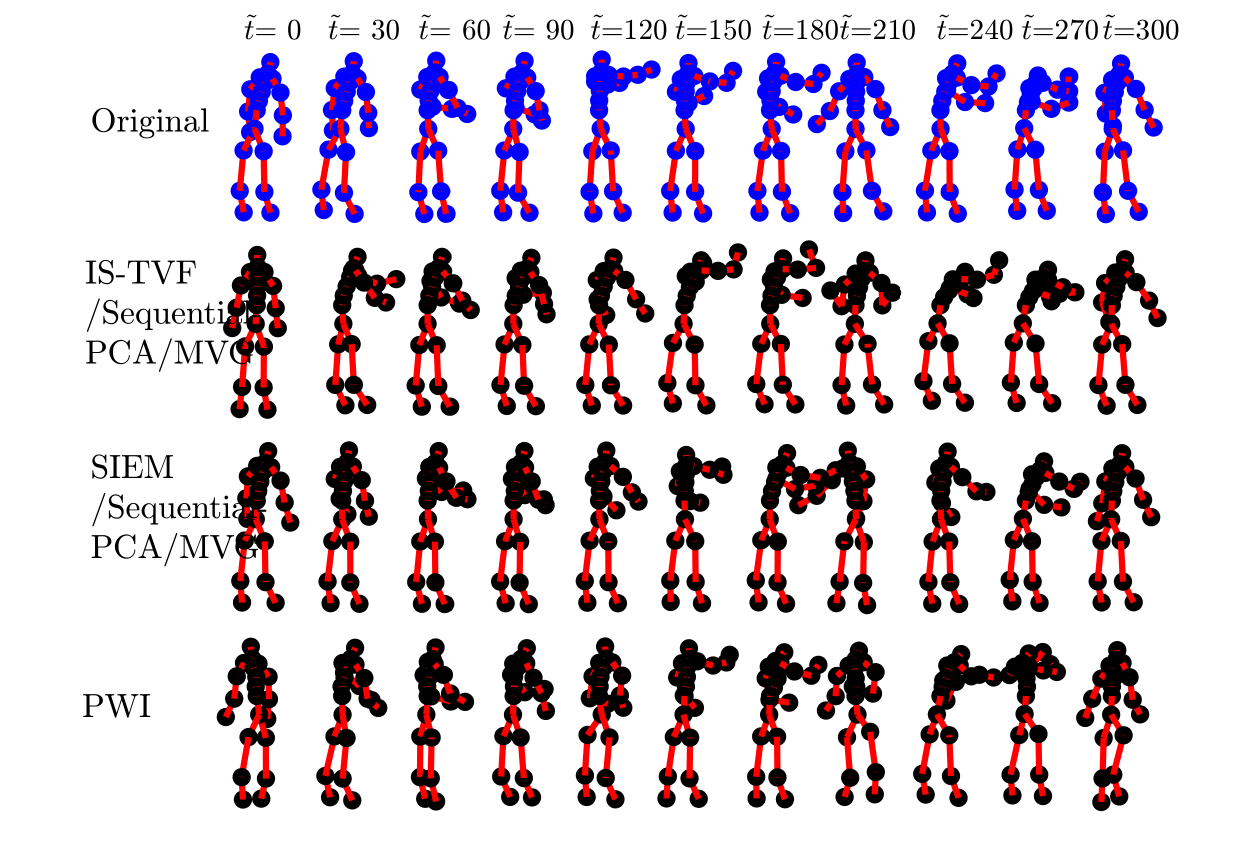}}
        \subfloat{\includegraphics[width=0.45\textwidth]{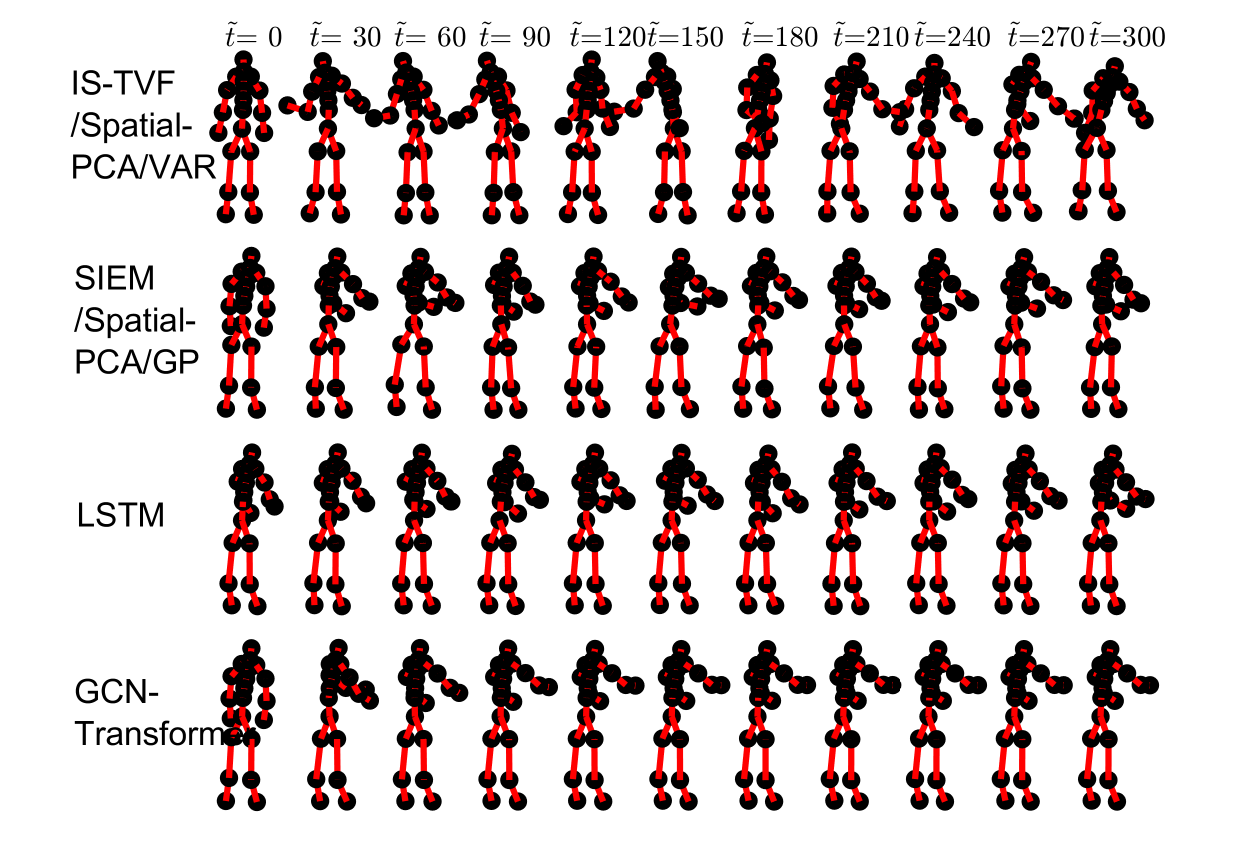}}
    \caption{Examples of simulated sequences $\{\alpha_i^{*}\}$ of Worker Motion using different methods and an original sequence.}
    \label{fig:5-7}   
\end{figure}

\noindent {\bf Visual Comparisons}:
Fig.~\ref{fig:5-7} shows simulated sequences from the first motion class using different approaches. The {\it IS-TVF/SequentialPCA/MVG} and {\it SIEM/SequentialPCA/MVG} approaches simulate realistic sequences by retaining broad spatial and temporal patterns. The PWI model ignores the temporal correlations in sequences, and, as a result, the sequences lack temporal coherence. The shapes are essentially independent samples around their means without any connection between successive shapes. The {\it IS-TVF/SpatialPCA/VAR} is a short-term stationary model that cannot capture the evolving dynamics of body shapes. As a result, the simulated postures tend to repeat similar short-term movements throughout, without exhibiting long-term movement trends. Perhaps a time-varying VAR model would perform better for the current application. The two deep-learning benchmarks also do not work well as the sequences ultimately become constant. One explanation is that training data is too small to train a complex deep-learning model.

\noindent {\bf Comparing the Distributions of the Original and Simulated Sequences}:
We apply several different metrics discussed above to compare the distributions of original and simulated sequences. Tab.~\ref{tab:1} shows the energy distances.  Although the PWI approach has the best result, it doesn't produce realistic human motions because it completely neglects the temporal relationship. The {\it SIEM/SequentialPCA/MVG} model consistently achieves the second-best results across all motion classes.
Tab.~\ref{tab:11} shows the mean of the cross-sectional variance over time with the reference numbers of the original data. The PWI model achieves the best result because it directly samples the cross-sectional covariance. The sequences simulated by the {\it SIEM/SequentialPCA/MVG} model are the second best and very close to the original variance. The {\it SIEM/SpatialPCA/VAR} model performs inconsistently. In contrast, the {\it SIEM/SpatialPCA/GP}, LSTM, and GCN-Transformer models display small cross-sectional variances, indicating that the simulated sequences lack the individual variability typically found in human motion. 


\begin{table}[htbp]
    \footnotesize
    \centering
    \resizebox{0.9\textwidth}{!}{%
    \makebox[\linewidth][c]{
    \setlength{\tabcolsep}{2pt}
    \begin{tabular}{|c|P{2.6cm}|P{2.4cm}|c|P{2.2cm}|P{1.9cm}|c|P{1.8cm}|} 
        \hline
        \multirow{3}{*}{Motion Type}  & \multicolumn{7}{c|}{Model} \\ 
        \cline{2-8}
        & {\it IS-TVF/Sequen-tialPCA/MVG} & {\it SIEM/Sequen-tialPCA/MVG} & \multirow{2}{*}{PWI} & {\it IS-TVF/Spa-tialPCA/VAR} & {\it SIEM/Spa-tialPCA/GP} &\multirow{2}{*}{LSTM} & GCN-Transformer\\ 
        \hline
        Worker Motion 1 & 1.32 & \underline{0.35} & \textbf{0.25} & 6.94 & 6.20 & 6.38 & 6.49\\ 
        \hline
        Worker Motion 2 & 1.90 & \underline{0.35} & \textbf{0.21} & 10.82 & 7.30 & 6.05 & 7.41\\    
        \hline
        Worker Motion 3 & 1.91 & \underline{0.40} & \textbf{0.28} & 5.40 & 5.04 & 5.22 & 5.48\\ 
        \hline
        Worker Motion 4 & 1.42 & \underline{0.78} & \textbf{0.62} & 4.84 & 6.50 & 6.11 & 6.88\\     
        \hline
        Worker Motion 5 & 1.84 & \underline{0.47} & \textbf{0.31} & 9.03 & 6.42 & 5.11 & 6.00\\  
        \hline
        Exercise Motion & 5.58 & \underline{0.57}& \textbf{0.29} & 44.03 & 14.73 & 16.41 & 13.18 \\
        \hline
        \multicolumn{7}{c}{\scriptsize Note: Best result for each motion class is in \textbf{bold} and the second best is \underline{underlined}.}
    \end{tabular}}}
        \caption{Energy distances of the simulated sequences compared against the original sequences.}
        \label{tab:1}
\end{table}

\begin{table}[htbp]
    \footnotesize
    \centering
    \resizebox{0.9\textwidth}{!}{%
    \setlength{\tabcolsep}{2pt}
    \begin{tabular}{|c|P{2.6cm}|P{2.4cm}|c|P{2.2cm}|P{1.9cm}|c|P{1.8cm}|} 
        \hline
        \multirow{3}{*}{Motion Type}  & \multicolumn{7}{c|}{Model} \\ 
        \cline{2-8}
        & {\it IS-TVF/Sequen-tialPCA/MVG} & {\it SIEM/Sequen-tialPCA/MVG} & \multirow{2}{*}{PWI} & {\it IS-TVF/Spa-tialPCA/VAR} & {\it SIEM/Spa-tialPCA/GP} & \multirow{2}{*}{LSTM} & GCN-Transformer\\
        \hline
        Worker Motion 1 (1.66) & 4.36 & \underline{1.47} & \textbf{1.66} & 3.92 & 0.08 & 0.18 & 0.11\\ 
        \hline
        Worker Motion 2 (1.30) & 5.17 & \underline{1.19} & \textbf{1.31} & 2.83 & 0.09 & 0.49 & 0.26\\    
        \hline
        Worker Motion 3 (1.22) & 4.40 & \underline{1.07} & \textbf{1.22} & 3.31 & 0.08 & 0.14 & 0.11\\ 
        \hline
        Worker Motion 4 (2.67) & 4.90 & \underline{2.61} & \textbf{2.67}& 3.38 & 0.10 & 0.73 & 0.41\\     
        \hline
        Worker Motion 5 (1.47) & 4.80 & \underline{1.34} & \textbf{1.47} & 1.70 & 0.08 & 0.39 & 0.23\\  
        \hline
        Exercise Motion (2.16) & 5.87 & \underline{1.48} & \textbf{2.15} & 15.94 & 0.15 & 0.34 & 0.25\\
        \hline
        \multicolumn{7}{c}{\scriptsize Note: Best result for each motion class is in \textbf{bold} and the second best is \underline{underlined}.}
    \end{tabular}}
    \caption{Mean of the cross-sectional variance over time.}
    \label{tab:11}
\end{table}

\noindent {\bf Evaluating Roughness of Simulated Sequences}: 
Tab.~\ref{tab:6} shows the jerk energy and acceleration energy. Both {\it IS-TVF/SequentialPCA/MVG} and {\it SIEM/SequentialPCA/MVG} have a jerk/acceleration energy close to the original motion and effectively capture the temporal evolution of the motion, producing smooth and realistic sequences that accurately follow the procedure. PWI sequences fluctuate around the mean sequence, which is rough and unrealistic for operating motions that have low jerk/acceleration energy. Although the {\it IS-TVF/SpatialPCA/VAR} model achieves jerk/acceleration energy similar to the original motion, it fails in other ways to capture the structure of the long sequences. The {\it SIEM/SpatialPCA/GP} and LSTM models produce overly smooth simulations, leading to a small values  of jerk/acceleration energy.

\begin{table}[htbp]
    \footnotesize
    \centering
    \resizebox{0.9\textwidth}{!}{%
    \setlength{\tabcolsep}{2pt}
    \begin{tabular}{|c|c|P{2.6cm}|P{2.4cm}|c|P{2.2cm}|P{1.9cm}|c|P{1.8cm}|} 
        \hline
        \multirow{3}{*}{Motion Type}  &\multirow{3}{*}{Original} & \multicolumn{7}{c|}{Model} \\ 
        \cline{3-9}
        &  & {\it IS-TVF/Sequen-tialPCA/MVG} & {\it SIEM/Sequen-tialPCA/MVG} & \multirow{2}{*}{PWI} & {\it IS-TVF/Spa-tialPCA/VAR} & {\it SIEM/Spa-tialPCA/GP}& \multirow{2}{*}{LSTM} & {GCN-Transformer}\\ 
        \hline
        Worker Motion 1 & 1.07 \textbar 0.42 & 0.63 \textbar 0.26 & \underline{0.69} \textbar \underline{0.28} & 31.65 \textbar 9.56 & \textbf{1.18} \textbar \textbf{0.47} & 0.11 \textbar 0.04 & 0.10 \textbar 0.04 & 2.32 \textbar 0.71 \\ 
        \hline
        Worker Motion 2 & 0.98 \textbar 0.38 & 0.57 \textbar 0.23 & \underline{0.63} \textbar \underline{0.25} & 24.58 \textbar 7.45 & \textbf{0.96} \textbar \textbf{0.40} & 0.12 \textbar 0.04 & 0.20 \textbar 0.08 & 2.36 \textbar 0.73 \\    
        \hline
        Worker Motion 3 & 1.41 \textbar 0.54 & 0.84 \textbar 0.34 & \underline{0.93} \textbar 0.37 & 23.65 \textbar 7.15 & \textbf{1.31} \textbar \textbf{0.53} & 0.11 \textbar 0.04 & 0.13 \textbar 0.05 & 2.24 \textbar \underline{0.68} \\ 
        \hline
        Worker Motion 4 & 1.22 \textbar 0.50 & 0.71 \textbar 0.32 & \underline{0.94} \textbar \underline{0.40} & 48.97 \textbar 14.85 & \textbf{1.12} \textbar \textbf{0.45} & 0.15 \textbar 0.05 & 0.17 \textbar 0.07 & 2.47 \textbar 0.76 \\     
        \hline
        Worker Motion 5 & 0.73 \textbar 0.31 & 0.44 \textbar 0.20 & \textbf{0.87} \textbar \textbf{0.34} & 27.48 \textbar 8.33 & \underline{0.58} \textbar \underline{0.24} & 0.11\textbar 0.04 & 0.07 \textbar 0.04 & 2.50 \textbar 0.77 \\  
        \hline
        Exercise Motion & 4.22 \textbar 1.58 & 1.42 \textbar 0.58 & \underline{2.42} \textbar \underline{0.95} & 37.79 \textbar 11.62 & \textbf{3.72} \textbar \textbf{1.36} & 0.04 \textbar 0.01 & 0.68 \textbar 0.26 & 7.56 \textbar 2.30 \\
        \hline
        \multicolumn{8}{c}{\scriptsize Note: Best result for each motion class is in \textbf{bold} and the second best is \underline{underlined}.}
    \end{tabular}}
    \caption{Jerk\textbar acceleration energy of the simulated sequences compared against the original sequences.}
    \label{tab:6}
\end{table}

\noindent {\bf Evaluating Quality of the Simulated Posture}:
We also evaluate the quality of individual postures using the posture validity score defined in Sec.~\ref{subsec4-4} and of the posture sequences using quantization variability. 

Tab.~\ref{tab:7} shows the posture validity scores and the integrity rates of the simulated sequences. Since both metrics are determined by the critical value of a kernel distribution, the original sequences do not achieve a perfect score. Thus, a score closest to that of the original sequences is considered the best. For the validity score, most methods can simulate sequences with good average posture quality, whereas {\it SIEM/SequentialPCA/MVG} achieves the best results on this metric. Regarding the stricter integrity rate, \textit{SIEM/SequentialPCA/MVG} also yields the most robust results. Other methods either lack sufficient variation ({\it e.g.,} {\it SIEM/SpatialPCA/GP}, which has near-perfect rates) or present notably lower rates, indicating a higher probability of producing implausible postures.

\color{black}
Tab.~\ref{tab:8} shows the mean variability of quantization experiments.
We obtain 12 clusters for Worker Motion and seven for Exercise Motion. The same quantization process is applied to both the original and simulated sequences. The quantized sequences for the observed sequences and the simulated sequences are compared in terms of the variability measure defined in Sec.~\ref{subsec4-4}. A smaller difference in variability measures between the observed and simulated ones indicates a better match. The {\it SIEM/SequentialPCA/MVG} simulations are usually the closest to the original sequences. The {\it IS-TVF/SequentialPCA/MVG} simulations have considerable variability, which means the simulated sequences include much noise and can become slightly unrealistic. In comparison, the PWI simulations have much smaller variability because the simulated sequences tend to be mostly repetitions of the mean sequences with small variability; {\it SIEM/SpatialPCA/GP}, {\it IS-TVF/SpatialPCA/VAR}, LSTM, and GCN-Transformer methods failed to capture the long-term temporal patterns and have unstable performance with higher variability than the proposed methods. Overall, the {\it SIEM/SequentialPCA/MVG} simulations replicate the structure of the original data best, and both proposed methods outperform the other approaches.

The proposed framework {\it SIEM/SequentialPCA/MVG} consistently achieves the best or the second-best result across all the metrics and motion classes. As detailed via a ranking methodology in Supplementary Materials Sec.~S4.3, {\it SIEM/\allowbreak SequentialPCA/\allowbreak MVG} produces the top average rank of $1.79$ across all baseline models. The result is further validated by a Friedman's test ($p<0.001$), indicating statistical significance.

\begin{table}[htbp]
    \footnotesize
    \centering
    \resizebox{0.95\textwidth}{!}{%
    \setlength{\tabcolsep}{2pt}
    \begin{tabular}{|c|c|P{2.6cm}|P{2.4cm}|c|P{2.2cm}|P{1.9cm}|c|P{1.8cm}|} 
        \hline
        \multirow{3}{*}{Motion Type}  &\multirow{3}{*}{Original} & \multicolumn{7}{c|}{Model} \\ 
        \cline{3-9}
        &  & {\it IS-TVF/Sequen-tialPCA/MVG} & {\it SIEM/Sequen-tialPCA/MVG} & \multirow{2}{*}{PWI} & {\it IS-TVF/Spa-tialPCA/VAR} & {\it SIEM/Spa-tialPCA/GP}& \multirow{2}{*}{LSTM} & GCN-Transformer\\ 
        \hline
        Worker Motion 1 & 0.97 \textbar 0.86 & 0.92 \textbar 0.70 & \underline{0.98} \textbar \textbf{0.90} & 0.86 \textbar 0.27 & \textbf{0.97} \textbar \underline{0.78} & 1.00 \textbar 1.00 & 0.99 \textbar 0.95 & 1.00 \textbar 0.98\\ 
        \hline
        Worker Motion 2 & 0.96 \textbar 0.85 & 0.88 \textbar 0.53 & 0.97 \textbar \textbf{0.92} & 0.88 \textbar 0.42 & \textbf{0.97} \textbar \underline{0.79}  & 1.00 \textbar 1.00 & 0.97 \textbar 0.95 & \underline{0.97} \textbar 0.94 \\    
        \hline
        Worker Motion 3 & 0.97 \textbar 0.86 & 0.89 \textbar 0.70 & \textbf{0.98} \textbar \textbf{0.90} & \underline{0.90} \textbar 0.46 & \underline{0.98} \textbar \underline{0.92} & 1.00 \textbar 1.00 & 1.00 \textbar 0.98 & 1.00 \textbar 0.97 \\ 
        \hline
        Worker Motion 4 & 0.94 \textbar 0.82 & 0.89 \textbar 0.64 & \textbf{0.93} \textbar \textbf{0.81} & 0.79 \textbar 0.18 & \underline{0.97} \textbar \underline{0.85} & 1.00 \textbar 1.00 & 0.98 \textbar 0.97 & 0.99 \textbar 0.97 \\     
        \hline
        Worker Motion 5 & 0.90 \textbar 0.70 & 0.80 \textbar 0.42 & \textbf{0.91} \textbar \textbf{0.75} & 0.79 \textbar 0.24 & 1.00 \textbar 0.97 & 1.00 \textbar 1.00 & \underline{0.96} \textbar 0.92 & 0.96 \textbar \underline{0.89} \\  
        \hline
        Exercise Motion & 0.95 \textbar 0.67 & 0.81 \textbar 0.14 & \textbf{0.96} \textbar \textbf{0.71} & 0.87 \textbar 0.31 & 0.32 \textbar 0.02 & 1.00 \textbar 1.00 & \underline{0.94} \textbar \underline{0.47} & 0.99 \textbar 0.88  \\
        \hline
        \multicolumn{8}{c}{\scriptsize Note: Best result for each motion class is in \textbf{bold} and the second best is \underline{underlined}.}
    \end{tabular}}
    \caption{Posture validity scores\textbar integrity rate of the simulated sequences compared against the original sequences.}
    \label{tab:7}
\end{table}

\begin{table}[htbp]
    \footnotesize
    \centering
    \resizebox{0.9\textwidth}{!}{%
    \setlength{\tabcolsep}{2pt}
    \begin{tabular}{|c|c|P{2.6cm}|P{2.4cm}|c|P{2.2cm}|P{2.2cm}|c|P{1.8cm}|} 
        \hline
        \multirow{3}{*}{Motion Type}  &\multirow{3}{*}{Original} & \multicolumn{7}{c|}{Model} \\ 
        \cline{3-9}
        &  & {\it IS-TVF/Sequen-tialPCA/MVG} & {\it SIEM/Sequen-tialPCA/MVG} & \multirow{2}{*}{PWI} & {\it IS-TVF/Spa-tialPCA/VAR} & {\it SIEM/Spa-tialPCA/GP}& \multirow{2}{*}{LSTM} & GCN-Transformer\\ 
        \hline
        Worker Motion 1 & 0.40 & \underline{0.54} & \textbf{0.41} & 0.24 & 0.89 & 0.74 & 0.67 & 0.67\\ 
        \hline
        Worker Motion 2 & 0.32 & 0.60 & \textbf{0.33} & \underline{0.16} & 0.97 & 0.94 & 0.52 & 0.92\\    
        \hline
        Worker Motion 3 & 0.35 & 0.63 & 0.41 & 0.18 & 0.76 & 0.43 & \underline{0.41} & \textbf{0.40}\\ 
        \hline
        Worker Motion 4 & 0.47 & \underline{0.64} & \textbf{0.56} & 0.28 & 0.89 & 0.82 & 0.73 & 0.72 \\     
        \hline
        Worker Motion 5 & 0.32 & 0.53 & \textbf{0.34} & \underline{0.16} & 0.88 & 0.68 & 0.51 & 0.63 \\  
        \hline
        Exercise Motion & 0.19 & 0.37 & \textbf{0.15} & \underline{0.08}& 0.88 & 0.74 & 0.72 & 0.65 \\
        \hline
        \multicolumn{8}{c}{\scriptsize Note: Best result for each motion class is in \textbf{bold} and the second best is \underline{underlined}.}
    \end{tabular}}
    \caption{Quantization variability of the simulated sequences compared against the original sequences.}
    \label{tab:8}
\end{table}

\section{Data Emulation and A Downstream Application}\label{sec7}

\begin{figure}[htbp]
    \centering  
    \subfloat[]{\includegraphics[width=0.2\textwidth]{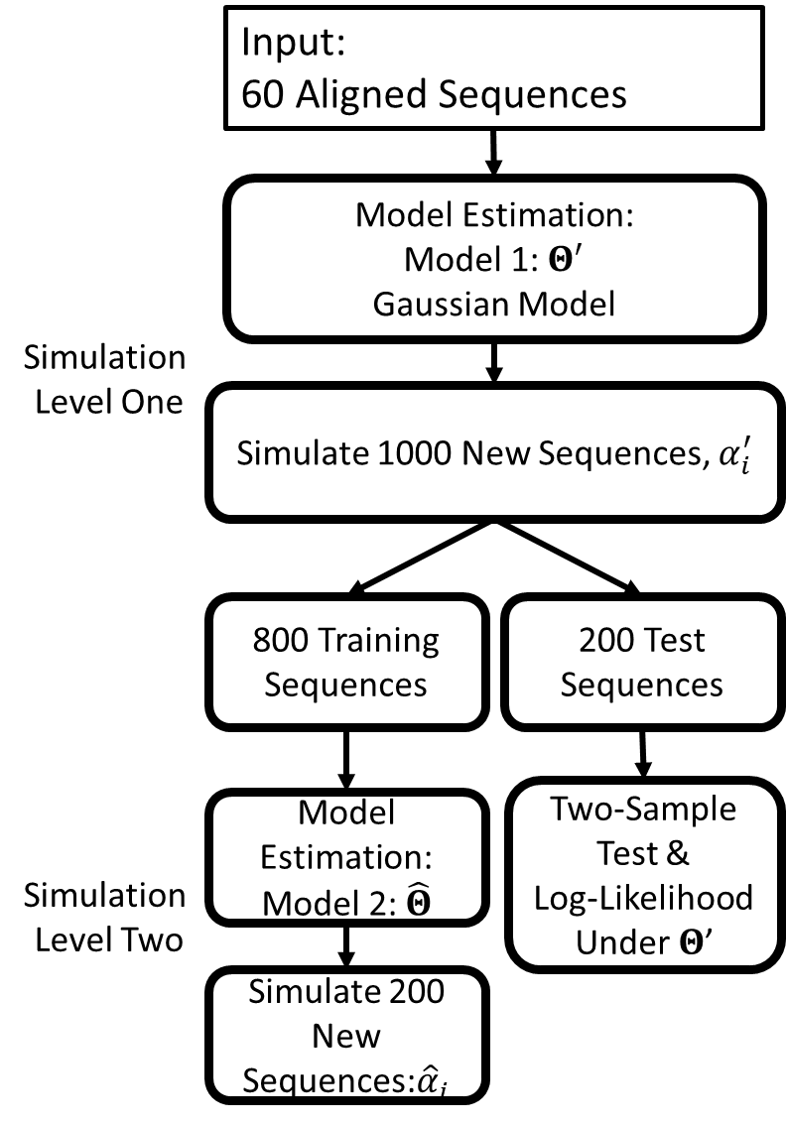}}    
    \subfloat[IS-TVF]{\includegraphics[width=0.37\textwidth]{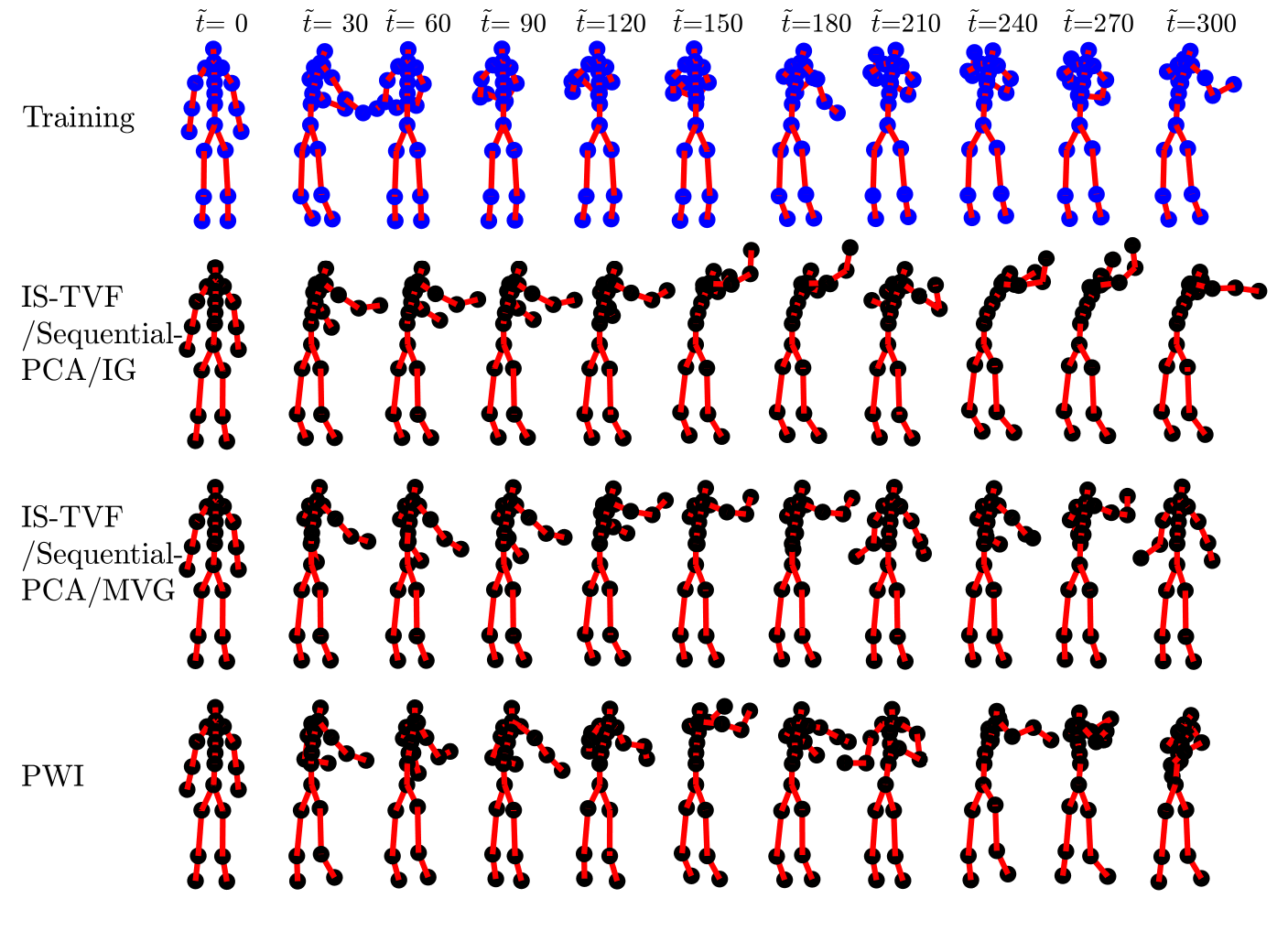}}
    \subfloat[SIEM]{\includegraphics[width=0.37\textwidth]{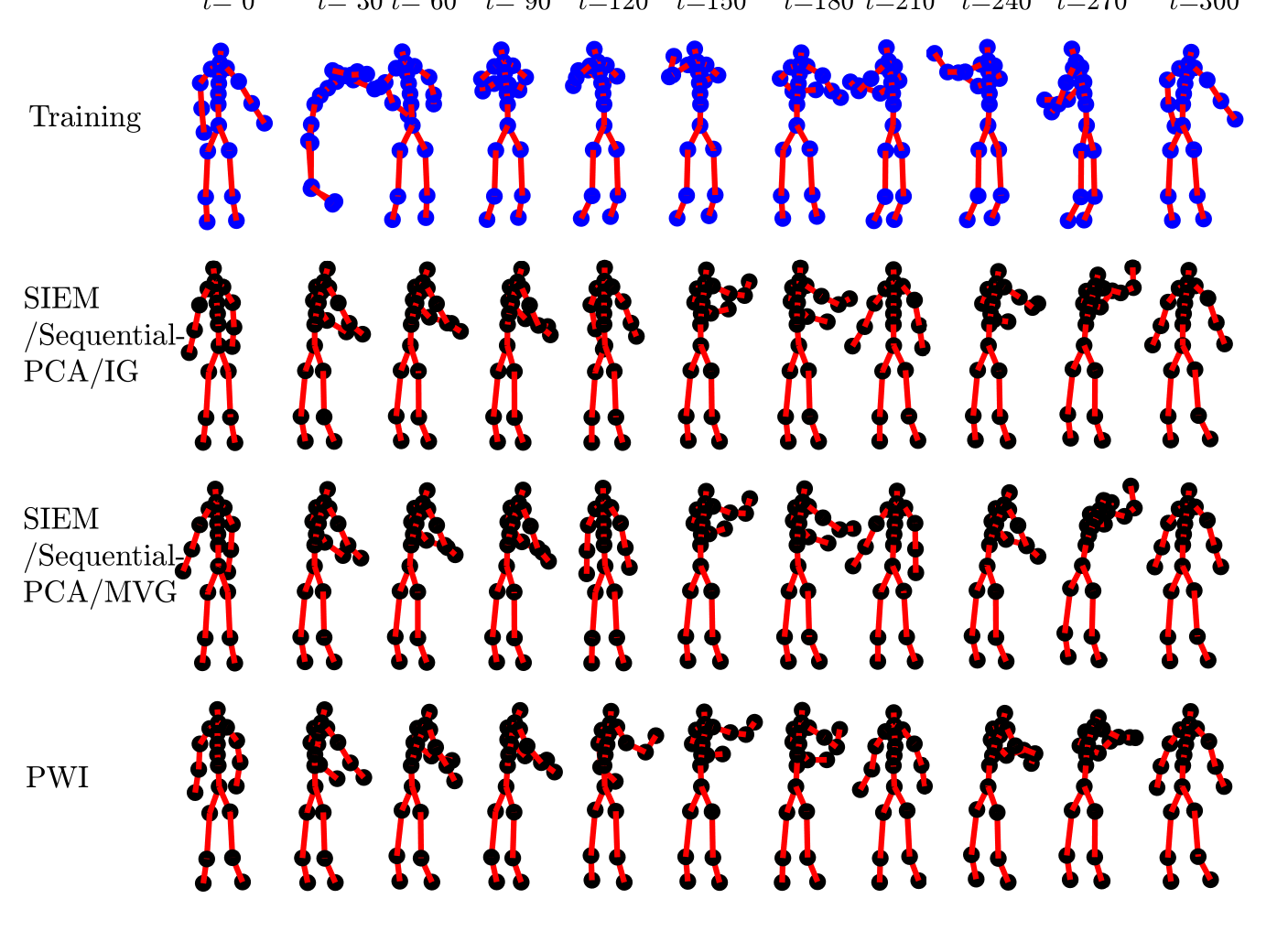}}
    \caption{(a): Simulation pipeline. (b) and (c): Examples of the simulation. Level one simulation (top rows of (b) and (c) estimates a model $\tilde{\Theta}$ from the 60 training sequences and simulates 1000 new sequences (800 training, 200 test). The second-level estimates a model $\hat{\Theta}$ (from 800 training) and simulates 200 new sequences. These 200 are compared to the 200 test sequences from the first simulation using metrics outlined in Sec.~\ref{subsec4-4}.}
    \label{fig:5-1}
\end{figure}

One critical real-world application of the proposed framework is as a synthetic data augmentation tool for downstream machine learning (ML) tasks where human motion data is scarce. Training ML models on small sample sizes inevitably leads to severe overfitting. While Sec.~\ref{sec5} demonstrates the model performance on the original training data, this section evaluates its effectiveness as a data augmentation tool in a controlled environment. To analyze this data augmentation, we establish a synthetic environment where the underlying distribution is fully controlled and known. In the first-level simulation, we use a restricted set of parameters and low dimensionality to generate a relatively large, controlled dataset that serves as the original training. As illustrated in Fig.~\ref{fig:5-1}(a), the second-level simulation builds on this training to generate an additional dataset for evaluation, allowing for a direct statistical comparison of the second-simulation from the first.  

\color{black}
Specifically, we use $60$ observed sequences in motion class one to fit a model and then sample this fitted model to generate $1000$ new sequences $\{\alpha_i'\}$. We randomly divide them into a training set (800 sequences) and a test set (200 sequences). Here we focus only on  {\it IS-TVF/SequentialPCA/IG} and {\it SIEM/SequentialPCA/IG} models with dimension $d_1\times d_2=4\times5$ to simulate the $1000$ new sequences. We refer to this step as the level-one simulation. 

Subsequently, we use the training set and test set as benchmark datasets to evaluate different statistical emulators, including the PWI model, {\it IS-TVF/SequentialPCA/MVG}, {\it IS-TVF/SequentialPCA/IG}, {\it SIEM/SequentialPCA/MVG}, and {\it SIEM/SequentialPCA/IG} models. To evaluate the performance under a realistic constrained scenario, the second-level simulations are configured with slightly lower PCA dimensions of $d_1\times d_2=4\times4$.
Each statistical emulator is fitted using the training set of the benchmark dataset, and the fitted emulator is used to simulate 200 sequences $\{\hat\alpha_i\}$, which we refer to as the level-two simulation data. The level-two simulated sequences $\{\hat\alpha_i\}$ are compared to the test set of the benchmark dataset, using log-likelihood based on the underlying known distribution and the two-sample test mentioned in Sec.~\ref{subsec4-4}.

\noindent {\bf Visual Inspections of Level-Two Simulations}: Fig.~\ref{fig:5-1} shows some examples of the two-level simulations:
$\alpha'$ represents the first level simulation (for training) and $\hat\alpha$ represents the second level simulation (for test). Fig.~\ref{fig:5-1}(b) uses {\it IS-TVF/SequentialPCA/IG} for the level-one simulation, and Fig.~\ref{fig:5-1}(c) uses {\it SIEM/SequentialPCA/IG} for the level-one simulation. Similar to the visual comparisons of the sub-sequences experiments (Sec.~\ref{subsec5-1}), the simulated sequences from the proposed models are realistic and reasonable.

\begin{figure}[htbp]
    \begin{minipage}[t]{0.6\textwidth}
    \vspace{-6.5\baselineskip}    
    \resizebox{0.9\columnwidth}{!}{%
    \setlength{\tabcolsep}{2pt}
    \footnotesize    
    \centering
    \begin{tabular}   {|c|P{1.5cm}|P{1.5cm}|c|P{1.5cm}|P{1.5cm}|c|} 
    \hline
    Level One Model $\rightarrow$ & \multicolumn{3}{c|}{\it IS-TVF/SequentialPCA/IG} & \multicolumn{3}{c|}{\it SIEM/SequentialPCA/IG}\\ 
    \hline
    \multirow{4}{*}{Level Two Model $\rightarrow$}   & {\it IS-TVF /Seque-ntialPCA /IG} & {\it IS-TVF /Seque-ntialPCA /MVG} & \multirow{4}{*}{PWI} & {\it SIEM /Sequen-tialPCA /IG} & {\it SIEM /Sequen-tialPCA /MVG} & \multirow{4}{*}{PWI}\\ 
    \hline
    Worker Motion 1 & 0.0583 & \textbf{0.1745} & 0.1117 & \textbf{0.0192} & 0.0034 & 0.0142\\ 
                              \hline
    Worker Motion 2 & 0.0327 & \textbf{0.2540} & 0.0167 & 0.0012 & 0.0038 & \textbf{0.3956}\\   
    \hline
    Worker Motion 3 & 0.3404 & \textbf{0.5452} & 0.2246 & 0.0619 & 0.2006 & \textbf{0.3901}\\  
    \hline
    Worker Motion 4 & 0.5927 & \textbf{0.7371} & 0.4248 & 0.2154 & 0.0832 & \textbf{0.5075}\\    
    \hline
    Worker Motion 5 & 0.0979 & \textbf{0.3618} &  0.0020 & 0.1294 & 0.0775 &  \textbf{0.2792}\\    
    \hline
    \multicolumn{7}{c}{\scriptsize Note: Best result for each motion class is in \textbf{bold}.}
    \end{tabular}}
    \captionsetup{belowskip=10pt}
    \captionof{table}{$p$-values of the two-sample tests on the two-level simulation. The first simulation uses {\it IS-TVF/SequentialPCA/IG} (left side) and {\it SIEM/SequentialPCA/IG} (right side). The models for simulation level two include IG and MVG with corresponding flattening methods (IS-TVF or SIEM) as well as PWI.}
    \label{tab:9}
    \end{minipage}
    \hspace{1em}
    \vspace{-1\baselineskip}
    \begin{minipage}[t]{.35\linewidth}
        \centering
        \includegraphics[width=\textwidth]{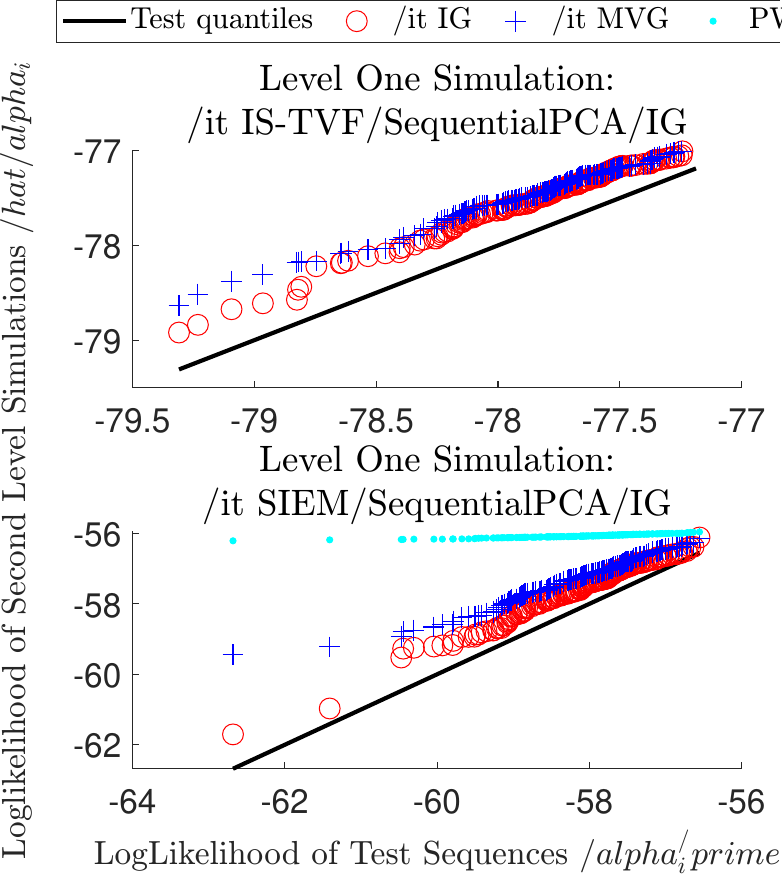}
        \captionsetup{belowskip=10pt}
        \caption{Q-Q plots of the sequences $\{\hat\alpha_i\}$ of level two simulation compared to the test data from the first simulation $\{\tilde\alpha_i\}$.}
    \label{fig:5-11}
    \end{minipage}
\end{figure}

\noindent {\bf Two-sample test}:
Besides visualization, we also use a two-sample test (Sec.~\ref{subsec4-4}) to compare a sample of 200 sequences in the benchmark test dataset versus a sample of 200 sequences simulated by statistical emulators. Tab.~\ref{tab:9} shows the $p$-value of the two-sample tests for various simulated configurations. When the benchmark data is generated by the {\it IS-TVF/sequentialPCA/IG} model, the statistical emulators with {\it IS-TVF/sequentialPCA/MVG} model have the highest $p$-values in most of the simulation configurations. The $p$-values are significantly higher than popular statistical significance levels such as 0.01 or 0.05. This implies that the sequences emulated by the proposed {\it IS-TVF/sequentialPCA/MVG} emulators and {\it IS-TVF/sequentialPCA/IG} emulators are indistinguishable from the benchmark data with a high statistical significance level. 
When the benchmark dataset is generated by the {\it IS-TVF/sequentialPCA/IG} model, most $p$-values remain higher than commonly used statistical significance thresholds, such as 0.01 or 0.05. 

\noindent {\bf Q-Q Plot of Likelihood}:
Fig.~\ref{fig:5-11} shows the Q-Q plots of the likelihood of the further simulated sequences $\{\hat{\alpha}_i\}$ compared with those of the test benchmark sequences from $\{\alpha_i'\}$, where the likelihood is calculated using Eqn.~\ref{eq:7}. As for the {\it IS-TVF/SequentialPCA/IG} experiments (the left plot of Fig.~\ref{fig:5-11}), the sequences simulated by different models perform similarly and follow the identity line except for the PWI model. The likelihood of the PWI model-simulated sequences is too small, so it is not visible in these charts. The results of {\it SIEM/SequentialPCA/IG} experiments (the right plot of Fig.~\ref{fig:5-11}) are similar. Both flattening approaches outperform the PWI model. The PWI model is closer to the ground truth than those in the {\it IS-TVF/SequentialPCA/IG} experiments, but still performs the worst and does not follow the identity line. In terms of likelihood, the proposed models work well for modeling and simulating using the training data. 

\section{Summary \& Conclusion}\label{sec6}
This paper develops an end-to-end pipeline to represent (sensor-captured) human operation motion data and provide a statistical tool to simulate realistic motions. A bijective mapping is used to flatten motion sequences into multivariate functions, followed by a sequential PCA to transform functions into a set of Euclidean coefficients. Then, it imposes statistical models on the coefficients and generates new sequences. We conduct extensive experiments to evaluate different flattening and modeling ideas. We evaluate simulated sequences using several evaluation metrics, including a novel quantization metric. For our datasets, both {\it integrated single-hop transported vector field} (IS-TVF) and {\it single inverse exponential map} (SIEM) successfully preserve salient features and simulate realistic sequences. In particular, the SIEM representation with sequential PCA and a Gaussian model yields the top overall average rank of $1.67$. See the Supplementary Material Sec.~S4.3 for details.

We return to scientific questions posed in Section 2. For capturing the essential statistical structure of long human skeletal motion sequences from limited observations, this paper successfully develops a pipeline for doing so. In situations where only limited motion data are available, the paper demonstrates that analytically tractable, geometry-based statistical models provide more reliable and interpretable emulators than deep-learning frameworks. 
It also provides viable low-dimensional representations that can represent variability in human motion.
Finally, it successfully compares summaries of dynamical features, such as jerk and acceleration, in simulated sequences with that of real sequences.

\color{black}

This framework can provide useful tools: (1) for planning and optimizing human efforts, (2) for a generative model on human shapes in general. Future directions include incorporating certain physical restrictions and utilizing weighting factors for different body parts to improve the simulations further.

\section*{Data Availability}
The data that support the findings of this study are available from the author, C.P.\ (chiwpark@uw.edu), upon reasonable request.


\section*{Acknowledgments}
This work was supported by NSF (grants \# 2132311, 2428742, 2413748) and AFOSR (grant \# FA9550-23-1-0673). The authors would also like to thank Daniel Qiu and Yang Wu from the University of Washington for their valuable assistance with data collection.

\color{black}
\spacingset{1}
\bibliographystyle{agsm}
\bibliography{main}

\clearpage
\spacingset{1.8}
\begin{center}
{\large\bf SUPPLEMENTARY MATERIAL}
\end{center}
This supplementary material provides some details of the simulation. Sec.~\ref{secs1} shows the detailed description of the geometry operations on the shape manifold $\mathcal{Y}$. Sec.~\ref{secs2} shows the details of the dimension reduction process. Sec.~\ref{secs3} presents the intermediate steps of the simulation process. Sec.~\ref{secs4} presents the full evaluation results, including experimental uncertainty, several additional metrics, and a comprehensive ranking system. Sec.~\ref{secs5} presents a sensitivity test under perturbed conditions. Sec.~\ref{secs6} describes videos of the simulated motion in the supplementary material. 

\section{Details of Posture Representation}\label{secs1}

The posture space $\mathcal{Y}$ is defined as the cross product of $n-1$ unit spheres $\mathbb{S}^2$, which is a well-studied manifold. Therefore, we can define the geometry operations element-wise on the posture space $\mathcal{Y}$. The tangent space of $\mathcal{Y}$ at any $\mathbf{Y} \in \mathcal{Y}$ is defined elemen-twise as $T_{\mathbf{Y}}(\mathcal{Y})=T_{\mathbf{y}_1}(\mathbb{S}^2)\times T_{\mathbf{y}_2}(\mathbb{S}^2)\times\ldots\times T_{\mathbf{y}_{n-1}}(\mathbb{S}^2)$, 
where $T_{\mathbf{y}_i}(\mathbb{S}^2)=\{\mathbf{v}\in\mathbb{R}^3| \mathbf{y}_i^{\top} \mathbf{v}=0\}$ is the tangent space of $\mathbb{S}^2$ at point $\mathbf{y}_i$. We can project any point $\mathbf{Z}\in\mathcal{Y}$ to the tangent space $T_{\mathbf{Y}}(\mathcal{Y})$ using the element-wise inverse exponential map 
\begin{equation}\label{eq:2}
    \exp^{-1}_{\mathbf{Y}}(\mathbf{Z})=(\exp_{\mathbf{y}_1}^{-1}(\mathbf{z}_1),\exp_{\mathbf{y}_2}^{-1}(\mathbf{z}_2),\ldots,\exp_{\mathbf{y}_{n-1}}^{-1}(\mathbf{z}_{n-1})), 
\end{equation}
where $\exp_{\mathbf{y}_i}^{-1}(\mathbf{z}_i)=\frac{\theta_i}{\sin(\theta_i)}(\mathbf{z}_i-\mathbf{y}_i \cos(\theta_i))$ and where $\theta_i=\cos^{-1}(\mathbf{y}_i^{\top}\mathbf{z}_i)$. Since $T_{\mathbf{y}_i}(\mathbb{S}^2)$ is a 2D vector space, any tangent vector $\mathbf{v}\in_{\mathbf{y}_i}(\mathbb{S}^2)$ can be expressed as a 2D vector {\it w.r.t} to a fixed orthonormal basis. As a result, the inverse exponential map $\exp^{-1}_{\mathbf{Y}}(\mathbf{Z})$ yields a $2\times(n-1)$ matrix $\mathbf{V}\in T_{\mathbf{Y}}(\mathcal{Y})$, which can be vectorized as $\mathbf{w}=\mathrm{vec}(\mathbf{V})\in\mathbb{R}^{2(n-1)}$. Similarly, the exponential map $\exp_{\mathbf{y}_i}(\mathbf{v}_i)=\cos(\|\mathbf{v}_i\|)\mathbf{y}_i+\sin((\|\mathbf{v}_i\|)\frac{\mathbf{v}_i}{\|\mathbf{v}_i\|}$ brings any point $\mathbf{v}_i\in T_{\mathbf{y}_i}(\mathbb{S}^2)$ back to the unit sphere. 
For any point $\mathbf{V}\in T_{\mathbf{Y}}(\mathcal{Y})$, the exponential map is
\begin{equation}\label{eq:3}
    \exp_{\mathbf{Y}}(\mathbf{V})=(\exp_{\mathbf{y}_1}(\mathbf{v}_1),\exp_{\mathbf{y}_2}(\mathbf{v}_2),\ldots,\exp_{\mathbf{y}_{n-1}}(\mathbf{v}_{n-1})).
\end{equation}
We also define a parallel transport that maps a tangent vector $\mathbf{U}\in T_{\mathbf{Y}}(\mathcal{Y})$ to $T_{\mathbf{Z}}(\mathcal{Y})$ along the geodesic between $\mathbf{Y}$ and $\mathbf{Z}$ element-wise as $\mathbf{v}_{\mathbf{y}_i\rightarrow\mathbf{z}_i}=\mathbf{u}_i-\frac{2\mathbf{u}_i^{\top}\mathbf{z}_i}{\|\mathbf{y}_i+\mathbf{z}_i\|^2}(\mathbf{y}_i+\mathbf{z}_i)$,
where $\mathbf{y}_i,\mathbf{z}_i(\neq\mathbf{y}_i)\in\mathbb{S}^2$ and $\mathbf{u}_i\in T_{\mathbf{y}_i}(\mathbb{S}^2)$.

\section{Details of PCA}\label{secs2}

\begin{figure}[htbp]
    \centering
    \subfloat[]{
    \includegraphics[width=0.24\textwidth]{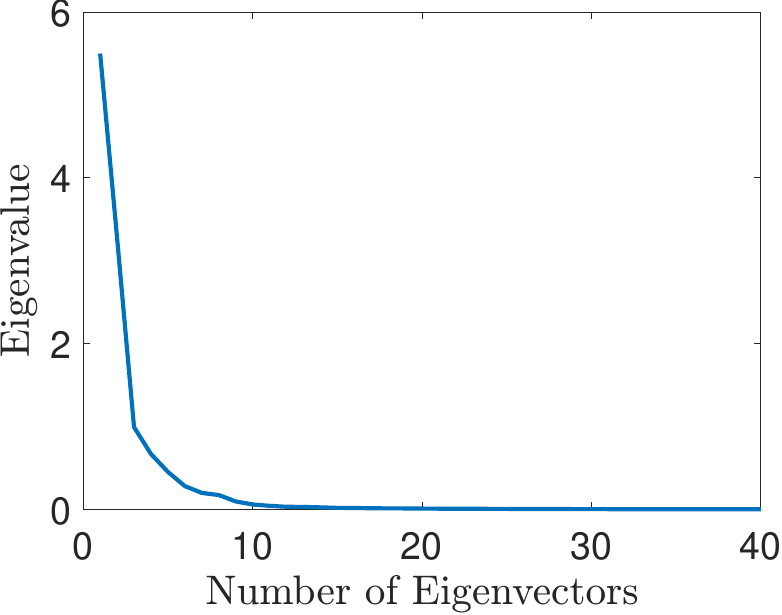}}
    \subfloat[]{
    \includegraphics[width=0.26\textwidth]{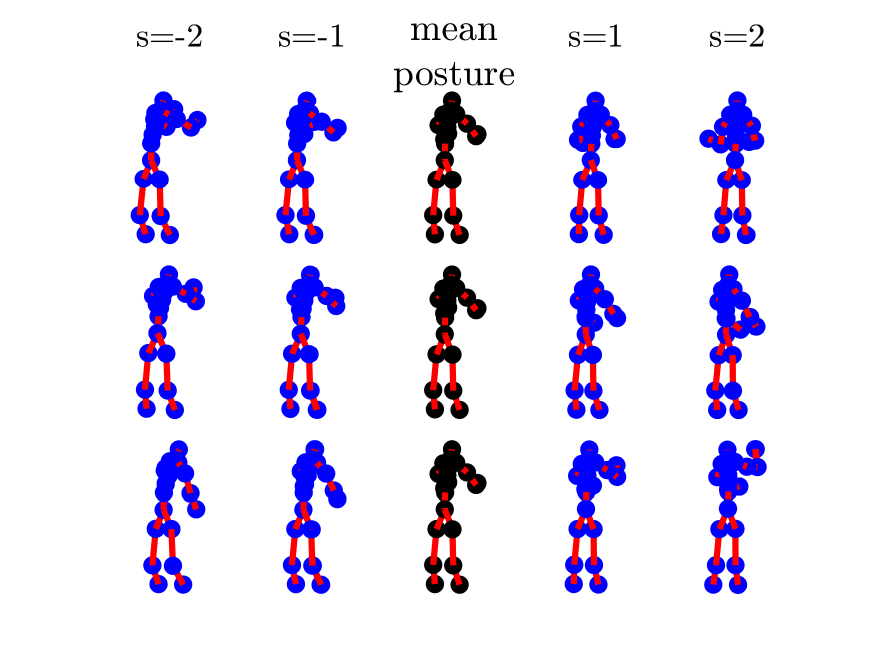}}
    \subfloat[]{
    \includegraphics[width=0.48\textwidth]{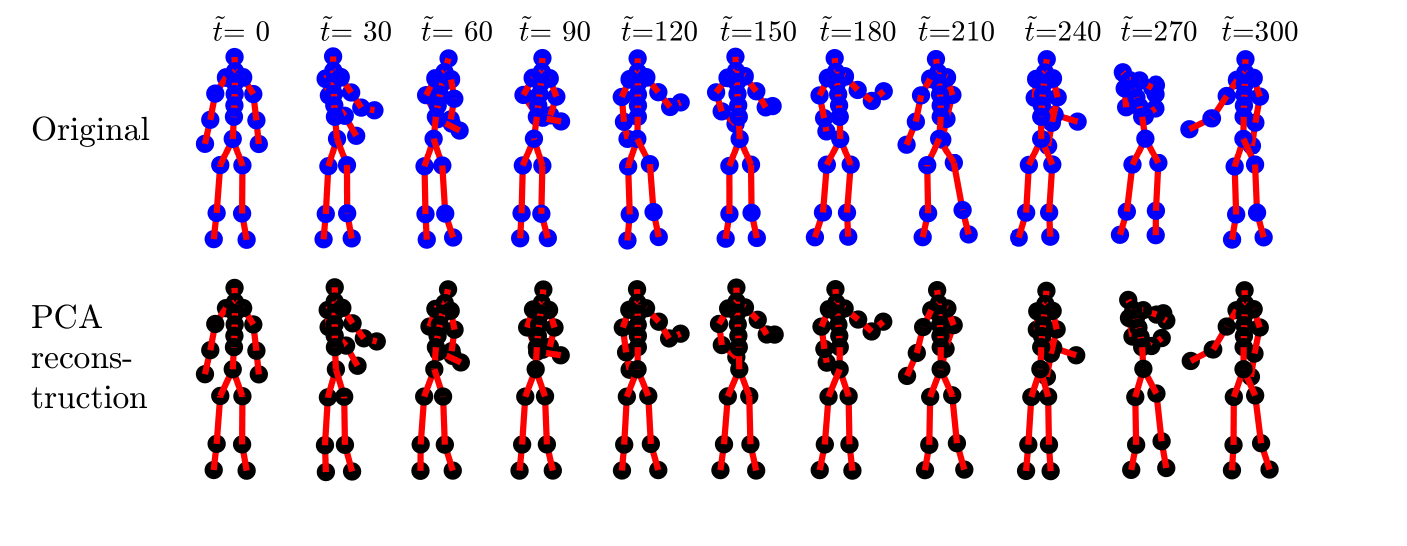}}
    \caption{Results of spatial PCA. (a): the plot of eigenvalues; (b): a display of the first three PCA directions; (c): an example of spatial PCA reconstruction using the first five components in the IS-TVF space.}
    \label{fig:s1}   
\end{figure}

\noindent {\bf Sequential Spatial and Temporal PCA}: We will describe this process for IS-TVF, but the process is the same for SIEM. To reduce spatial dimensions of an IS-TVF sequence, $\mathbf{G}_\alpha\in\mathbb{R}^{2(n-1)\times(T-1)}$, we treat the vectors $\mathbf{G}_\alpha(t)$ for all $\alpha$ and $t$ as points in  $\mathbb{R}^{2(n-1)}$. For $M$ sequences of length $T-1$, the total number of vector observations is $M(T-1)$. We perform PCA of these vectors by computing the mean vector $\mathbf{m}$
and the covariance matrix $\mathbf{S}_{\mathcal{Y}}$.
Then, the first (dominant) $d_1$  eigenvectors of $\mathbf{S}_{\mathcal{Y}}$ 
result in projection $\mathbf{U}\in\mathbb{R}^{2(n-1)\times d_1}$. Using this projection, each IS-TVF $\mathbf{G}_{\alpha_m}$ can now be approximated by a $d_1$-dimensional Euclidean sequence $\mathbf{H}_{\alpha_m}\in\mathbb{R}^{d_1\times (T-1)} = \mathbf{U}^{\top} (\mathbf{G}_{\alpha_m}-\mathbf{m})$. Given a coefficient sequence $\mathbf{H}_{\alpha_m}$, the  mean vector $\mathbf{m}$, and the projection matrix $\mathbf{U}$, one can reconstruct the IS-TVF sequence $\mathbf{G}_{\alpha_m}$ according to $\tilde{\mathbf{G}}_{\alpha_m}(t) =\mathbf{m}+\mathbf{U}\mathbf{H}_{\alpha_m}(t)$. 
Fig.~\ref{fig:s1} shows an example of this spatial PCA. 
Fig.~\ref{fig:s1}(c) shows some sequences reconstructed using the five spatial PCA components, {\it i.e.}, the spatial dimension is reduced from $2(n-1)=40$ to $d_1=5$. 

\begin{figure}[htbp]
    \centering
    \subfloat[]{
    \includegraphics[width=.27\textwidth]{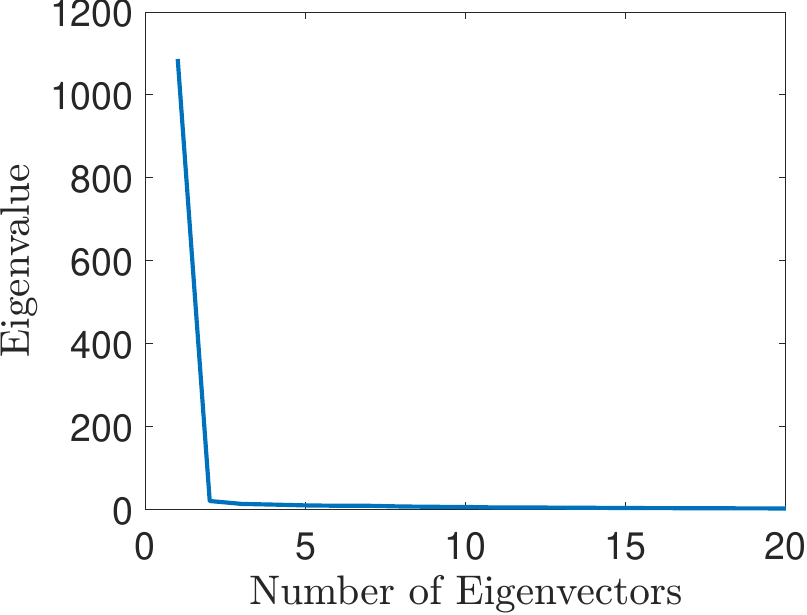}}
    \hspace{1em}   
    \subfloat[]{
    \includegraphics[width=.39\textwidth]{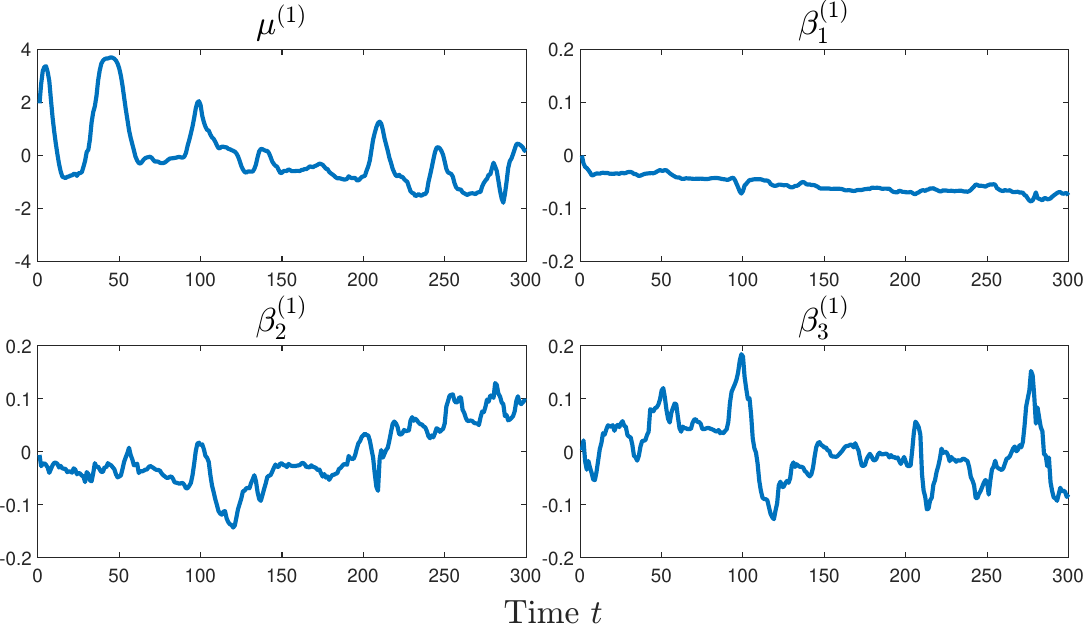}}

    \subfloat[]{
    \includegraphics[width=.7\textwidth]{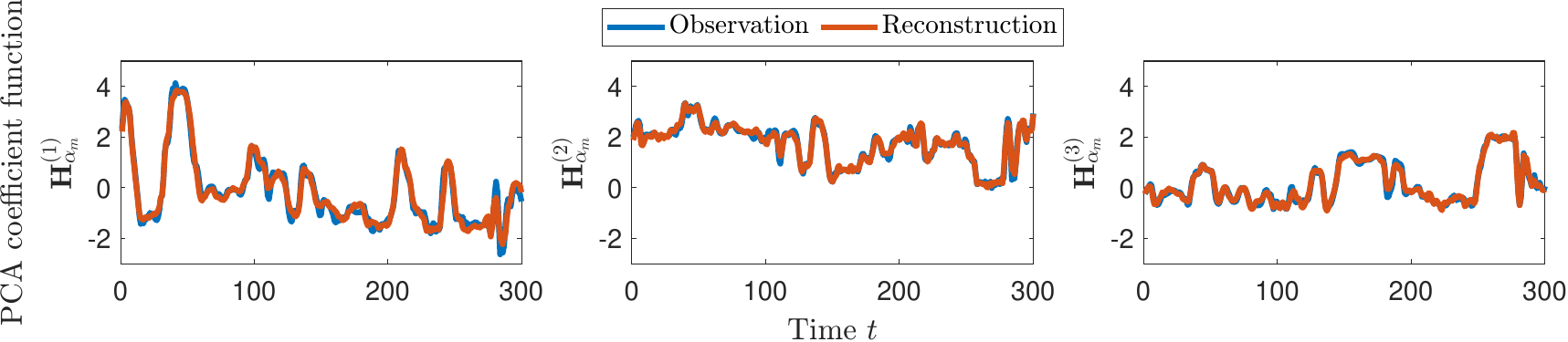}}
    \caption{Results of FPCA for functional data $\{\mathbf{H}_{\alpha_m}^{(1)}\}$. (a): eigenvalues; (b): mean and first three basis functions of $\mathbf{H}_{\alpha_m}^{(1)}$; (c): three examples of $\mathbf{H}_{\alpha_m}^{(i)}$  reconstruction.}
    \label{fig:s2}   
\end{figure}

\begin{figure}[htb]
\centering
    \subfloat[]{
    \includegraphics[width=0.4\textwidth]{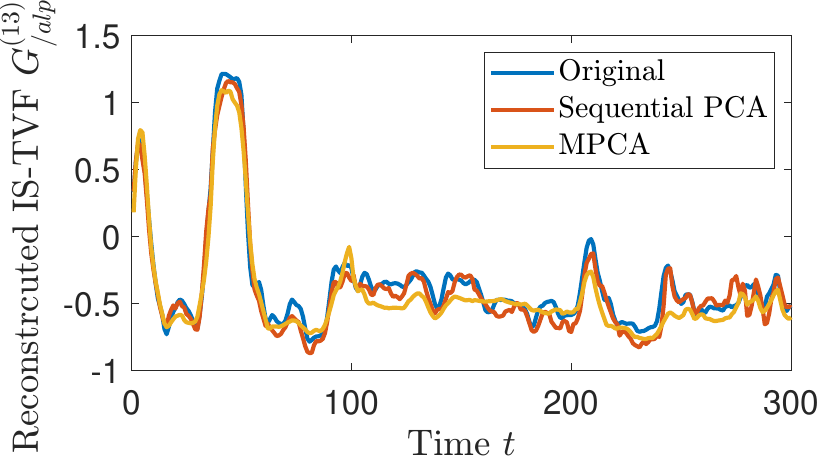}}
    \subfloat[]{\includegraphics[width=0.55\textwidth]{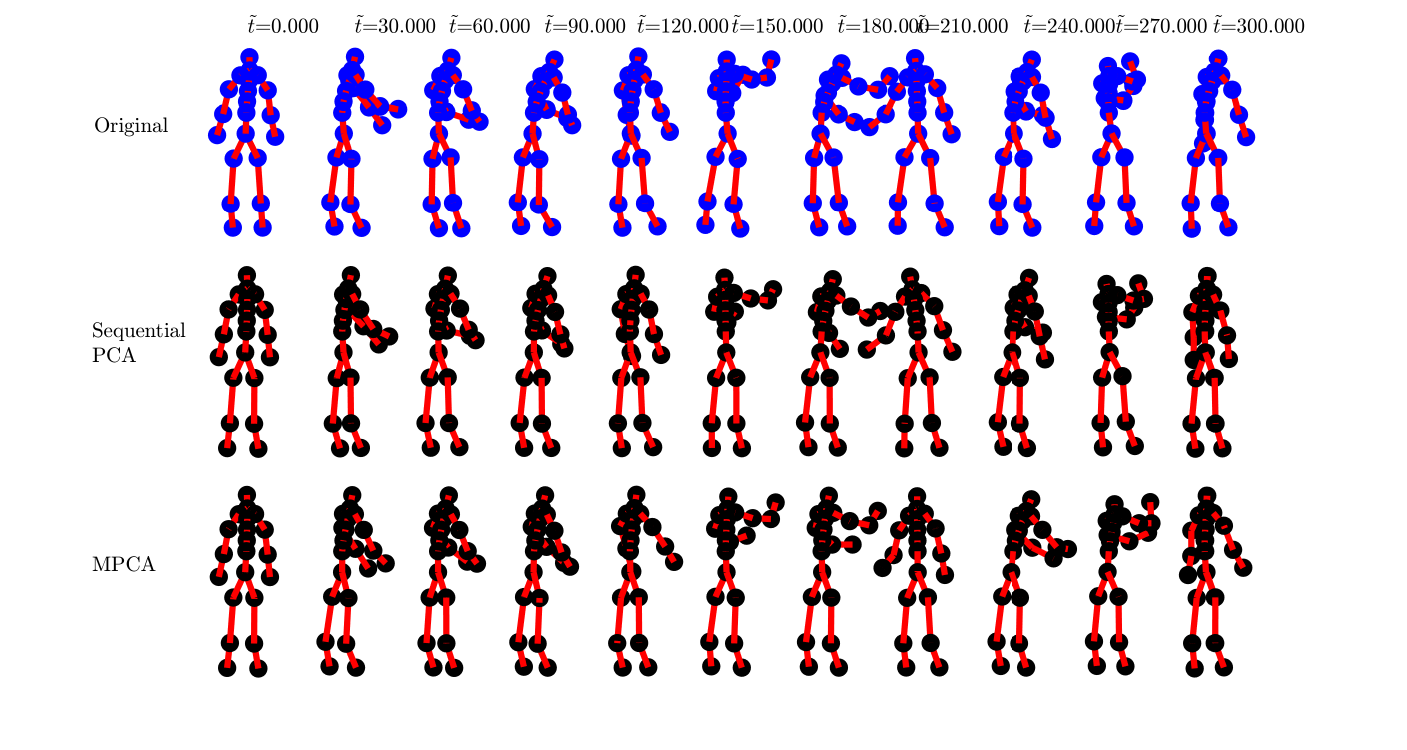}}    
    \caption{Comparison of the MPCA and sequential PCA. (a): an example of reconstructed IS-TVF elements. They are the same IS-TVF elements reconstructed by MPCA and sequential PCA. (b): the comparison between the reconstruction of sequential PCA and MPCA with the original sequence as a reference. (Note: this Figure is identical to Fig. 6(a) and 6(c) in the main manuscript. It is reproduced here to facilitate self-contained reading of the discussion).}
    \label{fig:s3}   
\end{figure}

Next, we view $\mathbf{H}_{\alpha_m}^{(i)},\,i=1,2,3,\ldots,d_1,\,m=1,2,3,\ldots,M$ as scalar functions over the observation interval, and perform functional PCA.
Using discrete time samples, we compute the functional mean $\boldsymbol{\mu}_{\mathbf{H}}^{(i)}=\frac{1}{M}\sum_{m=1}^M \mathbf{H}_{\alpha_m}^{(i)}\in\mathbb{R}^{T-1}$ and the functional covariance $\boldsymbol{\Sigma}_f^{(i)}=\frac{1}{M-1}\sum_{m=1}^M(\mathbf{H}_{\alpha_m}^{(i)}-\boldsymbol{\mu}_{\mathbf{H}}^{(i)})(\mathbf{H}_{\alpha_m}^{(i)}-\boldsymbol{\mu}_{\mathbf{H}}^{(i)})^\top\in\mathbb{R}^{(T-1)\times (T-1)}$. The eigen-decomposition of $\boldsymbol{\Sigma}_f^{(i)}$ leads to eigenvectors $\boldsymbol{\beta}_j^{(i)}$ and eigenvalues $\lambda_j^{(i)}$. Fig.~\ref{fig:s2}(a) shows the eigenvalues of the first spatial dimension $\mathbf{H}_{\alpha_m}^{(1)}$ as an example. For $d_2$ dominant vectors, a function $\mathbf{H}_{\alpha_m}^{(i)}$ can be approximated by
$\tilde{\mathbf{H}}_{\alpha_m}^{(i)}=\boldsymbol{\mu}_{\mathbf{H}}^{(i)}+\sum_{j=1}^{d_2} \phi_{ij} \boldsymbol{\beta}_j^{(i)}$,
where $\boldsymbol{\mu}_{\mathbf{H}}^{(i)}$ is the mean function for the $i$th dimension, $\boldsymbol{\beta}_{j}^{(i)}$ is the $j$th basis function for $i$th dimension, and $a_{ij}$ is the corresponding coefficient of $\boldsymbol{\beta}_{j}^{(i)}$. The mean function and the first three basis functions of $\mathbf{H}_{\alpha_m}^{(1)}$ are shown in Fig.~\ref{fig:s2}(b) as an example. Given a set of coefficients $\boldsymbol{\Phi}_m=\{\phi_{ij}\}$, the sequence $\mathbf{H}_{\alpha_m}$ could be rebuilt by using corresponding mean $\boldsymbol{\mu}_{\mathbf{H}}^{(i)},\ i=1,2,\ldots,d_1$ and basis functions $\boldsymbol{\beta}_{j}^{(i)},\ i=1,2,\ldots,d_{1},\  j=1,2\ldots,d_2$. This step reduces the temporal dimension from $(T-1)=300$ to $d_2=10$. Fig.~\ref{fig:s2}(c) shows examples of the reconstructed scalar functions $\tilde{\mathbf{H}}_{\alpha_m}^{(i)}$, and Fig.~\ref{fig:s3}(a) show an example of the reconstructed IS-TVF sequences $\tilde{\mathbf{G}}_{\alpha_m}^{(i)}$. The reconstructed posture sequences differ slightly from the original sequence, but the main patterns are captured well for both scalar functions $\mathbf{H}_{\alpha_m}^{(i)}$ and IS-TVF functions $\mathbf{G}_{\alpha_m}^{(i)}$. After this process, the dimension of an IS-TVF sequence is effectively reduced from $12000$ to $50$. Besides shape PCA, one can also use other nonlinear dimension reduction methods such as {\it autoencoder} (AE) and {\it variational autoencoder} (VAE) for the spatial reduction, followed by FPCA for the temporal part. The performance comparisons between shape PCA and AE/VAE for spatial dimension reduction are presented later.

\noindent {\bf Joint Dimension Reduction}:
A more comprehensive albeit complex approach for dimension reduction is to reduce spatial and temporal dimensions simultaneously rather than sequentially. Here we view IS-TVF sequences $\mathbf{G}_{\alpha_m}$ (or SIEM $\mathbf{W}_{\alpha_m}$) as a set of $N$th order tensors, $\{\mathcal{X}_m\in\mathbb{R}^{I_1\times I_2\times\ldots\times I_N},\ m=1,2,\ldots,M\}$, in spatial and time dimensions. Then we apply the multilinear principal component analysis (MPCA) framework~\citep{Lu2008MPCA}, essentially a tensor extension of the standard PCA. One can refer to \cite{Lu2008MPCA}'s paper for a detailed optimization algorithm. After the optimization, the feature tensor is obtained by a set of $N$ projection matrices $\tilde{\mathbf{U}}^{(i)}$ as:
$\mathcal{Z}_m=\mathcal{X}_m\times_1 \tilde{\mathbf{U}}^{(1)^{\top}} \times_2 \tilde{\mathbf{U}}^{(2)^{\top}}\times\ldots\times_N\tilde{\mathbf{U}}^{(N)^{\top}}\in\mathbb{R}^{d_1}\bigotimes\mathbb{R}^{d_2}\ldots\bigotimes\mathbb{R}^{d_N}$, where $\times_s$ is the $s$-mode product, $\bigotimes$ is the Kronecker product, and $d_s < I_s$ are the reduced dimensions. Conversely, the original tensor $\mathcal{X}_m$ can be reconstructed by the features $\mathcal{Z}_m$ as 
$\mathcal{X}_m=\mathcal{Z}_m\times_1 \tilde{\mathbf{U}}^{(1)} \times_2 \tilde{\mathbf{U}}^{(2)}\times\ldots\times_N\tilde{\mathbf{U}}^{(N)}$.

The IS-TVF sequences $\mathbf{G}_{\alpha_m}\in\mathbb{R}^{2(n-1)\times(T-1)}$ (alternatively SIEM $\mathbf{W}_{\alpha_m}$) can be viewed as $2$nd order tensor objects. As a result, we can obtain feature tensors $\mathcal{Z}_{\alpha_m}\in\mathbb{R}^{d_1}\bigotimes\mathbb{R}^{d_2}$ (with total dimension of $d_1\times d_2=5\times 4=20$ in this experiment), which is similar to the coefficient matrix $\boldsymbol{\Phi}_m\in\mathbb{R}^{d_1\times d_2}$ obtained by the sequential approach. Fig.~\ref{fig:s3} shows a reconstructed IS-TVF element $\tilde{\mathbf{G}}_{\alpha_m}^{(i)}$ and a reconstructed sequence $\tilde{\alpha}_m$ from its MPCA projection.

\section{Intermediate Steps of the Simulation Process}\label{secs3}
\begin{figure}[htbp]
    \centering
    \subfloat{
        \includegraphics[width=0.8\textwidth]{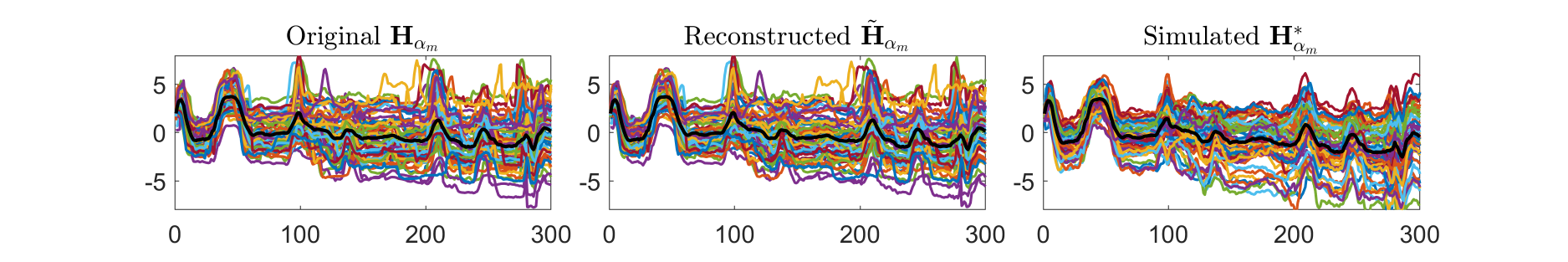}}

    \subfloat{
        \includegraphics[width=0.8\textwidth]{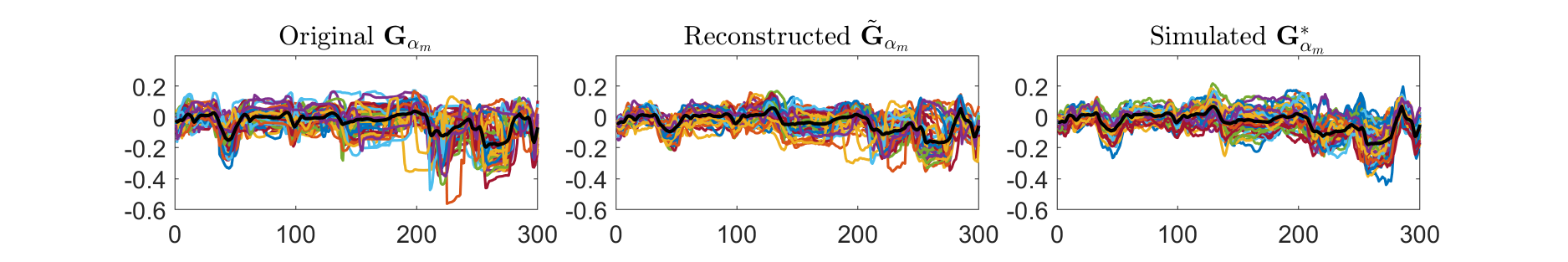}}

    \subfloat{
        \includegraphics[width=0.8\textwidth]{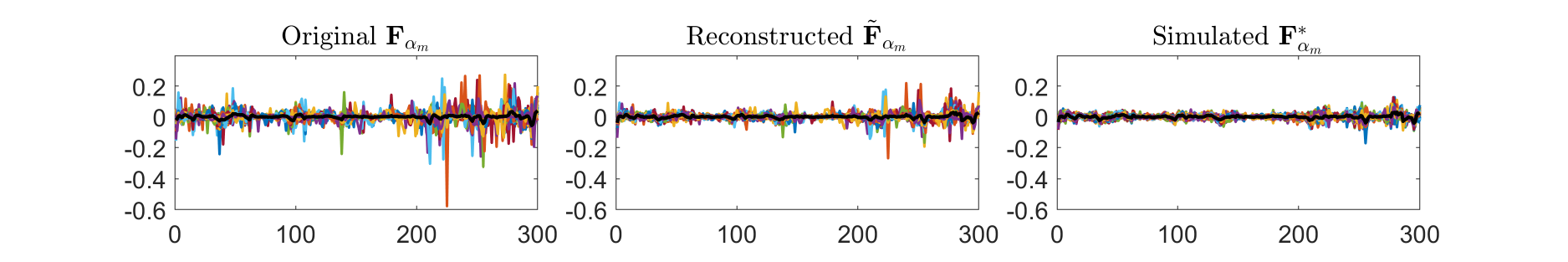}}

    \subfloat{
        \includegraphics[width=0.75\textwidth]{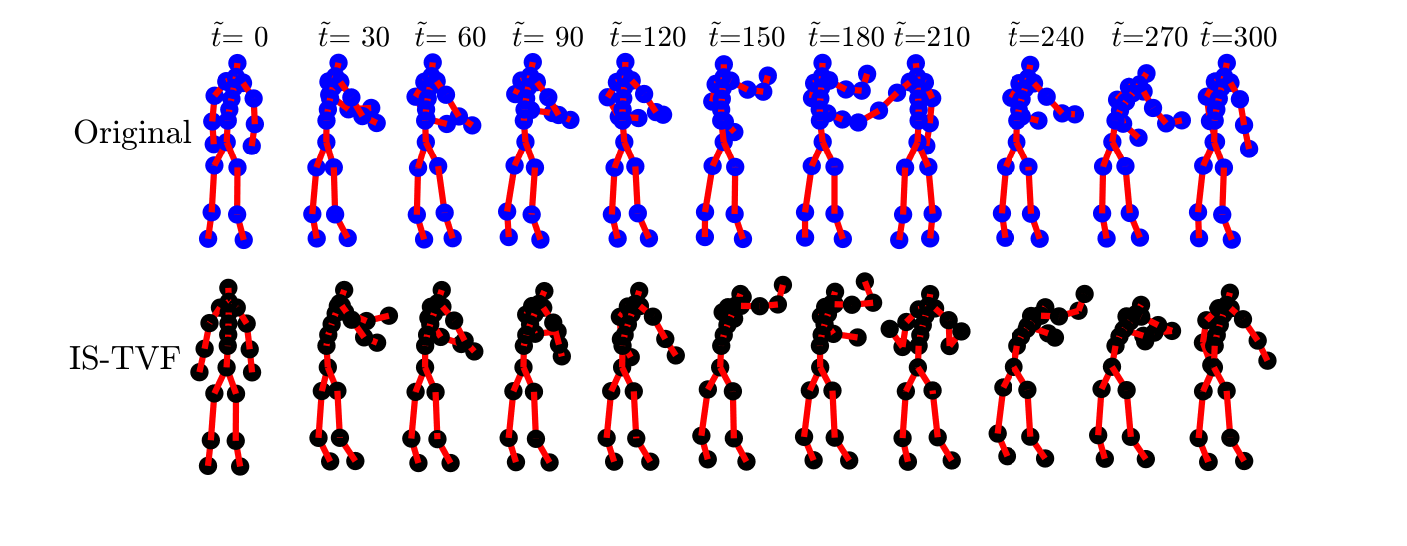}}
    
    \caption{Simulated functions and motion sequences of motion class one using {\it IS-TVF/SequentialPCA/MVG}. From the top to the bottom are one element of the PCA scalar functions $\mathbf{H}_{\alpha_m}$, one element of the IS-TVF function $\mathbf{G}_{\alpha_m}$, the same element of the S-TVF function $\mathbf{G}_{\alpha_m}$, and examples of simulated motion sequences.}
    \label{fig:s4}   
\end{figure}

\begin{figure}[htb]
    \centering
    \subfloat{
        \includegraphics[width=0.8\textwidth]{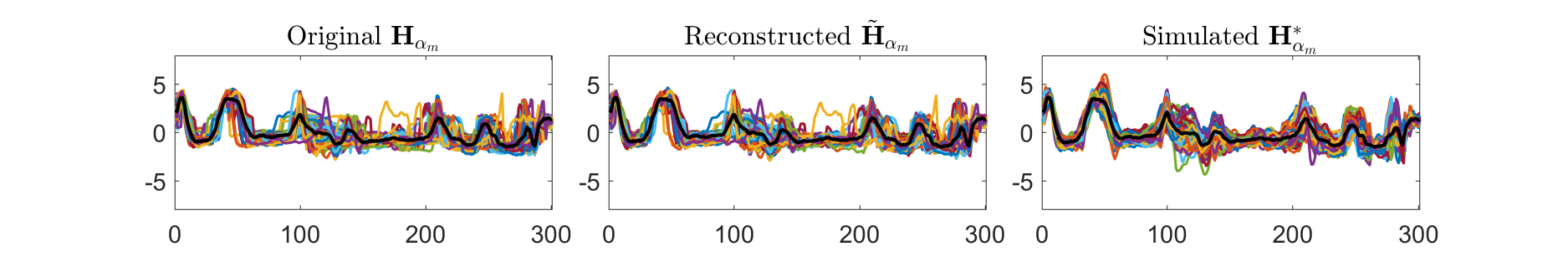}}
    
    \subfloat{
    \includegraphics[width=0.8\textwidth]{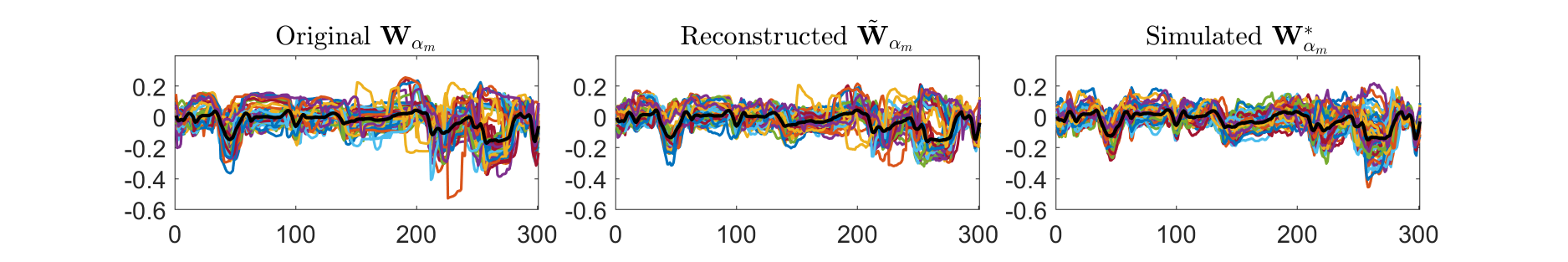}}

    \subfloat{
    \includegraphics[width=0.75\textwidth]{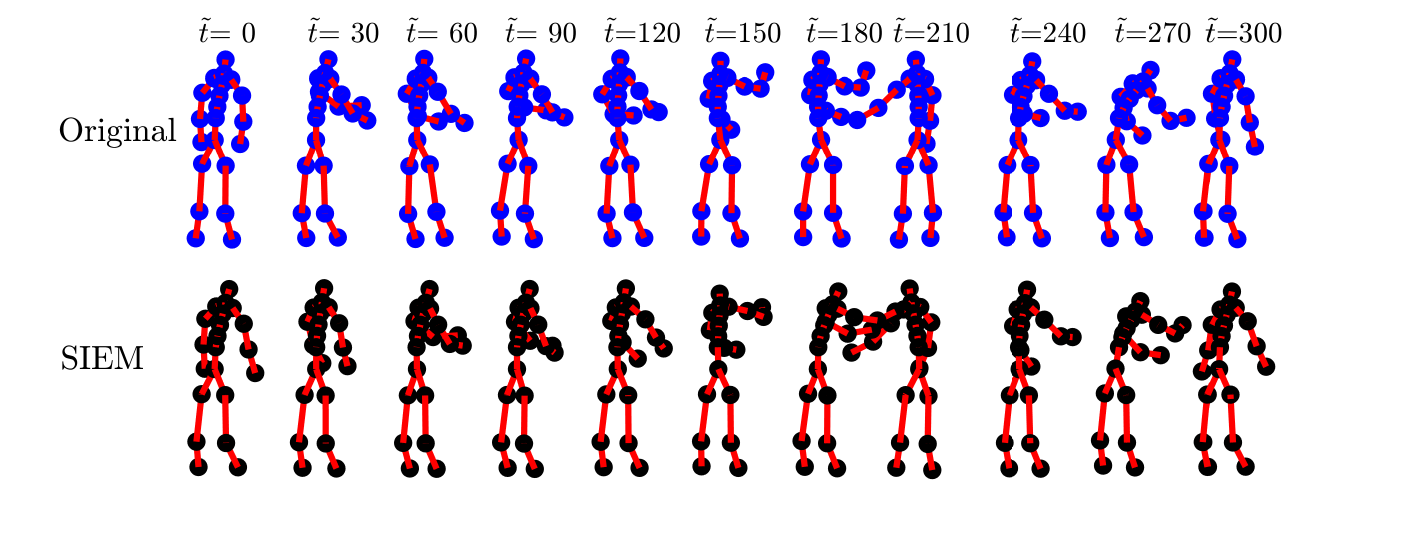}}
    \caption{Simulated functions and motion sequences of motion class one using {\it SIEM/SequentialPCA/MVG}. From the top to the bottom are one element of the PCA scalar functions $\mathbf{H}_{\alpha_m}$, one element of the SIEM function $\mathbf{W}_{\alpha_m}$, and an example of simulated sequences.}
    \label{fig:s5}   
\end{figure}

Here, we use the data from Worker Data motion class one as an example. The simulation uses {\it IS-TVF/SequentialPCA/MVG} (Fig.~\ref{fig:s4}) and {\it SIEM/SequentialPCA/MVG} (Fig.~\ref{fig:s5}) with a set of $d_1\times d_2=10\times 30$ PCA coefficients. Once a set of PCA coefficients $\mathbf{A}^*$ is generated, one can create a vector-valued function of the spatial PCA coefficients $\mathbf{H}^*$ by $\mathbf{H}^{*(i)}=\boldsymbol{\mu}^{(i)}+\sum_{j=1}^{d_2} a_{ij}^* \boldsymbol{beta}_j^{(i)}$, where $a_{ij}^*$ is one element of $\mathbf{A}^*$, $\boldsymbol{\mu}^{(i)}$ and $\boldsymbol{beta}_j^{(i)}$ are the mean and basis functions of the FPCA as discussed in Sec.~4-2. Subsequently, one can construct the IS-TVF (or SIEM) by $\mathbf{G}^*(t)(\text{or }\mathbf{W}^*(t))=\mathbf{m}+\mathbf{U} \mathbf{H}^*(t)$, where $\mathbf{m}$ is the mean vector and $\mathbf{U}$ is the spatial PCA projection matrix as discussed in Sec.~4-2. If IS-TVF is used, one can use finite differences to get S-TVF $\mathbf{G}^*$ and then compute the motion sequence $\alpha^*$; if SIEM is used, one can directly map $\mathbf{W}^*$ to the posture space to get $\alpha^*$ (Sec.~4-1). The following figures show some examples of these intermediate functions of the simulation process.

\noindent 1. {\bf \it IS-TVF/SequentialPCA/MVG}:
Fig.~\ref{fig:s4} shows the detailed simulation process for the {\it IS-TVF/SequentialPCA/MVG} model. While the simulated spatial PCA coefficient trajectories $\mathbf{H}_{\alpha_m}^*$ (the first panel) and IS-TVF functions $\mathbf{G}_{\alpha_m}^*$ (the second panel) are similar to the original functions, the simulated S-TVF functions $\mathbf{G}_{\alpha_m}^*$ (the third panel) are much smoother than the original functions. Fig.~\ref{fig:s4} (bottom) shows examples of the simulated sequences $\alpha^*$. One can see that the simulated sequences are realistic and reasonable compared to the real sequences. 

\noindent 2. {\bf \it SIEM/SequentialPCA/MVG}:
Similar to the previous case, Fig.~\ref{fig:s5} shows the simulation process of {\it SIEM/SequentialPCA/MVG} model. The top panel shows the simulated spatial PCA coefficient trajectories $\mathbf{H}_{\alpha_m}^*$, and the middle shows the SIEM functions $\mathbf{W}_{\alpha_m}^*$. In contrast to the IS-TVF approach, the simulated SIEM functions $\mathbf{W}_{\alpha_m}^*$ have a similar smoothness as the original SIEM functions $\mathbf{W}_{\alpha_m}$ (as shown in the second panel of Fig.~\ref{fig:s5}) because the SIEM mapping doesn't have extra numerical steps (the differentiating and integrating steps used in IS-TVF reconstructions). Fig.~\ref{fig:s5} (bottom) shows an example of the simulated sequences $\alpha^*$. The simulated sequences are also realistic and reasonable compared to the real sequences.

\section{Full Evaluation Results}\label{secs4}
\subsection{Detailed Experimental Results}\label{subsecs4-1}
This section provides the complete experimental results for all evaluation metrics presented in the main manuscript. While the main text reports the aggregated mean performance across 10 independent trials to maintain brevity, Tab.\ref{tab:s1} through \ref{tab:s7} detail both the averages and standard deviations ($\text{Mean}\pm\text{SD}$) for each motion class and method. Consistent with the main text, bold entries in these tables represent the best performance, while the second-best is indicated by an underline. For the four selected metrics: \textbf{Energy Distance}, \textbf{Jerk Energy}, \textbf{Posture Validity Score}, and \textbf{Quantization Variability}, the corresponding method rank is additionally provided as a subscript next to the performance values.
\begin{table}[ht]
    \footnotesize
    \centering
    \resizebox{0.95\textwidth}{!}{%
    \setlength{\tabcolsep}{2pt}
    \begin{tabular}{|c|P{2.6cm}|P{2.4cm}|c|P{2.2cm}|P{2.2cm}|c|P{2.2cm}|} 
        \hline
        \multirow{3}{*}{Motion Type}  & \multicolumn{7}{c|}{Model} \\ 
        \cline{2-8}
        & {\it IS-TVF/Sequen-tialPCA/MVG} & {\it SIEM/Sequen-tialPCA/MVG} & \multirow{2}{*}{PWI} & {\it IS-TVF/Spa-tialPCA/VAR} & {\it SIEM/Spa-tialPCA/GP} &\multirow{2}{*}{LSTM} & GCN-Transformer\\ 
        \hline
        Worker Motion 1 & \makecell{$1.32_{(3)}$\\$\pm 0.0566$} & \makecell{$\underline{0.35}_{(2)}$ \\ $\pm 0.264$} & \makecell{$\mathbf{0.25}_{(1)}$\\ $\pm 0.0007$} & \makecell{$ 6.94_{(7)}$ \\ $\pm 0.1186$} & \makecell{$ 6.20_{(4)}$ \\ $\pm 0.0117$} & \makecell{$ 6.38_{(5)}$ \\ $\pm 0.7521$} & \makecell{$ 6.49_{(6)}$ \\ $\pm 0.0575$}\\ 
        \hline
        Worker Motion 2 & \makecell{$ 1.90_{(3)}$ \\ $\pm 0.1040$} & \makecell{$ \underline{0.35}_{(2)}$ \\ $\pm 0.0280$} & \makecell{$ \mathbf{0.21}_{(1)}$ \\ $\pm 0.0015$} & \makecell{$ 10.82_{(7)}$ \\ $\pm 0.2122$} & \makecell{$ 7.30_{(5)}$ \\ $\pm 0.0195$} & \makecell{$ 6.05_{(4)}$ \\ $\pm 1.0205$} & \makecell{$ 7.41_{(6)}$ \\ $\pm 0.0458$}\\    
        \hline
        Worker Motion 3 & \makecell{$1.91_{(3)}$ \\ $\pm 0.0792$} & \makecell{$\underline{0.40}_{(2)}$ \\ $\pm 0.0208$} & \makecell{$ \mathbf{0.28}_{(1)}$ \\ $\pm 0.0012$} & \makecell{$ 5.40_{(6)}$ \\ $\pm 0.0582$} & \makecell{$ 5.04_{(4)}$ \\ $\pm 0.0116$} & \makecell{$ 5.22_{(5)}$ \\ $\pm 0.2437$} & \makecell{$ 5.48_{(7)}$ \\ $\pm 0.0658$}\\ 
        \hline
        Worker Motion 4 & \makecell{$ 1.42_{(3)}$ \\ $\pm 0.0739$} & \makecell{$ \underline{0.78}_{(2)}$ \\ $\pm 0.1192$} & \makecell{$\mathbf{0.62}_{(1)}$ \\ $\pm 0.0045$} & \makecell{$ 4.84_{(4)}$ \\ $\pm 0.0847$} & \makecell{$ 6.50_{(6)}$ \\ $\pm 0.0160$} & \makecell{$ 6.11_{(5)}$ \\ $\pm 0.6640$} & \makecell{$ 6.88_{(7)}$ \\ $\pm 0.0577$}\\     
        \hline
        Worker Motion 5 & \makecell{$ 1.84_{(3)}$ \\ $\pm 0.1318$} & \makecell{$ \underline{0.47}_{(2)}$ \\ $\pm 0.0186$} & {\makecell{$\mathbf{0.31}_{(1)}$ \\ $\pm 0.0011$}} & \makecell{$ 9.03_{(7)}$ \\ $\pm 0.0517$} & \makecell{$ 6.42_{(6)}$ \\ $\pm 0.0160$} & \makecell{$ 5.11_{(4)}$ \\ $\pm 0.0.3481$} & \makecell{$ 6.00_{(5)}$ \\ $\pm 0.0588$}\\  
        \hline
        Exercise Motion & \makecell{$ 5.57_{(3)}$ \\ $\pm 0.0069$} & \makecell{$\underline{0.57}_{(2)}$ \\ $\pm 0.0149$}& \makecell{$\mathbf{0.29}_{(1)}$ \\ $\pm 0.0008$} & \makecell{$ 44.03_{(7)}$ \\ $\pm 0.8538$} & \makecell{$ 14.75_{(5)}$ \\ $\pm 0.0242$} & \makecell{$ 16.41_{(6)}$ \\ $\pm 0.2.0285$} & \makecell{$ 13.18_{(4)}$ \\ $\pm 1.3798$} \\
        \hline
        \multicolumn{7}{c}{\scriptsize Note: Best result for each motion class is in \textbf{bold} and the second best is \underline{underlined}.}
    \end{tabular}}
        \caption{Energy distances of the simulated sequences compared against the original sequences.}
        \label{tab:s1}
\end{table}

\begin{table}[htbp]
    \footnotesize
    \centering
    \resizebox{0.95\textwidth}{!}{%
    \setlength{\tabcolsep}{2pt}
    \setlength{\tabcolsep}{2pt}
    \begin{tabular}{|c|P{2.6cm}|P{2.4cm}|c|P{2.2cm}|P{2.2cm}|c|P{2.2cm}|} 
        \hline
        \multirow{3}{*}{Motion Type}  & \multicolumn{7}{c|}{Model} \\ 
        \cline{2-8}
        & {\it IS-TVF/Sequen-tialPCA/MVG} & {\it SIEM/Sequen-tialPCA/MVG} & \multirow{2}{*}{PWI} & {\it IS-TVF/Spa-tialPCA/VAR} & {\it SIEM/Spa-tialPCA/GP} & \multirow{2}{*}{LSTM} & GCN-Transformer\\
        \hline
        Worker Motion 1 (1.66) & \makecell{$ 4.36$ \\ $\pm 0.2822$} & \makecell{$\underline{ 1.47}$ \\ $\pm 0.0747$} & \makecell{$\mathbf{1.66}$ \\ $\pm 0.0059$} & \makecell{$ 3.93$ \\ $\pm 0.1323$} & \makecell{$ 0.08$ \\ $\pm 0.0011$} & \makecell{$ 0.18$ \\ $\pm 0.0455$} & \makecell{$ 0.11$ \\ $\pm 0.0014$} \\ 
        \hline
        Worker Motion 2 (1.30) & \makecell{$ 5.17$ \\ $\pm 0.4878$} & \makecell{$ \underline{1.19}$ \\ $\pm 0.0537$} & \makecell{$\mathbf{1.31}$ \\ $\pm 0.0034$} & \makecell{$ 2.83$ \\ $\pm 0.0833$} & \makecell{$ 0.09$ \\ $\pm 0.0012$} & \makecell{$ 0.49$ \\ $\pm 0.2639$} & \makecell{$ 0.26$ \\ $\pm 0.0013$} \\   
        \hline
        Worker Motion 3 (1.22) & \makecell{$ 4.40$ \\ $\pm 0.2365$} & \makecell{$\underline{1.07}$ \\ $\pm 0.0548$} & \makecell{$\mathbf{1.22}$ \\ $\pm 0.0034$} & \makecell{$ 3.31$ \\ $\pm 0.1161$} & \makecell{$ 0.08$ \\ $\pm 0.0010$} & \makecell{$ 0.14$ \\ $\pm 0.0264$} & \makecell{$ 0.11$ \\ $\pm 0.0011$} \\
        \hline
        Worker Motion 4 (2.67) & \makecell{$ 4.90$ \\ $\pm 0.3612$} & \makecell{$\underline{2.61}$ \\ $\pm 0.1511$} & \makecell{$\mathbf{2.67}$ \\ $\pm 0.0071$} & \makecell{$ 3.38$ \\ $\pm 0.0817$} & \makecell{$ 0.10$ \\ $\pm 0.0013$} & \makecell{$ 0.73$ \\ $\pm 0.0937$} & \makecell{$0.41$ \\ $\pm 0.0016$} \\     
        \hline
        Worker Motion 5 (1.47) & \makecell{$ 4.80$ \\ $\pm 0.3758$} & \makecell{$\underline{1.34}$ \\ $\pm 0.0761$} & \makecell{$\mathbf{1.46}$ \\ $\pm 0.0031$} & \makecell{$ 1.70$ \\ $\pm 0.0529$} & \makecell{$ 0.08$ \\ $\pm 0.0011$} & \makecell{$ 0.39$ \\ $\pm 0.0532$} & \makecell{$ 0.23$ \\ $\pm 0.0011$} \\  
        \hline
        Exercise Motion (2.16) & \makecell{$ 5.87$ \\ $\pm 0.1153$} & \makecell{$\underline{1.48}$ \\ $\pm 0.0416$} & \makecell{$ \mathbf{2.15}$ \\ $\pm 0.0042$} & \makecell{$ 15.94$ \\ $\pm 0.2621$} & \makecell{$ 0.15$ \\ $\pm 0.0023$} & \makecell{$0.34$ \\ $\pm 0.0814$} & \makecell{$ 0.25$ \\ $\pm 0.0026$} \\
        \hline
        \multicolumn{7}{c}{\scriptsize Note: Best result for each motion class is in \textbf{bold} and the second best is \underline{underlined}.}
    \end{tabular}}
    \caption{Mean of the cross-sectional variance over time.}
    \label{tab:s2}
\end{table}

\begin{table}[htbp]
    \footnotesize
    \centering
    \resizebox{0.95\textwidth}{!}{%
    \setlength{\tabcolsep}{2pt}
    \begin{tabular}{|c|c|P{2.6cm}|P{2.4cm}|c|P{2.2cm}|P{2.2cm}|c|P{2.2cm}|} 
        \hline
        \multirow{3}{*}{Motion Type}  &\multirow{3}{*}{Original} & \multicolumn{7}{c|}{Model} \\ 
        \cline{3-9}
        &  & {\it IS-TVF/Sequen-tialPCA/MVG} & {\it SIEM/Sequen-tialPCA/MVG} & \multirow{2}{*}{PWI} & {\it IS-TVF/Spa-tialPCA/VAR} & {\it SIEM/Spa-tialPCA/GP}& \multirow{2}{*}{LSTM} & {GCN-Transformer}\\ 
        \hline
        Worker Motion 1 & $1.07$ & \makecell{$ 0.63_{(3)}$ \\ $\pm 0.0230$} & \makecell{$\underline{0.69}_{(2)}$ \\ $\pm 0.0239$} & \makecell{$ 31.65_{(7)}$ \\ $\pm 0.0724$} & \makecell{$ \mathbf{1.18}_{(1)}$ \\ $\pm 0.0058$} & \makecell{$ 0.11_{(4)}$ \\ $\pm 0.0003$} & \makecell{$ 0.10_{(5)}$ \\ $\pm 0.0002$} & \makecell{$2.32_{(6)}$ \\ $\pm 0.5111$} \\ 
        \hline
        Worker Motion 2 & $0.98$ & \makecell{$ 0.57_{(3)}$ \\ $\pm 0.0100$} & \makecell{$ \underline{0.63}_{(2)}$ \\ $\pm 0.0110$} & \makecell{$ 24.58_{(7)}$ \\ $\pm 0.0731$} & \makecell{$ \mathbf{0.96}_{(1)}$ \\ $\pm 0.0033$} & \makecell{$ 0.12_{(5)}$ \\ $\pm 0.0003$} & \makecell{$ 0.20_{(4)}$ \\ $\pm 0.0002$} & \makecell{$2.36_{(6)}$ \\ $\pm 0.4939$} \\    
        \hline
        Worker Motion 3 & $1.41$ & \makecell{$ 0.84_{(3)}$ \\ $\pm 0.0140$} & \makecell{$\underline{0.93}_{(2)}$ \\ $\pm 0.0176$} & \makecell{$ 23.65_{(7)}$ \\ $\pm 0.0996$} & \makecell{$ \mathbf{1.31}_{(1)}$ \\ $\pm 0.0071$} & \makecell{$ 0.11_{(6)}$ \\ $\pm 0.0003$} & \makecell{$ 0.13_{(5)}$ \\ $\pm 0.0002$} & \makecell{$2.24_{(4)}$ \\ $\pm 0.5032$} \\ 
        \hline
        Worker Motion 4 & $1.22$ & \makecell{$ 0.71_{(3)}$ \\ $\pm 0.0231$} & \makecell{$\underline{0.94}_{(2)}$ \\ $\pm 0.0263$} & \makecell{$ 48.97_{(7)}$ \\ $\pm 0.1413$} & \makecell{$ \mathbf{1.12}_{(1)}$ \\ $\pm 0.0036$} & \makecell{$ 0.15_{(5)}$ \\ $\pm 0.0004$} & \makecell{$ 0.17_{(4)}$ \\ $\pm 0.0004$} & \makecell{$2.47_{(6)}$ \\ $\pm 0.5068$} \\     
        \hline
        Worker Motion 5 & $0.73$ & \makecell{$ 0.44_{(3)}$ \\ $\pm 0.0143$} & \makecell{$ \mathbf{0.87}_{(1)}$ \\ $\pm 0.0364$} & \makecell{$ 27.48_{(7)}$ \\ $\pm 0.1117$} & \makecell{$ \underline{0.58}_{(2)}$ \\ $\pm 0.0030$} & \makecell{$ 0.11_{(4)}$ \\ $\pm 0.0003$} & \makecell{$ 0.07_{(5)}$ \\ $\pm 0.0007$} & \makecell{$2.50_{(6)}$ \\ $\pm 0.4692$} \\  
        \hline
        Exercise Motion & $4.22$ & \makecell{$ 1.42_{(3)}$ \\ $\pm 0.0262$} & \makecell{$\underline{2.42}_{(2)}$ \\ $\pm 0.0501$} & \makecell{$ 37.79_{(7)}$ \\ $\pm 0.1171$} & \makecell{$ \mathbf{3.72}_{(1)}$ \\ $\pm 0.1149$} & \makecell{$ 0.04_{(6)}$ \\ $\pm 0.0001$} & \makecell{$ 0.68_{(5)}$ \\ $\pm 0.0037$} & \makecell{$7.56_{(4)}$ \\ $\pm 5.2129$} \\
        \hline
        \multicolumn{8}{c}{\scriptsize Note: Best result for each motion class is in \textbf{bold} and the second best is \underline{underlined}.}
    \end{tabular}}
    \caption{Jerk energy of the simulated sequences compared against the original sequences.}
    \label{tab:s3}
\end{table}

\begin{table}[htbp]
    \footnotesize
    \centering
    \resizebox{0.95\textwidth}{!}{%
    \setlength{\tabcolsep}{2pt}
    \begin{tabular}{|c|c|P{2.6cm}|P{2.4cm}|c|P{2.2cm}|P{2.2cm}|c|P{2.2cm}|} 
        \hline
        \multirow{3}{*}{Motion Type}  &\multirow{3}{*}{Original} & \multicolumn{7}{c|}{Model} \\ 
        \cline{3-9}
        &  & {\it IS-TVF/Sequen-tialPCA/MVG} & {\it SIEM/Sequen-tialPCA/MVG} & \multirow{2}{*}{PWI} & {\it IS-TVF/Spa-tialPCA/VAR} & {\it SIEM/Spa-tialPCA/GP}& \multirow{2}{*}{LSTM} & {GCN-Transformer}\\ 
        \hline
        Worker Motion 1 & $0.42$ & \makecell{$ 0.26$ \\ $\pm 0.0089$} & \makecell{$\underline{0.28}$ \\ $\pm 0.0093$} & \makecell{$ 9.56$ \\ $\pm 0.0214$} & \makecell{$ \mathbf{0.47}$ \\ $\pm 0.0026$} & \makecell{$ 0.04$ \\ $\pm 0.0001$} & \makecell{$ 0.04$ \\ $\pm 0.0001$} & \makecell{$0.71$ \\ $\pm 0.1528$} \\ 
        \hline
        Worker Motion 2 & $0.38$ & \makecell{$ 0.23$ \\ $\pm 0.0036$} & \makecell{$\underline{0.25}$ \\ $\pm 0.0037$} & \makecell{$ 7.45$ \\ $\pm 0.0190$} & \makecell{$ \mathbf{0.40}$ \\ $\pm 0.0010$} & \makecell{$ 0.04$ \\ $\pm 0.0001$} & \makecell{$ 0.08$ \\ $\pm 0.0001$} & \makecell{$0.73$ \\ $\pm 0.1475$} \\    
        \hline
        Worker Motion 3 & $0.54$ & \makecell{$ 0.34$ \\ $\pm 0.0059$} & \makecell{$0.37$ \\ $\pm 0.0066$} & \makecell{$ 7.15$ \\ $\pm 0.0275$} & \makecell{$ \mathbf{0.53}$ \\ $\pm 0.0026$} & \makecell{$ 0.04$ \\ $\pm 0.0001$} & \makecell{$ 0.05$ \\ $\pm 0.0000$} & \makecell{$\underline{0.68}$ \\ $\pm 0.1506$} \\ 
        \hline
        Worker Motion 4 & $0.50$ & \makecell{$ 0.32$ \\ $\pm 0.0106$} & \makecell{$\underline{0.40}$ \\ $\pm 0.0114$} & \makecell{$ 14.85$ \\ $\pm 0.0410$} & \makecell{$ \mathbf{0.45}$ \\ $\pm 0.0013$} & \makecell{$ 0.05$ \\ $\pm 0.0001$} & \makecell{$ 0.07$ \\ $\pm 0.0001$} & \makecell{$0.76$ \\ $\pm 0.1517$} \\     
        \hline
        Worker Motion 5 & $0.31$ & \makecell{$ 0.20$ \\ $\pm 0.0060$} & \makecell{$ \mathbf{0.34}$ \\ $\pm 0.0123$} & \makecell{$ 8.33$ \\ $\pm 0.0302$} & \makecell{$ \underline{0.24}$ \\ $\pm 0.0014$} & \makecell{$ 0.04$ \\ $\pm 0.0001$} & \makecell{$ 0.04$ \\ $\pm 0.0003$} & \makecell{$ 0.77$ \\ $\pm 0.1404$} \\  
        \hline
        Exercise Motion & $1.58$ & \makecell{$ 0.58$ \\ $\pm 0.0102$} & \makecell{$\underline{0.95}$ \\ $\pm 0.0172$} & \makecell{$ 11.62$ \\ $\pm 0.0323$} & \makecell{$ \mathbf{1.36}$ \\ $\pm 0.0404$} & \makecell{$ 0.01$ \\ $\pm 0.0000$} & \makecell{$ 0.26$ \\ $\pm 0.0108$} & \makecell{$2.30$ \\ $\pm 1.5258$} \\
        \hline
        \multicolumn{8}{c}{\scriptsize Note: Best result for each motion class is in \textbf{bold} and the second best is \underline{underlined}.}
    \end{tabular}}
    \caption{Acceleration energy of the simulated sequences compared against the original sequences.}
    \label{tab:s4}
\end{table}

\begin{table}[htbp]
    \footnotesize
    \centering
    \resizebox{0.95\textwidth}{!}{%
    \setlength{\tabcolsep}{2pt}
    \begin{tabular}{|c|c|P{2.6cm}|P{2.4cm}|c|P{2.2cm}|P{2.2cm}|c|P{2.2cm}|} 
        \hline
        \multirow{3}{*}{Motion Type}  &\multirow{3}{*}{Original} & \multicolumn{7}{c|}{Model} \\ 
        \cline{3-9}
        &  & {\it IS-TVF/Sequen-tialPCA/MVG} & {\it SIEM/Sequen-tialPCA/MVG} & \multirow{2}{*}{PWI} & {\it IS-TVF/Spa-tialPCA/VAR} & {\it SIEM/Spa-tialPCA/GP}& \multirow{2}{*}{LSTM} & GCN-Transformer\\ 
        \hline
        Worker Motion 1 & $0.97$ & \makecell{$ 0.92_{(6)}$ \\ $\pm 0.0117$} & \makecell{$ \underline{0.98}_{(2)}$ \\ $\pm 0.0013$} & \makecell{$ 0.86_{(7)}$ \\ $\pm 0.0007$} & \makecell{$ \mathbf{0.97}_{(1)}$ \\ $\pm 0.0021$} & \makecell{$ 1.00_{(5)}$ \\ $\pm 0.0000$} & \makecell{$ 0.99_{(3)}$ \\ $\pm 0.0148$} & \makecell{$ 1.00_{(4)}$ \\ $\pm 0.0013$} \\ 
        \hline
        Worker Motion 2 & $0.96$ & \makecell{$ 0.88_{(6)}$ \\ $\pm 0.0124$} & \makecell{$0.97_{(3)}$ \\ $\pm 0.0022$} & \makecell{$ 0.88_{(7)}$ \\ $\pm 0.0002$} & \makecell{$ \mathbf{0.97}_{(1)}$ \\ $\pm 0.0023$} & \makecell{$ 1.00_{(5)}$ \\ $\pm 0.0000$} & \makecell{$ 0.97_{(4)}$ \\ $\pm 0.0004$} & \makecell{$ \underline{0.97}_{(2)}$ \\ $\pm 0.0018$} \\    
        \hline
        Worker Motion 3 & $0.97$ & \makecell{$ 0.89_{(7)}$ \\ $\pm 0.0147$} & \makecell{$ \mathbf{0.98}_{(1)}$ \\ $\pm 0.0016$} & \makecell{$ 0.90_{(6)}$ \\ $\pm 0.0006$} & \makecell{$ \underline{0.98}_{(2)}$ \\ $\pm 0.0016$} & \makecell{$ 1.00_{(5)}$ \\ $\pm 0.0000$} & \makecell{$ 1.00_{(4)}$ \\ $\pm 0.0000$} & \makecell{$ 1.00_{(3)}$ \\ $\pm 0.0012$} \\ 
        \hline
        Worker Motion 4 & $0.94$ & \makecell{$ 0.89_{(4)}$ \\ $\pm 0.0117$} & \makecell{$ \mathbf{0.93}_{(1)}$ \\ $\pm 0.0051$} & \makecell{$ 0.79_{(7)}$ \\ $\pm 0.0008$} & \makecell{$\underline{0.97}_{(2)}$ \\ $\pm 0.0017$} & \makecell{$ 1.00_{(6)}$ \\ $\pm 0.0000$} & \makecell{$ 0.98_{(3)}$ \\ $\pm 0.0014$} & \makecell{$0.99_{(5)}$ \\ $\pm 0.0007$} \\     
        \hline
        Worker Motion 5 & $0.90$ & \makecell{$0.80_{(6)}$ \\ $\pm 0.0187$} & \makecell{$ \mathbf{0.91}_{(1)}$ \\ $\pm 0.0031$} & \makecell{$ 0.79_{(7)}$ \\ $\pm 0.0008$} & \makecell{$ 1.00_{(4)}$ \\ $\pm 0.0005$} & \makecell{$ 1.00_{(5)}$ \\ $\pm 0.0000$} & \makecell{$\underline{0.96}_{(2)}$ \\ $\pm 0.0030$} & \makecell{$ 0.96_{(3)}$ \\ $\pm 0.0055$} \\  
        \hline
        Exercise Motion & $0.95$ & \makecell{$ 0.81_{(3)}$ \\ $\pm 0.0053$} & \makecell{$ \mathbf{0.96}_{(1)}$ \\ $\pm 0.0014$} & \makecell{$ 0.87_{(5)}$ \\ $\pm 0.0003$} & \makecell{$ 0.32_{(7)}$ \\ $\pm 0.0092$} & \makecell{$1.00_{(4)}$ \\ $\pm 0.0000$} & \makecell{$\underline{0.94}_{(2)}$ \\ $\pm 0.0305$} & \makecell{$ 0.99_{(3)}$ \\ $\pm 0.0050$} \\
        \hline
        \multicolumn{8}{c}{\scriptsize Note: Best result for each motion class is in \textbf{bold} and the second best is \underline{underlined}.}
    \end{tabular}}
    \caption{Posture validity scores of the simulated sequences compared against the original sequences.}
    \label{tab:s5}
\end{table}

\begin{table}[htbp]
    \footnotesize
    \centering
    \resizebox{0.95\textwidth}{!}{%
    \setlength{\tabcolsep}{2pt}
    \begin{tabular}{|c|c|P{2.6cm}|P{2.4cm}|c|P{2.2cm}|P{2.2cm}|c|P{2.2cm}|} 
        \hline
        \multirow{3}{*}{Motion Type}  &\multirow{3}{*}{Original} & \multicolumn{7}{c|}{Model} \\ 
        \cline{3-9}
        &  & {\it IS-TVF/Sequen-tialPCA/MVG} & {\it SIEM/Sequen-tialPCA/MVG} & \multirow{2}{*}{PWI} & {\it IS-TVF/Spa-tialPCA/VAR} & {\it SIEM/Spa-tialPCA/GP}& \multirow{2}{*}{LSTM} & GCN-Transformer\\ 
        \hline
        Worker Motion 1 & $0.86$ & \makecell{$ 0.70$ \\ $\pm 0.0266$} & \makecell{$ \mathbf{0.90}$ \\ $\pm 0.0062$} & \makecell{$ 0.27$ \\ $\pm 0.0024$} & \makecell{$\underline{0.78}$ \\ $\pm 0.0132$} & \makecell{$ 1.00$ \\ $\pm 0.0000$} & \makecell{$ 0.95$ \\ $\pm 0.1330$} & \makecell{$ 0.98$ \\ $\pm 0.0114$} \\ 
        \hline
        Worker Motion 2 & $0.85$ & \makecell{$ 0.53$ \\ $\pm 0.0388$} & \makecell{$ \mathbf{0.92}$ \\ $\pm 0.0064$} & \makecell{$ 0.42$ \\ $\pm 0.0013$} & \makecell{$ \underline{0.79}$ \\ $\pm 0.0141$} & \makecell{$ 1.00$ \\ $\pm 0.0000$} & \makecell{$ 0.95$ \\ $\pm 0.0036$} & \makecell{$ 0.94$ \\ $\pm 0.0160$} \\    
        \hline
        Worker Motion 3 & $0.86$ & \makecell{$ 0.70$ \\ $\pm 0.0305$} & \makecell{$ \mathbf{0.90}$ \\ $\pm 0.0045$} & \makecell{$ 0.46$ \\ $\pm 0.0028$} & \makecell{$ \underline{0.92}$ \\ $\pm 0.0080$} & \makecell{$ 1.00$ \\ $\pm 0.0000$} & \makecell{$ 0.98$ \\ $\pm 0.0000$} & \makecell{$ 0.97$ \\ $\pm 0.0110$} \\ 
        \hline
        Worker Motion 4 & $0.82$ & \makecell{$ 0.64$ \\ $\pm 0.0316$} & \makecell{$ \mathbf{0.81}$ \\ $\pm 0.0139$} & \makecell{$ 0.18$ \\ $\pm 0.0021$} & \makecell{$ \underline{0.85}$ \\ $\pm 0.0074$} & \makecell{$ 1.00$ \\ $\pm 0.0000$} & \makecell{$ 0.97$ \\ $\pm 0.0031$} & \makecell{$ 0.97$ \\ $\pm 0.0066$} \\     
        \hline
        Worker Motion 5 & $0.70$ & \makecell{$ 0.42$ \\ $\pm 0.0297$} & \makecell{$ \mathbf{0.75}$ \\ $\pm 0.0112$} & \makecell{$ 0.24$ \\ $\pm 0.0017$} & \makecell{$0.97$ \\ $\pm 0.0037$} & \makecell{$ 1.00$ \\ $\pm 0.0000$} & \makecell{$0.92$ \\ $\pm 0.0087$} & \makecell{$ \underline{0.89}$ \\ $\pm 0.0438$} \\  
        \hline
        Exercise Motion & $0.67$ & \makecell{$ 0.14$ \\ $\pm 0.0103$} & \makecell{$ \mathbf{0.71}$ \\ $\pm 0.0056$} & \makecell{$ 0.31$ \\ $\pm 0.0011$} & \makecell{$ 0.02$ \\ $\pm 0.0037$} & \makecell{$1.00$ \\ $\pm 0.0000$} & \makecell{$ \underline{0.47}$ \\ $\pm 0.3228$} & \makecell{$ 0.88$ \\ $\pm 0.0425$} \\
        \hline
        \multicolumn{8}{c}{\scriptsize Note: Best result for each motion class is in \textbf{bold} and the second best is \underline{underlined}.}
    \end{tabular}}
    \caption{Posture integrity rate of the simulated sequences compared against the original sequences.}
    \label{tab:s6}
\end{table}

\begin{table}[htbp]
    \footnotesize
    \centering
    \resizebox{0.95\textwidth}{!}{%
    \setlength{\tabcolsep}{2pt}
    \begin{tabular}{|c|c|P{2.6cm}|P{2.4cm}|c|P{2.2cm}|P{2.2cm}|c|P{2.2cm}|} 
        \hline
        \multirow{3}{*}{Motion Type}  &\multirow{3}{*}{Original} & \multicolumn{7}{c|}{Model} \\ 
        \cline{3-9}
        &  & {\it IS-TVF/Sequen-tialPCA/MVG} & {\it SIEM/Sequen-tialPCA/MVG} & \multirow{2}{*}{PWI} & {\it IS-TVF/Spa-tialPCA/VAR} & {\it SIEM/Spa-tialPCA/GP}& \multirow{2}{*}{LSTM} & GCN-Transformer\\ 
        \hline
        Worker Motion 1 & $0.40$ & \makecell{$ \underline{0.54}_{(2)}$ \\ $\pm 0.0230$} & \makecell{$ \mathbf{0.41}_{(1)}$ \\ $\pm 0.0074$} & \makecell{$ 0.24_{(3)}$ \\ $\pm 0.0021$} & \makecell{$ 0.89_{(7)}$ \\ $\pm 0.0049$} & \makecell{$ 0.74_{(6)}$ \\ $\pm 0.0020$} & \makecell{$ 0.67_{(5)}$ \\ $\pm 0.0697$} & \makecell{$ 0.67_{(4)}$ \\ $\pm 0.0018$} \\ 
        \hline
        Worker Motion 2 & $0.32$ & \makecell{$ 0.60_{(4)}$ \\ $\pm 0.0107$} & \makecell{$ \mathbf{0.33}_{(1)}$ \\ $\pm 0.0097$} & \makecell{$\underline{0.16}_{(2)}$ \\ $\pm 0.0017$} & \makecell{$ 0.97_{(7)}$ \\ $\pm 0.0038$} & \makecell{$ 0.94_{(6)}$ \\ $\pm 0.0040$} & \makecell{$ 0.52_{(3)}$ \\ $\pm 0.1760$} & \makecell{$ 0.92_{(5)}$ \\ $\pm 0.0058$} \\    
        \hline
        Worker Motion 3 & $0.35$ & \makecell{$ 0.63_{(6)}$ \\ $\pm 0.0132$} & \makecell{$0.41_{(3)}$ \\ $\pm 0.0093$} & \makecell{$ 0.18_{(5)}$ \\ $\pm 0.0021$} & \makecell{$ 0.76_{(7)}$ \\ $\pm 0.0056$} & \makecell{$ 0.43_{(4)}$ \\ $\pm 0.0015$} & \makecell{$\underline{0.41}_{(2)}$ \\ $\pm 0.0050$} & \makecell{$\mathbf{0.40}_{(1)}$ \\ $\pm 0.0000$} \\ 
        \hline
        Worker Motion 4 & $0.47$ & \makecell{$\underline{0.64}_{(2)}$ \\ $\pm 0.0157$} & \makecell{$ \mathbf{0.56}_{(1)}$ \\ $\pm 0.0115$} & \makecell{$ 0.28_{(3)}$ \\ $\pm 0.0019$} & \makecell{$ 0.89_{(7)}$ \\ $\pm 0.0038$} & \makecell{$ 0.82_{(6)}$ \\ $\pm 0.0053$} & \makecell{$ 0.73_{(5)}$ \\ $\pm 0.0502$} & \makecell{$0.72_{(4)}$ \\ $\pm 0.0041$} \\     
        \hline
        Worker Motion 5 & $0.32$ & \makecell{$ 0.53_{(4)}$ \\ $\pm 0.0148$} & \makecell{$ \mathbf{0.34}_{(1)}$ \\ $\pm 0.0085$} & \makecell{$\underline{0.16}_{(2)}$ \\ $\pm 0.0013$} & \makecell{$ 0.88_{(7)}$ \\ $\pm 0.0045$} & \makecell{$ 0.68_{(6)}$ \\ $\pm 0.0024$} & \makecell{$ 0.51_{(3)}$ \\ $\pm 0.0553$} & \makecell{$0.63_{(5)}$ \\ $\pm 0.0051$} \\  
        \hline
        Exercise Motion & $0.19$ & \makecell{$ 0.37_{(3)}$ \\ $\pm 0.0058$} & \makecell{$ \mathbf{0.15}_{(1)}$ \\ $\pm 0.0018$} & \makecell{$\underline{0.08}_{(2)}$ \\ $\pm 0.0006$} & \makecell{$ 0.88_{(7)}$ \\ $\pm 0.0039$} & \makecell{$ 0.74_{(6)}$ \\ $\pm 0.0001$} & \makecell{$ 0.72_{(5)}$ \\ $\pm 0.0774$} & \makecell{$0.65_{(4)}$ \\ $\pm 0.0515$} \\
        \hline
        \multicolumn{8}{c}{\scriptsize Note: Best result for each motion class is in \textbf{bold} and the second best is \underline{underlined}.}
    \end{tabular}}
    \caption{Quantization variability of the simulated sequences compared against the original sequences.}
    \label{tab:s7}
\end{table}

\color{black}
\subsection{Additional Evaluation Metrics}\label{subsecs4-2}
We use several additional metrics to evaluate the simulated sequences. This section describes these metrics and presents the results.

\noindent 1. {\bf KNN Classifier}: 
One can evaluate samples through classification. We pool the original and simulated sequences and use a KNN classifier to perform one-versus-rest classification for each sequence. If the two samples are from the same distribution, they should be indistinguishable and have an accuracy of $0.5$. On the contrary, if the two samples are from very different distributions, then the classification should be perfect at $1.0$.

Tab.~\ref{tab:s8} shows the classification accuracy of the KNN classifier. Despite the PWI model achieving an ideal accuracy of $0.5$ on all motion classes, the simulated sequences are tightly clustered around the mean sequences, making them easily distinguishable from the real motion. The proposed {\it SIEM/SequentialPCA/MVG} model yields the second-best results across all motion classes.

\renewcommand{\arraystretch}{0.7}
\begin{table}[htbp]
    \footnotesize
    \centering
    \resizebox{0.95\textwidth}{!}{%
    \setlength{\tabcolsep}{4pt}
    \begin{tabular}{|l|p{2.6cm}|p{2.4cm}|l|p{2.2cm}|p{2.2cm}|l|p{2.2cm}|} 
        \hline
        \multirow{3}{*}{Motion Type}  & \multicolumn{7}{c|}{Model} \\ 
        \cline{2-8}
        & {\it IS-TVF/Sequen-tialPCA/MVG} & {\it SIEM/Sequen-tialPCA/MVG} & \multirow{2}{*}{PWI} & {\it IS-TVF/Spa-tialPCA/VAR} & {\it SIEM/Spa-tialPCA/GP} & \multirow{2}{*}{LSTM} & GCN-Transformer\\
        \hline
        Worker Motion 1 & \makecell{$ 0.91$ \\ $\pm 0.0204$} & \makecell{$ \underline{0.85}$ \\ $\pm 0.0131$} & \makecell{$ \mathbf{0.38}$ \\ $\pm 0.0000$} & \makecell{$ 1.00$ \\ $\pm 0.0000$} & \makecell{$ 1.00$ \\ $\pm 0.0000$} & \makecell{$ 1.00$ \\ $\pm 0.0026$} & \makecell{$1.00$ \\ $\pm 0.0000$} \\ 
        \hline
        Worker Motion 2 & \makecell{$ 0.93$ \\ $\pm 0.0181$} & \makecell{$\underline{0.90}$ \\ $\pm 0.0173$} & \makecell{$ \mathbf{0.38}$ \\ $\pm 0.0000$} & \makecell{$ 1.00$ \\ $\pm 0.0000$} & \makecell{$ 1.00$ \\ $\pm 0.0000$} & \makecell{$ 1.00$ \\ $\pm 0.0026$} & \makecell{$1.00$ \\ $\pm 0.0000$} \\    
        \hline
        Worker Motion 3 & \makecell{$ 0.92$ \\ $\pm 0.0206$} & \makecell{$\underline{0.90}$ \\ $\pm 0.0137$} & \makecell{$ \mathbf{0.38}$ \\ $\pm 0.0000$} & \makecell{$ 1.00$ \\ $\pm 0.0000$} & \makecell{$ 0.99$ \\ $\pm 0.0020$} & \makecell{$ 0.98$ \\ $\pm 0.0069$} & \makecell{$0.98$ \\ $\pm 0.0000$} \\
        \hline
        Worker Motion 4 & \makecell{$ 0.91$ \\ $\pm 0.0196$} & \makecell{$\underline{0.91}$ \\ $\pm 0.0199$} & \makecell{$\mathbf{0.38}$ \\ $\pm 0.0000$} & \makecell{$ 1.00$ \\ $\pm 0.0019$} & \makecell{$ 1.00$ \\ $\pm 0.0000$} & \makecell{$ 1.00$ \\ $\pm 0.0000$} & \makecell{$1.00$ \\ $\pm 0.0000$} \\     
        \hline
        Worker Motion 5 & \makecell{$ 0.91$ \\ $\pm 0.0170$} & \makecell{$\underline{0.91}$ \\ $\pm 0.0158$} & \makecell{$ \mathbf{0.38}$ \\ $\pm 0.0000$} & \makecell{$ 1.00$ \\ $\pm 0.0000$} & \makecell{$ 1.00$ \\ $\pm 0.0000$} & \makecell{$ 0.97$ \\ $\pm 0.0118$} & \makecell{$0.99$ \\ $\pm 0.0000$} \\  
        \hline
        Exercise Motion & \makecell{$ 1.00$ \\ $\pm 0.0000$} & \makecell{$\underline{0.94}$ \\ $\pm 0.0135$} & \makecell{$ \mathbf{0.50}$ \\ $\pm 0.0000$} & \makecell{$ 1.00$ \\ $\pm 0.0023$} & \makecell{$ 1.00$ \\ $\pm 0.0000$} & \makecell{$ 0.97$ \\ $\pm 0.0000$} & \makecell{$1.00$ \\ $\pm 0.0000$} \\
        \hline
        \multicolumn{7}{c}{\scriptsize Note: Best result for each motion class in \textbf{bold} and the second best is in \underline{underline}.}
    \end{tabular}}
    \caption{Classification accuracy of the simulated sequences compared against the original sequences.}
    \label{tab:s8}
\end{table}

\noindent 2. {\bf Average Nearest Neighbor Distance}:
One can also use the average nearest neighbor distance (ANND) to compare the two distributions. The nearest neighbor distance for a simulated sequence $\alpha_i^B$ is defined as $d_i=\min_j d_{\mathcal{A}}(\alpha_i^B, \alpha_j^A)$, where $\alpha_j^A$'s are training sequences. Then the average nearest neighbor distance is $\mathrm{ANND} = (\sum_{i=1}^{N}d_i)/N$. In addition to the sequence distance $d_{\mathcal{A}}$, we defined another distance $d_M(\alpha_1,\alpha_2)=\max_t d_{\mathcal{Y}}(\alpha_1(t),\alpha_2(t))$, where $d_{\mathcal{Y}}$ is the posture distance. Thus, we can have another average nearest neighbor distance $\mathrm{ANND_M}$ using this newly defined distance $d_M$.

Tab.~\ref{tab:s9} and Tab.~\ref{tab:s10} show the ANND of sequence distance and max-posture distance. The proposed {\it SIEM/SequentialPCA/MVG} model consistently yields the best or second-best results across all motion classes. 
\begin{table}[htbp]
    \footnotesize
    \centering
    \resizebox{0.95\textwidth}{!}{%
    \setlength{\tabcolsep}{4pt}
    \begin{tabular}{|l|p{2.6cm}|p{2.4cm}|l|p{2.2cm}|p{2.2cm}|l|p{2.2cm}|} 
        \hline
        \multirow{3}{*}{Motion Type}  & \multicolumn{7}{c|}{Model} \\ 
        \cline{2-8}
        & {\it IS-TVF/Sequen-tialPCA/MVG} & {\it SIEM/Sequen-tialPCA/MVG} & \multirow{2}{*}{PWI} & {\it IS-TVF/Spa-tialPCA/VAR} & {\it SIEM/Spa-tialPCA/GP} & \multirow{2}{*}{LSTM} & GCN-Transformer\\
        \hline
        Worker Motion 1 & \makecell{$ 6.65$ \\ $\pm 0.1670$} & \makecell{$ \mathbf{4.23}$ \\ $\pm 0.0519$} & \makecell{$\underline{5.02}$ \\ $\pm 0.0033$} & \makecell{$ 8.79$ \\ $\pm 0.0692$} & \makecell{$ 5.28$ \\ $\pm 0.0097$} & \makecell{$ 5.46$ \\ $\pm 0.5270$} & \makecell{$5.36$ \\ $\pm 0.0453$} \\ 
        \hline
        Worker Motion 2 & \makecell{$ 6.70$ \\ $\pm 0.2081$} & \makecell{$ \mathbf{3.65}$ \\ $\pm 0.0319$} & \makecell{$\underline{4.42}$ \\ $\pm 0.0037$} & \makecell{$ 10.29$ \\ $\pm 0.0842$} & \makecell{$ 5.70$ \\ $\pm 0.0102$} & \makecell{$ 5.10$ \\ $\pm 0.2357$} & \makecell{$5.34$ \\ $\pm 0.0478$} \\    
        \hline
        Worker Motion 3 & \makecell{$ 6.76$ \\ $\pm 0.1746$} & \makecell{$ \mathbf{3.69}$ \\ $\pm 0.0612$} & \makecell{$ \underline{4.23}$ \\ $\pm 0.0033$} & \makecell{$ 7.32$ \\ $\pm 0.0485$} & \makecell{$ 4.70$ \\ $\pm 0.0059$} & \makecell{$ 4.38$ \\ $\pm 0.0736$} & \makecell{$ 4.59$ \\ $\pm 0.0507$} \\ 
        \hline
        Worker Motion 4 & \makecell{$ 6.69$ \\ $\pm 0.1558$} & \makecell{$ \mathbf{4.95}$ \\ $\pm 0.1134$} & \makecell{$ 5.86$ \\ $\pm 0.0076$} & \makecell{$ 8.16$ \\ $\pm 0.0670$} & \makecell{$ 5.88$ \\ $\pm 0.0053$} & \makecell{$ 5.95$ \\ $\pm 0.3693$} & \makecell{$ \underline{5.85}$ \\ $\pm 0.0420$} \\     
        \hline
        Worker Motion 5 & \makecell{$ 6.82$ \\ $\pm 0.2379$} & \makecell{$ \mathbf{3.82}$ \\ $\pm 0.0684$} & \makecell{$\underline{4.39}$ \\ $\pm 0.0028$} & \makecell{$ 7.71$ \\ $\pm 0.0294$} & \makecell{$ 5.03$ \\ $\pm 0.0061$} & \makecell{$4.69$ \\ $\pm 0.2649$} & \makecell{$4.96$ \\ $\pm 0.0459$} \\  
        \hline
        Exercise Motion & \makecell{$ 9.34$ \\ $\pm 0.0606$} & \makecell{$ \mathbf{4.38}$ \\ $\pm 0.0225$} & \makecell{$ \underline{5.11}$ \\ $\pm 0.0029$} & \makecell{$ 33.00$ \\ $\pm 0.3628$} & \makecell{$ 10.60$ \\ $\pm 0.0084$} & \makecell{$ 10.65$ \\ $\pm 0.9509$} & \makecell{$9.08$ \\ $\pm 0.6473$} \\
        \hline
        \multicolumn{7}{c}{\scriptsize Note: Best result for each motion class in \textbf{bold} and the second best in \underline{underline}.}
    \end{tabular}}
        \caption{Average nearest neighbor sequence distance of the simulated sequences compared against the original sequences.}
        \label{tab:s9}
\end{table}

\begin{table}[htbp]
    \footnotesize
    \centering
    \resizebox{0.95\textwidth}{!}{%
    \setlength{\tabcolsep}{4pt}
    \begin{tabular}{|l|p{2.6cm}|p{2.4cm}|l|p{2.2cm}|p{2.2cm}|l|p{2.2cm}|} 
        \hline
        \multirow{3}{*}{Motion Type}  & \multicolumn{7}{c|}{Model} \\ 
        \cline{2-8}
        & {\it IS-TVF/Sequen-tialPCA/MVG} & {\it SIEM/Sequen-tialPCA/MVG} & \multirow{2}{*}{PWI} & {\it IS-TVF/Spa-tialPCA/VAR} & {\it SIEM/Spa-tialPCA/GP} & \multirow{2}{*}{LSTM} & GCN-Transformer\\
        \hline
        Worker Motion 1 & \makecell{$ 11.64$ \\ $\pm 0.2587$} & \makecell{$\underline{9.23}$ \\ $\pm 0.1382$} & \makecell{$ \mathbf{9.00}$ \\ $\pm 0.0395$} & \makecell{$ 17.42$ \\ $\pm 0.1426$} & \makecell{$ 14.25$ \\ $\pm 0.0329$} & \makecell{$ 13.11$ \\ $\pm 0.6141$} & \makecell{$14.62$ \\ $\pm 0.1325$} \\ 
        \hline
        Worker Motion 2 & \makecell{$ 11.37$ \\ $\pm 0.1892$} & \makecell{$\underline{9.78}$ \\ $\pm 0.1760$} & \makecell{$ \mathbf{9.55}$ \\ $\pm 0.0832$} & \makecell{$ 17.56$ \\ $\pm 0.1035$} & \makecell{$ 17.10$ \\ $\pm 0.0514$} & \makecell{$ 12.08$ \\ $\pm 0.9709$} & \makecell{$11.16$ \\ $\pm 0.0850$} \\    
        \hline
        Worker Motion 3 & \makecell{$ 11.19$ \\ $\pm 0.2267$} & \makecell{$\underline{8.71}$ \\ $\pm 0.2330$} & \makecell{$ \mathbf{8.21}$ \\ $\pm 0.0431$} & \makecell{$ 15.97$ \\ $\pm 0.0935$} & \makecell{$ 14.30$ \\ $\pm 0.0348$} & \makecell{$ 15.15$ \\ $\pm 0.2352$} & \makecell{$14.82$ \\ $\pm 0.0873$} \\ 
        \hline
        Worker Motion 4 & \makecell{$ 11.81$ \\ $\pm 0.2552$} & \makecell{$\underline{10.55}$ \\ $\pm 0.1361$} & \makecell{$ \mathbf{10.37}$ \\ $\pm 0.0662$} & \makecell{$ 16.63$ \\ $\pm 0.1010$} & \makecell{$ 12.23$ \\ $\pm 0.0360$} & \makecell{$ 13.97$ \\ $\pm 1.1558$} & \makecell{$13.94$ \\ $\pm 0.1183$} \\     
        \hline
        Worker Motion 5 & \makecell{$ 11.13$ \\ $\pm 0.2564$} & \makecell{$ \mathbf{9.58}$ \\ $\pm 0.2735$} & \makecell{$ \underline{9.75}$ \\ $\pm 0.0245$} & \makecell{$ 16.47$ \\ $\pm 0.0984$} & \makecell{$ 12.75$ \\ $\pm 0.0272$} & \makecell{$ 12.74$ \\ $\pm 0.3481$} & \makecell{$15.20$ \\ $\pm 0.0428$} \\  
        \hline
        Exercise Motion & \makecell{$ 16.90$ \\ $\pm 0.1585$} & \makecell{$ \underline{13.22}$ \\ $\pm 0.1025$} & \makecell{$ \mathbf{12.99}$ \\ $\pm 0.0430$} & \makecell{$ 48.41$ \\ $\pm 0.2879$} & \makecell{$ 16.52$ \\ $\pm 0.0588$} & \makecell{$ 21.98$ \\ $\pm 1.2006$} & \makecell{$ 16.56$ \\ $\pm 0.4328$} \\
        \hline
        \multicolumn{7}{c}{\scriptsize Note: Best result for each motion class in \textbf{bold} and the second best in \underline{underline}.}
    \end{tabular}}
        \caption{Average nearest neighbor max-posture distance of the simulated sequences compared against the original sequences.}
        \label{tab:s10}
\end{table}

\noindent 3. {\bf Roughness}: 
The original motion sequences are naturally temporally smooth as they are generated by human workers. To evaluate simulated sequences, we can compute a measure of roughness using the shape differences between successive postures, $R(t)=d_{\mathcal{Y}}(\alpha(t),\alpha(t+1))$. We can compare roughness of two sequences using the $\mathbb{L}^2$ norm  $d_R(\alpha_1,\alpha_2)=\|R_{\alpha_1}-R_{\alpha_2}\|$. Then, we can use the average nearest neighbor distance of the roughness difference, $\mathrm{ANND}_R$, to compare the similarity between the original and simulated sequences.

Tab.~\ref{tab:s11} shows $\mathrm{ANND}_R$ between original and simulated sequences. The proposed {\it IS-TVF/SequentialPCA/MVG} and {\it SIEM/SequentialPCA/MVG} models achieve the best results. In contrast, the PWI, LSTM, and GCN-Transformer models yield higher roughness distances, indicating that they fail to capture the temporal pattern of the motion sequences.

\begin{table}[htbp]
    \footnotesize
    \centering
    \resizebox{0.95\textwidth}{!}{%
    \setlength{\tabcolsep}{4pt}
    \begin{tabular}{|l|p{2.6cm}|p{2.4cm}|l|p{2.2cm}|p{2.3cm}|l|p{2.3cm}|} 
        \hline
        \multirow{3}{*}{Motion Type}  & \multicolumn{7}{c|}{Model} \\ 
        \cline{2-8}
        & {\it IS-TVF/Sequen-tialPCA/MVG} & {\it SIEM/Sequen-tialPCA/MVG} & \multirow{2}{*}{PWI} & {\it IS-TVF/Spa-tialPCA/VAR} & {\it SIEM/Spa-tialPCA/GP}& \multirow{2}{*}{LSTM} & GCN-Transformer\\
        \hline
        Worker Motion 1 & \makecell{$ \mathbf{15.55}$ \\ $\pm 0.2587$} & \makecell{$ \underline{16.07}$ \\ $\pm 0.2051$} & \makecell{$ 82.19$ \\ $\pm 0.1216$} & \makecell{$ 21.23$ \\ $\pm 0.0818$} & \makecell{$ 19.94$ \\ $\pm 0.0048$} & \makecell{$ 22.62$ \\ $\pm 0.1410$} & \makecell{$ 21.95$ \\ $\pm 2.1124$} \\ 
        \hline
        Worker Motion 2 & \makecell{$ \mathbf{12.45}$ \\ $\pm 0.1521$} & \makecell{$ \underline{13.10}$ \\ $\pm 0.1501$} & \makecell{$ 73.47$ \\ $\pm 0.1055$} & \makecell{$ 19.12$ \\ $\pm 0.0453$} & \makecell{$ 15.44$ \\ $\pm 0.0103$} & \makecell{$ 19.25$ \\ $\pm 0.0761$} & \makecell{$22.73$ \\ $\pm 2.8559$} \\   
        \hline
        Worker Motion 3 & \makecell{$ \mathbf{15.89}$ \\ $\pm 0.1230$} & \makecell{$ \underline{16.38}$ \\ $\pm 0.1564$} & \makecell{$ 66.28$ \\ $\pm 0.1068$} & \makecell{$ 20.59$ \\ $\pm 0.0484$} & \makecell{$ 22.94$ \\ $\pm 0.0087$} & \makecell{$ 27.16$ \\ $\pm 0.0168$} & \makecell{$20.22$ \\ $\pm 1.2009$} \\
        \hline
        Worker Motion 4 & \makecell{$ \mathbf{17.64}$ \\ $\pm 0.2318$} & \makecell{$ \underline{18.48}$ \\ $\pm 0.2299$} & \makecell{$ 106.52$ \\ $\pm 0.1598$} & \makecell{$ 24.99$ \\ $\pm 0.0910$} & \makecell{$ 26.21$ \\ $\pm 0.0076$} & \makecell{$30.78$ \\ $\pm 0.0857$} & \makecell{$ 25.68$ \\ $\pm 1.5239$} \\     
        \hline
        Worker Motion 5 & \makecell{$ \mathbf{13.93}$ \\ $\pm 0.2069$} & \makecell{$ \underline{15.37}$ \\ $\pm 0.2377$} & \makecell{$ 74.72$ \\ $\pm 0.1521$} & \makecell{$ 18.22$ \\ $\pm 0.0414$} & \makecell{$ 18.45$ \\ $\pm 0.0097$} & \makecell{$ 21.12$ \\ $\pm 0.2334$} & \makecell{$25.839$ \\ $\pm 1.6471$} \\  
        \hline
        Exercise Motion & \makecell{$ \underline{33.74}$ \\ $\pm 0.1440$} & \makecell{$ \mathbf{33.14}$ \\ $\pm 0.2474$} & \makecell{$ 126.33$ \\ $\pm 0.1514$} & \makecell{$ 78.10$ \\ $\pm 3.3141$} & \makecell{$ 65.02$ \\ $\pm 0.0136$} & \makecell{$ 63.692 $ \\ $\pm 6.2893$} & \makecell{$64.58$ \\ $\pm 18.9613 $} \\
        \hline
        \multicolumn{7}{c}{\scriptsize Note: Best result for each motion class in \textbf{bold} and the second best in \underline{underline}.}
    \end{tabular}}
    \caption{Average nearest neighbor of roughness distances of the simulated sequences compared against the original sequences.}
    \label{tab:s11}
\end{table}

\subsection{Method Ranking Aggregation}\label{subsecs4-3}
This section presents methodology and final results for comparing our proposed methods against the baseline methods using an aggregate ranking system. To ensure a balanced evaluation, we selected four representative metrics: Energy Distance, capturing statistical similarity; Jerk Energy, capturing dynamics; Posture Validity Score, capturing physical plausibility; and Quantization Variability, capturing the quality of the underlying tasks.

The final ranking table shown in Tab.~\ref{tab:s12} is computed using a hierarchical aggregate system:
\begin{enumerate}
    \item \textbf{Individual Ranking}: For each motion class, all the methods are ranked from 1 (best) to 7 (worst) independently for each selected metric as reported in the subscripts of Tab.~\ref{tab:s1}, \ref{tab:s3}, \ref{tab:s5}, and~\ref{tab:s7}.
    \item \textbf{Dataset-Level Averaging}: These four metric ranks were averaged to generate a single comprehensive rank per method for each motion class, as shown in the first six rows of Tab.~\ref{tab:s12}.
    \item \textbf{Global Average}: The last row of Tab.~\ref{tab:s12} shows the overall mean ranks for each method across all motion classes.
\end{enumerate}

\begin{table}[htbp]
    \footnotesize
    \centering
    \resizebox{0.95\textwidth}{!}{%
    \setlength{\tabcolsep}{4pt}
    \begin{tabular}{|l|p{2.6cm}|p{2.4cm}|l|p{2.2cm}|p{2.3cm}|l|p{2.3cm}|} 
        \hline
        \multirow{3}{*}{Motion Type}  & \multicolumn{7}{c|}{Model} \\ 
        \cline{2-8}
        & {\it IS-TVF/Sequen-tialPCA/MVG} & {\it SIEM/Sequen-tialPCA/MVG} & \multirow{2}{*}{PWI} & {\it IS-TVF/Spa-tialPCA/VAR} & {\it SIEM/Spa-tialPCA/GP}& \multirow{2}{*}{LSTM} & GCN-Transformer\\
        \hline
        Worker Motion 1 & $3.50$ & $\mathbf{1.75}$ & $4.50$ & $ 4.00$ & $4.75$ & $4.50$ & $5.00$ \\ 
        \hline
        Worker Motion 2 & $4.00$ & $\mathbf{2.00}$ & $4.25$ & $4.00$ & $5.25$ & $3.75$ & $4.75$ \\   
        \hline
        Worker Motion 3 & $4.75$ & $ \mathbf{2.00}$ & $ 4.75$  & $ 4.00$ & $ 4.75$ & $4.00$ & $ 3.75$ \\
        \hline
        Worker Motion 4 & $3.00$ & $\mathbf{1.50}$  & $4.50$ & $3.50$ & $5.75$ & $4.25$ & $5.50$ \\     
        \hline
        Worker Motion 5 & $4.00$ & $\mathbf{1.25}$  & $4.25$ & $5.00$  & $5.25$ & $3.75$ & $4.50$ \\  
        \hline
        Exercise Motion & $3.75$ & $ \mathbf{1.50}$  & $3.75$ & $ 5.50$ & $5.25$ & $4.50$ & $3.75$ \\
        \hline
        Average Rank & $3.83$ & $\mathbf{1.67}$ & $4.33$ & $4.33$ & $5.17$ & $4.13$ & $4.54$ \\
        \hline
        \multicolumn{7}{c}{\scriptsize Note: Best result for each motion class in \textbf{bold}.}
    \end{tabular}}
    \caption{Average ranks of the comparison methods.}
    \label{tab:s12}
\end{table}

To verify the statistical validity of these rankings. We conducted a Friedman's test on the ranks across all the selected metrics ($6 \text{ Datasets }\times 4 \text{ Metrics}$). The test strongly rejects the null hypothesis ($\chi^2(6)=37.88,\ p<0.001$), indicating that the performance differences between the models are statistically significant. Subsequently, a Nemenyi post-hoc test for pairwise comparisons with strict multiplicity corrections is performed using an error rate of $\alpha=0.05$. The calculated Critical Difference (CD) is $1.84$. The proposed {\it SIEM/SequentialPCA/MVG} achieves an average rank of $1.67$, demonstrating a statistically significant improvement over all baseline methods. 

\color{black}
\section{Sensitivity Analysis}\label{secs5}
To investigate the stability of the proposed IS-TVF and SIEM representation, we conduct a sensitivity test using synthetic random noise. This test assesses how each representation handles noisy data.

\subsection{Experiment Setup}

We simulate potential signal errors by perturbing the clean skeleton sequences (denoted as $\{\alpha_0\}$) with additive white Gaussian noise to create a noisy dataset (denoted as $\{\alpha_n\}$). Each noisy posture $\mathbf{Y}_n=\alpha_n(t)$ is constructed element-wise as:
\[\mathbf{Y}_n=\exp_{\mathbf{Y}_0}(\mathbf{V}),\,\mathbf{V}_{i,j}\sim\mathcal{N}(\mathbf{0},\sigma^2)\]
where $\sigma$ is around $5\%$ perturbation the the average landmarks scales.

In real-world scenarios, noise is often heteroscedastic, i.e., it depends on movement characteristics, such as range of motion and velocity. However, given the high accuracy of our datasets, the exact structure of potential environmental noise remains latent. To simplify the analysis, we apply a universal random noise across all temporal steps and skeleton landmarks. Although this setup provides a rigorous evaluation of mathematical robustness, we note that sensitivity to such uncorrelated noise does not necessarily imply performance degradation in the presence of structured, real-world noise. 

The dimension reduction stage in our pipeline can serve as a noise reduction step as well. If latent coefficients remain stable despite input perturbations, the subsequent generative models remain unaffected. Therefore, we use the reconstruction errors as the evaluation metric for this sensitivity test. We first train the dimension reduction models exclusively on the clean data $\{\alpha_0\}$. These models are then applied to both clean and noisy datasets to generate the reconstructed sequences $\{\alpha_0'\}$ from the original clean data and $\{\alpha_n'\}$ from noisy data. We assess the robustness by comparing the average baseline errors $d_{\mathcal{A}}(\alpha_0, \alpha_0')$ with the average noisy reconstruction error $d_{\mathcal{Y}}(\alpha_0, \alpha_n')$. We also report the reconstruction errors for the other dimension-reduction methods discussed earlier.

\subsection{Experiment Results}

As shown in Tab.\ref{tab:s13}. The SIEM representation exhibits significant resilience to the added noise. The average reconstruction error for the noisy data remains nearly identical to the baseline.  In contrast, the IS-TVF representation exhibits greater sensitivity to unstructured white noise. Because the IS-TVF representation uses velocity and integral, it relies on the integrity of the local temporal patterns. High-frequency white noise is amplified during this transformation process.
These findings suggest that while the SIEM representation, as a simpler method, is more prone to geometric distortion, it is more robust to noisy observations.

\begin{table}[]
    \scriptsize
    \centering
    \begin{tabular}{c|c|c|c|c|c|c}
        \hline       
         \multirow{2}{*}{Method} & \multicolumn{3}{c|}{ISTVF} & \multicolumn{3}{c}{SIEM} \\
         \cline{2-7}
         & $d_{\mathcal{A}}(\alpha_0, \alpha_0')$ & $d_{\mathcal{A}}(\alpha_0, \alpha_n')$ & $\Delta$ & $d_{\mathcal{A}}(\alpha_0, \alpha_0')$ & $d_{\mathcal{A}}(\alpha_0, \alpha_n')$ & $\Delta$\\
         \hline
         PCA+FPCA($5\times10$) & 4.96 & 5.63 & 0.6708 & 3.00 & 3.01 & 0.0007\\
         MPCA($85\%$ Variance)  & 4.78 & 5.47 & 0.6893 & 2.46 & 2.46 & 0.0014 \\
         AE+FPCA($5\times10$) & 5.05 & 5.60 & 0.5490 & 3.02 & 3.03 & 0.0010 \\
         VAE+FPCA($5\times10$) & 4.59 & 5.31 & 0.7256 & 2.91 & 2.91 & 0.0000\\
         \hline
         \multicolumn{5}{c}{The mean distance between the original and noisy data $d_{\mathcal{S}}(\alpha_0, \alpha_n)$ is 4.20}
    \end{tabular}
    \caption{Comparison of the Mean Reconstruction Error Using the Original and Noisy Data. The differences $\Delta$ represent the sensitivity to the noise.}
    \label{tab:s13}
\end{table} 

\color{black}
\section{Videos of the Simulated Motion}\label{secs6}
The supplementary material contains videos of the simulated motion. For the Worker Motion, the naming of each video follows the rule of: {\it[{\it model}]\_Motion[$X$]\_[$Y$].mp4}, where {\it model} is the simulation model including {\it ISTVF} (stands for {\it IS-TVF/SequentialPCA/MVG}),  {\it SIEM} (stands for {\it SIEM/SequentialPCA/MVG}), {\it PWI}, {\it VAR} (stands for {\it IS-TVF/SpatialPCA/VAR}, {\it GP} (stands for {\it SIEM/SpatialPCA/GP}), {\it LSTM}, and {\it GCN-Trans} (stands for {\it GCN-Transformer}), $X$ is the number of the motion class, and $Y$ is the index of the simulated motion. {\it E.g.}, the file {\it ISTVF\_Motion1\_1.mp4} is the first simulated motion of the motion class one generated by the {\it IS-TVF/SequentialPCA/MVG} model. For the Exercise Motion, the naming of each video follows a similar pattern: {\it[{\it model}]\_ExerciseMotion\_[$Y$].mp4}, where {\it model} is the simulation model and $Y$ is the index of the simulated motion. Each video contains two motions: on the left (black) is the mean motion of the corresponding motion class, and on the right (blue) is the simulated motion. For {\it IS-TVF/SequentialPCA/MVG} and {\it SIEM/SequentialPCA/MVG} models, we use a set of $10\times 30$ sequential PCA coefficients.

\end{document}